# Nonstationary Spatial Process Models with Spatially Varying Covariance Kernels

Sébastien Coube-Sisqueille[a,1] *, Sudipto Banerjee[b,2], and Benoît Liquet[a,c,3]

[a] Laboratoire de Mathématiques et de leurs Applications,
Université de Pau et des Pays de l'Adour, E2S-UPPA, Pau, France

[b] Department of Biostatistics, University of California, Los Angeles, United States of America

[c] School of Mathematical and Physical Sciences, Macquarie University, Sydney, Australia

[1] sebastien.coube@univ.pau.fr [2] sudipto@ucla.edu [3] benoit.liquet-weiland@mq.edu.au

**Abstract**

Building spatial process models that capture nonstationary behavior while delivering computationally efficient inference is challenging. Nonstationary spatially varying kernels (see, e.g., Paciorek, 2003) offer flexibility and richness, but computation is impeded by high-dimensional parameter spaces resulting from spatially varying process parameters. Matters are exacerbated if the number of locations recording measurements is massive. With limited theoretical tractability, obviating computational bottlenecks requires synergy between model construction and algorithm development. We build a class of scalable nonstationary spatial process models using spatially varying covariance kernels. We implement a Bayesian modeling framework using Hybrid Monte Carlo with nested interweaving. We conduct experiments on synthetic data sets to explore model selection and parameter identifiability, and assess inferential improvements accrued from nonstationary modeling. We illustrate strengths and pitfalls with a data set on remote sensed normalized difference vegetation index.

# 1 Introduction

Bayesian hierarchical models for analyzing spatially and temporally oriented data are widely employed in scientific and technological applications in the physical, environmental and health sciences (Cressie and Wikle, 2015; Banerjee et al., 2014; Gelfand et al., 2019). Such

*The authors gratefully acknowledge *E2S: Energy and Environment Solutions*

models embed a spatial process within a hierarchical structure, which specifies a joint probability law for the data, an underlying spatial process, and the parameters. The process introduces spatial and/or temporal dependence, allows inference on the underlying data-generating mechanism, and yields probabilistic predictions over spatial domains. Point-referenced spatial data, which are our focus, refer to measurements at locations with fixed coordinates that are assumed to arise as a partial realization of a spatial process over the finite set of locations. Stationary Gaussian processes are a conspicuous choice, but impose simpler dependence structures, which are unlikely in most scientific applications.

Nonstationary spatial models attempt to relax such assumptions and deliver wide-ranging inferential benefits. If spatial variation is posited as a complex function of space composed of multiple locally varying processes, richer and more informative covariance structures in nonstationary processes, while adding complexity, can improve smoothing, goodness of fit and predictive inference. A comprehensive review of diverse approaches for nonstationary spatial modeling is beyond the scope of a single article (see, e.g., Higdon, 1998; Fuentes, 2002; Paciorek, 2003; Banerjee et al., 2008; Cressie and Johannesson, 2008; Lemos et al., 2009; Guhaniyogi et al., 2013; Yang and Bradley, 2021; Risser and Calder, 2015; Risser, 2016; Fuglstad et al., 2015a; Gelfand et al., 2010, and references therein). The richness of nonstationary models have been exemplified in the above references, with a key underlying theme being the use of basis function representations to capture nonstationarity.

Nonstationary process models are deterred by theoretical intractability so computational feasibility and efficiency are crucial considerations. Basis representations yielding low-rank nonstationary processes assist with dimension reduction, but cannot afford spatially varying behavior in the covariance kernel itself. Alternatively, richer models attempting to exploit information from predictors in spatial covariance matrices (see, e.g.,

Pourahmadi, 1999; Risser and Calder, 2015) have been extended and further developed in a number of different directions including regularized inference, but they do not scale to massive spatial data sets in the order of $10^5$+ locations. Recent developments have addressed data sets in the order of hundreds (Ingebrigtsen et al., 2015; Heinonen et al., 2016) or thousands (Guhaniyogi et al., 2013; Fuglstad et al., 2015a) of locations, but these are modest with respect to the size of commonly encountered spatial data (see the examples in Datta et al., 2016; Zhang et al., 2019; Heaton et al., 2019; Katzfuss and Guinness, 2021; Peruzzi et al., 2022). Bakka et al. (2019) proposed a novel computationally efficient "barrier" nonstationary model by using simultaneous autoregression represented through a stochastic partial differential equation. Nonstationary models also suffer from unidentifiable parameters arising from complex space-varying covariance kernels, which yield weakly identifiable models that are difficult to estimate and interpret. This also complicates model evaluation and selection as inference is very sensitive to the specifications of the model.

We devise a new class of arbitrarily rich nonstationary spatial models for massive data sets using spatially-varying Matérn covariance kernels (Paciorek, 2003; Guhaniyogi et al., 2011; Kleiber and Nychka, 2012). These are extremely rich models as they endow random fields on the spatial covariance kernel parameters, but are notoriously challenging to fit. We obviate these challenges by endowing the space-varying parameters with a low-dimensional Gaussian process, while the spatial process itself is modeled using directed acyclic graphs (DAGs) such as the Nearest Neighbor Gaussian Process (NNGP) (Datta et al., 2016) and, more generally, Vecchia processes (Katzfuss and Guinness, 2021). This achieves sparser representations than Paciorek (2003) or Risser and Calder (2015) while endowing the variances and the spatial range parameters with their own processes. We embed these processes within a coherent Bayesian hierarchical modeling framework (Heinonen et al., 2016), but tailor our method towards modeling large data sets and capturing local anisotropy.

Learning about the spatial covariance kernel processes is challenging and cannot be addressed solely using theoretical results (asymptotic theory for nonstationary processes are hardly accessible) or solely relying upon sophisticated algorithms (achieving convergence in high-dimensional parameter spaces will be unfeasible). Hence, we rely upon some novel consequences of sparser representations in conjunction with a Hamiltonian Monte Carlo (HMC) algorithm adapted from Heinonen et al. (2016) to overcome computational impediments and analyze spatial data at massive scales. More specifically, we hybridize the approach in Heinonen et al. (2016) with the interweaving strategies of Yu and Meng (2011) and Filippone et al. (2013). We achieve probabilistic inference on a latent nonstationary process which captures residual spatial behavior after accounting for spatial predictors. Our framework includes simpler models as special cases. While theoretical properties of model selection based upon nonstationary features are analytically inaccessible, we conduct an array of computer experiments on synthetic data to demonstrate that our devised sparsity-inducing process models do not over-fit data generated from a simpler model. Instead, they degenerate towards a state corresponding to a simpler model.

The balance of the article is as follows. Section 2 outlines the covariance and data models, and the properties of a nonstationary NNGP density, which are incorporated into a Bayesian hierarchical model presented in Section 3. Section 4 estimates the model based upon a Gibbs sampler using interweaving (Section 4.2) and upon a Hamiltonian Monte Carlo algorithm (Section 4.3). Section 5 focuses on applications: experiments on synthetic data to test the properties of the model and analysis of NDVI data. Section 6 summarizes and points to future research. An elaborate accompanying Supplement offers key derivations, further details on algorithms and numerical results from simulation experiments.

## 2 Spatially-Varying Covariance Kernel Models

Let $\mathcal{S} = \{s_1, s_2, \ldots, s_n\}$ be a set of $n$ spatial locations indexed in a spatial domain $\mathcal{D}$, where $\mathcal{D} \subset \mathbb{R}^d$ with $d \in \{1, 2, 3\}$. For any $s \in \mathcal{D}$ we envision a spatial regression model

$$z(s) = x(s)^{\mathrm{T}}\beta + w(s) + \epsilon(s) \ , \qquad (1)$$

where $z(s)$ represents a measurement at $s$, $x(s)$ is a $p \times 1$ vector of predictors, $\beta$ is the corresponding $p \times 1$ vector of slopes, $w(s)$ is a latent spatial process and $\epsilon(s)$ is white noise attributed to random disturbances. In full generality, the noise will be modeled as heteroskedastic so that $\epsilon(s) \overset{ind}{\sim} \mathcal{N}(0, \tau^2(s))$ while $w(s)$ is customarily modelled using a Gaussian process over $\mathcal{D}$. Therefore,

$$w(\mathcal{S}) := (w(s_1), w(s_2), \ldots, w(s_n))^{\mathrm{T}} \sim \mathcal{N}(0, \Sigma(\mathcal{S})) \ , \qquad (2)$$

where the elements of the $n \times n$ covariance matrix $\Sigma(\mathcal{S})$ are determined from a spatial covariance function $K(s, s')$ defined for any pair of locations $s$ and $s'$ in $\mathcal{D}$. In full generality the covariance function can accommodate spatially varying parameters to obtain nonstationarity, which is our current focus. Therefore, the $(i, j)$-th element of $\Sigma(\mathcal{S})$ is

$$\Sigma(s_i, s_j) = K(s_i, s_j) = \sigma(s_i)\sigma(s_j)K_0(s_i, s_j; \alpha(s_i), \alpha(s_j)), \qquad (3)$$

where $\sigma(s_1 \ldots s_n) := \{\sigma(s_i) : i = 1, \ldots, n\}$ is a collection of (positive) spatially varying marginal standard deviations, $K_0(s, s'; \{\alpha(s), \alpha(s')\})$ is a valid spatial correlation function defined for any pair of locations $s$ and $s'$ in $\mathcal{D}$ with two spatial range parameters $\alpha(s)$ and $\alpha(s')$ that vary with the locations. For two sets of locations $s \in \mathcal{S}$ and $s' \in \mathcal{S}$, $\Sigma(s, s')$ denotes the submatrix comprising the rows and columns of $\Sigma$ whose indices correspond to $s$ and $s'$ in $\mathcal{S}$, respectively. We also abbreviate $\Sigma(s, s)$ into $\Sigma(s)$. These parameters can be either positive-definite matrices offering a locally anisotropic nonstationary covariance structure or positive real numbers specifying a locally isotropic nonstationary range. For example, Paciorek (2003) proposed a valid class of nonstationary covariance functions

$$K_0(s,s';A(s),A(s')) = \frac{2^{d/2}|A(s)|^{1/4}|A(s')|^{1/4}}{|A(s)+A(s')|^{1/2}} K_i\left(d_M\left(s,s',(A(s)+A(s'))/2\right)\right), \quad (4)$$

where $A(s)$ and $A(s')$ are $d \times d$ anisotropic spatially-varying range matrices, $d$ is dimension of $\mathcal{D}$, $d_M(\cdot,\cdot,\cdot)$ denotes Mahalanobis distance and $K_i$ is an isotropic correlation function. If $A(\cdot)$ does not vary by location, the covariance structure is anisotropic but stationary. A nonstationary, locally isotropic correlation function is obtained by setting $A(s) = \alpha(s)I_d$,

$$K_0(s,s';\alpha(s),\alpha(s')) = \left(\frac{\sqrt{2}\alpha(s)^{1/4}\alpha(s')^{1/4}}{(\alpha(s)+\alpha(s'))^{1/2}}\right)^d K_i\left(d_E(s,s')/\left((\alpha(s)+\alpha(s'))/2\right)\right), \quad (5)$$

where $d_E(\cdot,\cdot)$ is the Euclidean distance (Mahalanobis distance with matrix $I_d$).

Spatial process parameters in isotropic covariance functions are not consistently estimable in fixed-domain asymptotic paradigms (Stein, 1999; Zhang, 2004; Tang et al., 2021). Therefore, irrespective of sample size, no function of the data can converge in probability to the value of the parameter from an oracle model. Irrespective of how many locations we sample, the effect of the prior on these parameters will not be eliminated in Bayesian inference. This can be addressed using penalized complexity priors to reduce the ridge of the equivalent range-marginal variance combinations to one of its points (Fuglstad et al., 2015b). The covariance function sharply drops to 0 so the observations that inform about the covariance parameters at a location tend to cluster around the site. Nonstationary models are significantly more complex. The parameters specifying the spatial covariance are functions over $\mathcal{D}$. These form uncountable collections and, hence, inference requires modeling them as spatial processes. This considerably exacerbates the challenges surrounding identifiability and inference for these completely unobserved processes. Asymptotic inference is precluded due to the lack of regularity conditions. Bayesian inference, while offering fully model-based solutions for completely unobserved processes, will also need to obviate the computational hurdles arising from weakly identified processes, which result in poorly behaved MCMC algorithms, and inference extensible to massive data sets.

The NNGP (Datta et al., 2016) specifies a Gaussian process in two steps. We begin with a "parent" process $GP(0, K(s, s'))$ so that $w(\mathcal{S})$ has the probability law in (2). Let $f(w(\mathcal{S}) \,|\, \theta)$ be the corresponding density, where $\theta$ denotes the parameters in $K(\cdot,\cdot)$. We build a sparse approximation of this joint density using a fixed topological order of the points in $\mathcal{S}$. We construct a nested sequence $\mathcal{S}_{i-1} \subset \mathcal{S}_i$, where $\mathcal{S}_i = \{s_1, s_2, \ldots, s_{i-1}\}$ for $i = 2, 3, \ldots, n$. The density of the NNGP is $\tilde{f}(w(\mathcal{S}) \,|\, \theta) = f(w(s_1) \,|\, \theta) \prod_{i=2}^{n} \tilde{f}(w(s_i) \,|\, w(\mathcal{S}_{i-1}), \theta)$ (also referred to as Vecchia's approximation Vecchia, 1988; Stein et al., 2004), where

$$\tilde{f}(w(s_i) \,|\, w(\mathcal{S}_{i-1}), \theta) = f(w(s_i) \,|\, w(pa(s_i)), \theta), \tag{6}$$

and $pa(s_i)$ comprises the parents of $s_i$ from a DAG over $\mathcal{S}$. The parents in the DAG are often chosen as the nearest neighbors of $s_i$ among $s_1, \ldots, s_{i-1}$. This approximation defines a Gaussian Process whose covariance matrix can be expressed as $(\tilde{R}^{\mathrm{T}}\tilde{R})^{-1}$. Each row of $\tilde{R}$ is computed straightforwardly from the conditional distribution $f(w(s_i) \,|\, w(pa(s_i)), \theta)$. It follows that $\tilde{R}$ is lower triangular and very sparse with its $i$-th row having nonzero entries only at the indices corresponding to $pa(s_i)$ (Pourahmadi, 1999, 2007; Datta et al., 2016; Katzfuss and Guinness, 2021). Prediction at unobserved locations is achieved by appending them sequentially as new nodes with edges connecting the new locations to its nearest neighbors from the preceding nodes already in the DAG. This results in the predictive distribution $\prod_{i=1}^{n_{pred}} f(w(\tilde{s}_i) \,|\, w(pa(\tilde{s}_i)), \theta)$, where $\tilde{s}_i$ are a topologically sorted sequence of locations where predictions are sought (see, e.g., Datta et al., 2016; Finley et al., 2019; Katzfuss et al., 2020, for scalable spatial interpolation and prediction).

We obtain $\tilde{f}(w(\mathcal{S}) \,|\, \theta(\mathcal{S})) = f(w(s_1) \,|\, \theta(s_1)) \prod_{i=2}^{n} \tilde{f}(w(s_i) \,|\, w(\mathcal{S}_{i-1}), \theta(\mathcal{S}_i))$ applying (6)

to spatially varying parameters, where $\theta(s)$ is a vector comprising the parameters of the covariance function in $s$ and $\theta(\mathcal{S}) = (\theta(s_1)^{\mathrm{T}}, \theta(s_2)^{\mathrm{T}}, \ldots, \theta(s_n)^{\mathrm{T}})^{\mathrm{T}}$ is a $(\sum_{i=1}^{n} |\theta(s_i)|) \times 1$ vector. We build scalable nonstationary processes by exploiting the following key property (see Section S1.1 of the Supplement for the derivation),

$$\tilde{f}(w(s_i) \mid w(\mathcal{S}_{i-1}), \theta(\mathcal{S})) = f(w(s_i) \mid w(pa(s_i)), \theta(s_i \cup pa(s_i))) \ , \tag{7}$$

where the NNGP density is derived using kernels (4) and (5). Both accommodate spatially varying parameters $\theta(\mathcal{S})$. Equation (7) reveals a crucial property that the NNGP reduces the dimension of conditional sets from $w(\mathcal{S}_{i-1})$ to $w(pa(s_i))$ and from $\theta(\mathcal{S})$ to $\theta(s_i \cup pa(s_i))$.

Another useful property relates the NNGP derived from (3) to its corresponding correlation function. Let $\tilde{R}_0(\mathcal{S})$ be the NNGP factor obtained from the precision matrix using the correlation function $K_0(\cdot)$ instead of the covariance function $K(\cdot)$ and let $\sigma(\mathcal{S}) = (\sigma(s_1), \sigma(s_2), \ldots, \sigma(s_n))^{\mathrm{T}}$. Then, $\tilde{R}(\mathcal{S}) = \tilde{R}_0(\mathcal{S})\mathrm{diag}(\sigma(\mathcal{S}))^{-1}$, where $\mathrm{diag}(\sigma(\mathcal{S}))$ is the diagonal matrix with elements $\sigma(s_i)$; see the Section S1.2 of the Supplement for the derivation. The logarithm $N(w(\mathcal{S}) \mid 0; \tilde{\Sigma}(\mathcal{S}, \theta))$ is computed up to a constant as

$$-\sum_{i=1}^{n} \log\left( (\tilde{R}_0(\mathcal{S}))_{i,i} / \sigma(s_i) \right) - \frac{1}{2} w^{\mathrm{T}} \mathrm{diag}(\sigma(\mathcal{S}))^{-1} \tilde{R}_0(\mathcal{S})^{\mathrm{T}} \tilde{R}_0(\mathcal{S}) \mathrm{diag}(\sigma(\mathcal{S}))^{-1} w, \tag{8}$$

exploiting cheap expressions for the determinant of $\tilde{R}(\mathcal{S})$. Furthermore, the topological ordering heuristics explored by Guinness (2018) hold for NNGP specifications in (4) and (5).

# 3 Hierarchical space-varying covariance models

## 3.1 General hierarchical framework

We build a hierarchical space-varying covariance model over a set of spatial locations $\mathcal{S} = \{s_1, \ldots, s_n\}$ with at least one observation per spatial location, but possibly more, so that $|z| \geq n$, where $|z|$ denotes the number of elements in $z := z(\mathcal{S}) = (z(s_1)^{\mathrm{T}}, \ldots, z(s_n)^{\mathrm{T}})^{\mathrm{T}}$ and $z(s_i)$ is a column vector (can be $1 \times 1$) of measurements at location $s_i$. The matrix $M$ is a $|z| \times n$ matching matrix with elements $M_{i,j} = 1$ if the $i$-th element of $z$ corresponds to the $j$-th spatial location and 0 otherwise. We preclude an observation from being obtained in two spatial locations at the same time. Hence, each row of $M$ has *exactly* one term equal to one. Also, since there is at least one observation in each location, each column of $M$ has *at least* one term equal to one. Henceforth, we do not index $z$, $\tau$, and $X$ on the spatial locations in $\mathcal{S}$ to ease the notation. We propose the following hierarchical framework,

$$\underbrace{[z \mid \beta, w(\mathcal{S}), \tau^2]}_{(a)} \times \underbrace{[w(\mathcal{S}) \mid \theta(\mathcal{S})]}_{(b)} \times \underbrace{[\theta(\mathcal{S}) \mid \beta_\theta, \gamma_\theta]}_{(c) \text{ and } (d)} \times \underbrace{[\tau^2 \mid \beta_\tau, \gamma_\tau]}_{(e) \text{ and } (f)} \times \underbrace{[\gamma_\theta, \gamma_\tau, \beta_\theta, \beta_\tau]}_{(g)}, \tag{9}$$

where $[\cdot]$ denotes probability distributions. These symbolic blocks are modeled as below:

$$\begin{aligned}
&(a)\ z \sim N(X\beta + Mw(\mathcal{S}), \mathrm{diag}(\tau^2))\ ; \quad (b)\ w(\mathcal{S}) \sim N(0, \tilde{\Sigma}(\mathcal{S}; \theta(\mathcal{S})))\ ;\\
&(c)\ \log(\theta(\mathcal{S})) = X_\theta(\mathcal{S})\beta_\theta + W_\theta(\mathcal{S})\ ; \quad (d)\ W_\theta(\mathcal{S}) \sim N(0, \zeta(\mathcal{S}; \gamma_\theta))\ ;\\
&(e)\ \log(\tau^2(\mathcal{S})) = X_\tau\beta_\tau + MW_\tau(\mathcal{S})\ ; \quad (f)\ W_\tau(\mathcal{S}) \sim N(0, \zeta(\mathcal{S}; \gamma_\tau))\ ;\\
&(g)\ \{\gamma_\theta, \gamma_\tau, \beta_\theta, \beta_\tau, \beta\} \sim p(\cdot, \cdot, \cdot, \cdot, \cdot)\ .
\end{aligned} \tag{10}$$

In (10($a$)) $z$ is the $|z| \times 1$ vector of all measurements, $X$ is $|z| \times p$, $\beta$ is the corresponding vector of regression coefficients, and $\mathrm{diag}(\tau^2)$ is a diagonal matrix with elements of the vector $\tau^2$ along the diagonal. Equation (10($a$)) yields (1) with $|z| = n$ and $M$ as a permutation matrix. In (10($b$)), the distribution of $w(\mathcal{S})$ is specified using (8) as described in Section 2.

Equations (10($c$)) and (10($e$)) specify models for $\theta(s)$ and $\tau^2(s)$, respectively. The vector $\tau^2(\mathcal{S})$ in the fully heteroskedastic specification is $|z| \times 1$, with elements corresponding to $z$. The process $\theta(s)$ realized over $\mathcal{S}$ is $\theta(\mathcal{S}) \in \{\alpha(\mathcal{S}), \sigma^2(\mathcal{S})\}$, where $\alpha(\mathcal{S})$ and $\sigma^2(\mathcal{S})$ are the realizations of $\alpha(s)$ and $\sigma^2(s)$ from Equation (3), respectively. The realization $\sigma^2(\mathcal{S}) = (\sigma^2(s_1), \ldots, \sigma^2(s_n))^{\mathrm{T}}$ is $n \times 1$, while $\alpha(\mathcal{S})$ follows from two special cases in our development: (i) $\alpha(s)$ is a scalar as in the locally isotropic case in (5) whence $\alpha(\mathcal{S})$ is $n \times 1$; and (ii) $\alpha(s) := A(s)$ in (4) whence $(10(c))$ is a matrix-variate linear model. We expound on these two cases in Sections 3.2 and 3.3, respectively. Our model allows additional flexibility not considered here (such as, e.g., Allard et al., 2022, for varying Matérn smoothness).

Spatial fields for $\theta(\mathcal{S})$ and $\tau(\mathcal{S})$ are specified by $W_\theta(\mathcal{S})$ and $W_\tau(\mathcal{S})$ in $(10(d))$ and $(10(f))$ with covariance kernels $\zeta(\cdot;\gamma_\theta)$ and $\zeta(\cdot;\gamma_\tau)$, respectively. The objects $\zeta_\theta(\mathcal{S};\gamma_\theta)$ and $\zeta_\tau(\mathcal{S};\gamma_\tau)$ are the corresponding $n \times n$ spatial covariance matrices. Here, $\gamma_\theta$ and $\gamma_\tau$ denote the set of all covariance kernel parameters, but here we simplify by specifying $\zeta(\cdot;\gamma_\theta) = \gamma_\theta\zeta(\cdot)$ and $\zeta(\cdot;\gamma_\tau) = \gamma_\tau\zeta(\cdot)$, where $\gamma_\theta$ and $\gamma_\tau$ are scalars (marginal variances) and $\zeta(\mathcal{S})$ an $n \times n$ correlation matrix (see Section 3.2). The matrix $X_\tau$ has $|z|$ rows and explains variation in $\tau^2(\mathcal{S})$, while $X_\theta(\mathcal{S})$ has $n$ rows corresponding to the spatial sites and ensures that $w(s)$ is well-identified. In $(10)(e)$, $W_\tau(\mathcal{S})$ is multiplied by $M$ to match observation indices.

We specify $\{\beta, \beta_\theta, \beta_\tau\}$ in (10) using independent Gaussian densities $N(\cdot \mid \mu, V)$, with mean $\mu$ and covariance matrix $V$. The logarithms of the marginal variance parameters $\gamma_\theta$ and $\gamma_\tau$ are assigned uniform priors $U(\cdot,\cdot)$ with known limits. The density $p$ in $(10(g))$ is a product of the prior densities $U(\gamma_\theta \mid a_{\gamma_\theta}, b_{\gamma_\theta}) \times U(\gamma_\tau \mid a_{\gamma_\tau}, b_{\gamma_\tau}) \times N(\beta \mid \mu_\beta, V_\beta) \times N(\beta_\theta \mid \mu_{\beta_\theta}, V_{\beta_\theta}) \times N(\beta_\tau \mid \mu_{\beta_\tau}, V_{\beta_\tau})$. The hyper-parameters in the prior distributions are fixed.

### 3.2 Specification for spatially varying covariance parameters

The distributions for $W_\theta(\mathcal{S})$ and $W_\tau(\mathcal{S})$ in $(10(c))$ and $(10(e))$ require attention. We expound the case of $W_\theta(\mathcal{S})$ with that of $W_\tau(\mathcal{S})$ being analogous. The NNGP in Section 2 is impeded by slow moving Monte-Carlo schemes due to the high dimension of the resulting field. An alternative is to introduce a low-rank process for $\theta(s)$, where $\theta(s)$ is a placeholder for $\alpha(s)$ and $\sigma^2(s)$. Thus, we specify (10)($d$) as

$$W_\theta(\mathcal{S}) = B(\mathcal{S})u_\theta\sqrt{\gamma_\theta}, \text{ inducing } W_\theta(\mathcal{S}) \overset{\text{a priori}}{\sim} \mathcal{N}(0, \gamma_\theta B(\mathcal{S})B(\mathcal{S})^{\mathrm{T}}) \tag{11}$$

where $W_\theta(\mathcal{S})$ is $n \times 1$, $B(\mathcal{S})$ is $n \times r$ whose elements are spatial basis functions with $B(\mathcal{S})_{i,j}$ the value of the $j^{th}$ function at the $i^{th}$ site of $\mathcal{S}$, $u_\theta$ is $r \times 1$ with $r << n$ basis coefficients with $u_\theta \overset{\text{a priori}}{\sim} \mathcal{N}(0, I_r)$, achieving dimension reduction and $\gamma_\theta$ is the scalar variance component. Hence, $\zeta(\mathcal{S}) = B(\mathcal{S})B(\mathcal{S})^{\mathrm{T}}$ is rank deficient and produces low-dimensional singular distributions for $W_\theta(\mathcal{S})$ and $W_\tau(\mathcal{S})$. Constructing $B(\mathcal{S})$ from a correlation function implies that $\gamma_\theta$ is the marginal variance. Our algorithms (Supplement S4) circumvent simulating directly from $W_\theta$. Instead, we update $u_\theta$ and then compute $W_\theta$ from (11). This choice of parametrization is further discussed in Supplement S3.3.

We construct the Gaussian predictive process (PP) implicitly from an NNGP, using fast sparse triangular solving instead of dense matrix multiplication; the construction of such an implicit PP is detailed in Supplement S3.1. Knot selection for PPs has been discussed in Banerjee et al. (2008), Finley et al. (2009), Guhaniyogi et al. (2011) and Gelfand et al. (2012). Knots are placed with K-means clustering, the user specifying the desired number of knots and the chosen Matérn range for the PP, as detailed in Supplement S3.2. Our package vignette explains the behavior of PP samples using our spatial plotting functions.

Explanatory variables $X_\theta(s) \in \{X_\alpha(s), X_{\sigma^2}(s)\}$ and $X_\tau(s)$ drive nonstationary behavior in $\theta(s)$ and $\tau(s)$ with fixed effects $\beta_\theta$ and $\beta_\tau$, respectively. If such variables are absent, $X_\theta$ and $X_\tau$ consist of an intercept only. We assign Gaussian priors for the associated regression coefficients. The structure of random effects is similar with $B$ being fixed and $u_\theta$ being unknown. The prior variance parameters of $W_\theta$ and $W_\tau$ are $\gamma_\theta$ and $\gamma_\tau$, respectively and are estimated while the prior variance of $\beta_\theta$ and $\beta_\tau$ is fixed.

## 3.3 Extension to anisotropic range parameters

For anisotropic range parameters in (4), $\alpha(s) = A(s)$ is $d \times d$ and needs to be modeled ensuring $A(s)$ is positive definite. We use a matrix logarithm (or exponential) to achieve this (Higham, 2008; Banerjee and Roy, 2014). We model $\log(A(s))$ in (10) as

$$\log(A(s)) = \text{SYMMAT}\left(\left(X_A(s)^{\mathrm{T}}\beta_A + W_A(s)^{\mathrm{T}}\right) H_d\right) \,, \tag{12}$$

where $X_A(s)^{\mathrm{T}}$ is a $1 \times p_{X_A}$ vector of possible explanatory variables for the range, $\beta_A$ is $p_{X_A} \times d(d+1)/2$ and $W_A(s)^{\mathrm{T}}$ is $1 \times d(d+1)/2$, so that $X_A(s)^{\mathrm{T}}\beta_A + W_A(s)^{\mathrm{T}}$ is $1 \times d(d+1)/2$, $s \in \mathbb{R}^d$, and $H_d$ is $d(d+1)/2 \times d(d+1)/2$. The function $\text{SYMMAT}(\cdot)$ is the inverse of the half-vectorization operator $\text{vech}(\cdot)$ that maps the elements of the vector in its argument to a $d \times d$ symmetric matrix by filling the elements in its upper and lower triangles.

The symmetry of $\log(A(s))$ leads to the positive-definiteness of $A(s)$. Since $\log A(s)$, as constructed above, is symmetric, we can write $\log(A(s)) = P(s)\Lambda(s)P(s)^{\mathrm{T}}$ using its spectral decomposition, where $\Lambda(s)$ is diagonal consisting of eigenvalues $\lambda_1(s), \lambda_2(s), \ldots, \lambda_d(s)$ and $P(s)$ is an orthogonal matrix of eigenvectors. From the definition of the matrix exponential, $A(s) = P(s)\exp(\Lambda(s))P(s)^{\mathrm{T}}$, where $\exp(\Lambda(s))$ is diagonal with elements $\exp(\lambda_i(s)) > 0$ being the eigenvalues of $A(s)$, thereby ensuring positive definiteness. The matrix logarithm also satisfies $(\log(A(s))^{-1}) = -\log(A(s))$, which is convenient for parametric specifications.

For easier interpretation, we apply a change of basis to the half-vectorization of $\log(A(s))$ using $H_d$. The first row of $H_d$ is $\text{vech}(I_{d\times(d+1)/2}/\sqrt{d \times (d+1)/2})$ and the other rows are obtained by applying $\text{vech}(\cdot)$ to any completion of basis of the symmetric matrices from $I_{d\times(d+1)/2}/\sqrt{d \times (d+1)/2}$. The first coefficient of $X_A(s)\beta_A + W_A(s)$ controls $\det(A(s))$, which yields the product of the range ellipse's axis lengths, and will inflate and deflate the range of the ellipse without changing its shape. The other components rotate and squash the ellipse without changing the product of the lengths of its axes. The above specification implies that these three components are linked to the same set of covariates $X_A$. We expand the PP prior from (11) to multivariate settings by defining $W_A(\mathcal{S})$ as a matrix of size $n \times d(d+1)/2$ with $W_A(\mathcal{S})_{i,\cdot} = W_A(s_i)$, for $1 \leq i \leq |\mathcal{S}|$. The left-hand side of (11) is modified as $W_A(\mathcal{S}) = B(\mathcal{S})u_A \text{chol}(\gamma_A)^{\mathrm{T}}$ with $\text{chol}(\gamma_A)\text{chol}(\gamma_A)^{\mathrm{T}} = \gamma_A$, where $\gamma_A$ is now a $d(d+1)/2 \times d(d+1)/2$ covariance matrix, $\text{chol}(\gamma_A)$ is the lower-triangular Cholesky factor of $\gamma_A$ and $u_A$ is $r \times d(d+1)/2$ with $\text{vec}(u_A) \overset{\text{a priori}}{\sim} \mathcal{N}(0, I_{rd(d+1)/2})$. Building $B(\mathcal{S})$ from a spatial correlation function interprets $\gamma_A = \text{Cov}(W_A(s))$. We persist with the hyperpriors presented earlier. A Gaussian prior is assigned to $\beta_A$, and Uniform priors with positive bounded support to the eigenvalues of $\gamma_A$. As in Section 3.2, we update $\text{vec}(u_A)$ to avoid working with the singular distribution, $\text{vec}(W_A) \overset{\text{a priori}}{\sim} \mathcal{N}(0, \gamma_A \otimes (B(\mathcal{S})B(\mathcal{S})^{\mathrm{T}}))$, where $\otimes$ is the Kronecker product. See Figure S1 in Section S3.4 of the Supplement for illustrations.

### 3.4 Model specification

There is little to argue about the complexity encoded in (9) and (10). Inference problems in even simpler stationary models are encumbered by well-known issues of model identifiability (Stein, 1999; Zhang, 2004; Tang et al., 2021). These issues are exacerbated in nonstationary models with almost no theoretical tractability, so it is crucial to model the latent processes with reasonable priors. Although a non-informative prior in $\beta$ is reasonable, informative priors for $\beta_\theta$ and $\beta_\tau$ are desirable based on $X_\theta$ and $X_\tau$. Even without such information, we are able to demonstrate fairly robust inference with only intercepts in these matrices (see Sections 5.1 and 5.3). The fixed effects and the spatially driven effects for the range and the field's marginal variance cannot be independent. Indeed, some spatial incoherence of these effects is very likely to cause identifiability issues between the parameter and the latent fields.

## 3.5 Posterior predictive inference

Predictive inference is achieved by sampling from the posterior predictive distribution

$$\begin{aligned}[Z(s_0), w(s_0), \theta(s_0), \tau(s_0) \,|\, z] = \int \Big\{ &[Z(s_0) \,|\, \beta, w(s_0), \tau(s_0)] \times [w(s_0) \,|\, \theta(s_0 \cup \mathcal{S}), w(\mathcal{S})] \\ \times [\theta(s_0) \,|\, u_\theta, \beta_\theta, \gamma_\theta] \times [\tau(s_0) \,|\, u_\tau, \beta_\tau, \gamma_\tau] &\times [\beta, u_\theta, \beta_\theta, \gamma_\theta, u_\tau, \beta_\tau, \gamma_\tau, w(\mathcal{S}) \,|\, z] \Big\},\end{aligned} \tag{13}$$

where $Z(s_0)$ is the unmeasured outcome, and $\{w(s_0), \theta(s_0), \tau(s_0)\}$ are values of the respective latent processes at $s_0 \notin \mathcal{S}$. Integration is with respect to all variables in the integrand that do not appear on the left.

The stochastic processes in (10) and (11) ensure that (13) is a legitimate probability distribution. This permits predictive inference for the outcome $Z(s_0)$ and the underlying spatial process $w(s_0)$, as in customary Bayesian geostatistics, but also for the spatially varying parameters $\theta(s_0)$ and $\tau(s_0)$. For each draw of $[\beta, u_\theta, \beta_\theta, \gamma_\theta, u_\tau, \beta_\tau, \gamma_\tau, w(\mathcal{S}) \,|\, z]$ obtained using the algorithm devised in Section 4, we draw a $\tau(s_0) \sim [\tau(s_0) \,|\, u_\tau, \beta_\tau, \gamma_\tau]$, a $\theta(s_0) \sim [\theta(s_0) \,|\, u_\theta, \beta_\theta, \gamma_\theta]$, and a $w(s_0) \sim [w(s_0) \,|\, w(\mathcal{S}), \theta(s_0 \cup \mathcal{S})]$.

For the low-rank processes in Section 3.2, $\zeta_\theta(\mathcal{S}; \gamma_\theta)$ (or $\zeta_\tau(\mathcal{S}; \gamma_\tau)$) are singular matrices so the conditional distributions $[\theta(s_0) \,|\, u_\theta, \beta_\theta, \gamma_\theta]$ and $[\tau(s_0) \,|\, u_\tau, \beta_\tau, \gamma_\tau]$ are well-defined in a lower-dimensional subspace. Here $\theta(s)$ and $\tau(s)$ are completely specified by the basis functions and their coefficients. Therefore, we directly compute $\log \theta(s_0) = x(s_0)^{\mathrm{T}} \beta_\theta + b(s_0; \gamma_\theta)^{\mathrm{T}} u_\theta$ for every posterior sample of $\{\beta_\theta, \gamma_\theta, u_\theta\}$ to obtain the posterior samples for $\theta(s_0)$. Posterior predictive samples of $\log \tau(s_0)$ are computed analogously.

Sampling from $[w(s_0) \,|\, w(\mathcal{S}), \theta(s_0), \Omega_\theta]$ is also straightforward as it is Gaussian and depends only on $\{w(pa(s_0)), \theta(s_0 \cup pa(s_0)), \Omega_\theta\}$ (recall the discussion on the predictive distribution for the NNGP described below (6)). Once the aforementioned samples are drawn, we finally draw $Z(s_0) \sim N(x(s_0)^{\mathrm{T}} \beta + w(s_0), \tau^2(s_0))$ for each sampled value of $\{\beta, w(s_0), \tau^2(s_0)\}$, thereby achieving full predictive inference from (13).

# 4 Markov chain Monte Carlo algorithms

## 4.1 Outline

We devise a Gibbs sampler, where each parameter is updated conditionally on the rest, which is further improved by introducing appropriate updating methods built from other works or developed in this article. The latent field $w(\mathcal{S})$ is sampled using the blocked update (Datta et al., 2016, Section 4.1) in order to mitigate spatial auto-correlation. The regression coefficients $\beta$ are updated using the "Interweaving" method proposed in Coube-Sisqueille and Liquet (2021). The low-rank formulation of $W_\theta(\mathcal{S})$ and $W_{\tau(\mathcal{S})}$ ensures that only the respective basis coefficients $u_\theta$ and $u_\tau$ need to be updated. These are block-sampled with their respective counterparts $\beta_\theta$ and $\beta_\tau$. The basis coefficients for $W_\theta(\mathcal{S})$ are updated using an interweaving strategy with respect to the latent field $w(\mathcal{S})$ inspired from (Filippone et al., 2013); see Section 4.2. In both cases, Hybrid Monte Carlo or HMC (see Section 4.3) is used yielding, respectively, a "HMC-within-interweaving" for $u_\theta$ and $\beta_\theta$ and "HMC-within-Gibbs" for $u_\tau$ and $\beta_\tau$. The higher-level parameters $\{\gamma_\theta, \gamma_\tau\}$, are successively updated using an interweaving strategy on their respective respective couples of latent fields parametrizations, $u_\theta$ and $W_\theta(\mathcal{S})$ on one hand, $u_\tau$ and $W_\tau(\mathcal{S})$ on the other hand. In the case of $\gamma_\theta$, this interacts with the interweaving on $w(\mathcal{S})$. The resulting algorithm is outlined in Section S4 of the Supplement.

## 4.2 Ancillary-Sufficient Interweaving Strategy

The problem of high correlation between latent fields and higher-level parameters is familiar in stationary spatial models, and several solutions exist such as blocking (Knorr-Held and Rue, 2002), collapsing (Finley et al., 2019) or Interweaving (Filippone et al., 2013).

Interweaving exploits the discordance between two parametrizations of a latent field to sample high-level parameters. When these two parametrizations are an ancillary-sufficient couple, we have an Ancillary-Sufficient Interweaving Strategy (AS-IS); see Yu and Meng (2011) and also Section S5.1 in the Supplement.

We interweave the whitened and natural parametrizations of the latent field $w(\mathcal{S})$ from (10($a$)) in order to update the higher-level parameters $u_\theta$ and $\beta_\theta$ impacting the covariance structure decomposed in (10($c$)). The so-called natural parametrization of the latent field is found in the decomposition (10($a$)) and is a sufficient parametrization. The whitened parametrization of the latent field is ancillary and is obtained by multiplying the natural parametrization with the right NNGP factor of its prior precision matrix from (10($b$)) $w^*(\mathcal{S}) = \tilde{R}(\mathcal{S})w(\mathcal{S})$. Hence, the prior distribution (10($b$)) is standard normal. The covariance parameters have no effect on the prior distribution of the latent field. In turn, they acquire a role in the decomposition of the data. In (10($a$)), $w(s_i)$ is replaced by the $i$-th element of $\tilde{R}(\mathcal{S})^{-1}w^*(\mathcal{S})$, while (10($b$) )is replaced by $w^*(\mathcal{S}) \sim \mathcal{N}(0, I_{|\mathcal{S}|})$.

The parameter of interest is sampled in two steps, one for each parametrization of the latent field. These individual steps can be full conditional draws, random walk Metropolis steps or, as in our case, Hybrid Monte-Carlo (HMC) steps. Hence, two potentials will be derived for the HMC steps of each parameter. Note that $\gamma_\theta$ and $\gamma_\tau$ are themselves marginal variance parameters for the latent fields $W_\theta(\mathcal{S})$ (10($d$)) and $W_\tau(\mathcal{S})$ (10($f$)). In order to update $\gamma_\theta$ and $\gamma_\tau$, interweaving is used by treating $W_\theta(\mathcal{S})$ and $W_\tau(\mathcal{S})$ as latent fields and using their respective whitened parametrizations $u_\theta$ and $u_\tau$. In the case of $\gamma_\theta$, nested interweaving (see S5.2 in the Supplement) allows us to use both whitened-natural couples parametrizations, $u_\theta$ and $W_\theta(\mathcal{S})$ one the one hand, $w^*(\mathcal{S})$ and $w(\mathcal{S})$ on the other, leading to 4 combinations.

### 4.3 Hybrid Monte-Carlo

Hybrid Monte-Carlo (HMC) (Neal et al., 2011) has been implemented successfully by Heinonen et al. (2016) for nonstationary Gaussian processes. We differ in two aspects. First, we use NNGP instead of full GP in (10($b$)) and, hence, differentiate the NNGP-induced potential, which is non-trivial. Second, we use an "HMC within AS-IS" algorithm and must find the gradients of the potential for the covariance parameters using both ancillary and sufficient parametrizations. We jointly update the Predictive Process coefficients and the regression coefficients $\beta_\tau$ and $u_\tau$ in (10($e$)), $\beta_\theta$ and $u_\theta$ in (10($c$)), or $\beta_A$ and $u_A$ in (12). This method is especially useful for the range parameters since it avoids a computationally onerous Metropolis-within-Gibbs sweep over $\beta_\theta$. The method applies to $\theta$ or $\tau$, so we use "$\lambda$" as a portmanteau. We differentiate $H = -g(\beta_\lambda, u_\lambda) - \log(N(\beta_\lambda, \mu_{\beta_\lambda}, V_{\beta_\lambda})) - \log(N(u \,|\, 0, I_r)_\lambda)$, where $g(\cdot,\cdot)$ depends on the role of $\lambda$ in the model and the chosen parameterization of the latent field in interweaving; the Normal prior distributions of $\beta_\lambda$ and $u_\lambda$ are introduced in 3.1. We express $g(\beta_\lambda, u_\lambda) = h \circ \log(\lambda)(\beta_\lambda, u_\lambda)$, $\log(\lambda)(\cdot,\cdot)$ being a function that links $\beta_\lambda, u_\lambda$ and the logarithm of $\lambda$ by transforming $u_\lambda$ into $W_\lambda(\mathcal{S})$ using (11), and plugging the resulting field into (10($c$)), (10($e$)), or (12). The Jacobian chain rule on $g(\beta_\lambda, u_\lambda)$ renders

$$\nabla_{\beta_\lambda, u_\lambda} H = -\nabla_{\beta_\lambda} \log(N(\beta_\lambda \,|\, \mu_\lambda, V_\lambda)) - \nabla_{u_\lambda} \log(N(u_\lambda \,|\, 0, I_r))) - J^{\mathrm{T}}_{\beta_\lambda, u_\lambda} \log(\lambda) \cdot \nabla_{\log(\lambda)} h(\log(\lambda)).$$

Recognizing that $\log(\lambda)$ varies linearly with respect to $\beta_\lambda$ and $u_\lambda$, we get $J^{\mathrm{T}}_{\beta_\lambda, u_\lambda} \log(\lambda) = (X_\lambda | B)^{\mathrm{T}}$. The cost of gradient computation depends on the specific parameter. While the noise variance boils down to a product of independent densities, the range and marginal variance of the latent field involve $\tilde{R}(\mathcal{S})$ for multiplication and solving. These operations, which would be unfeasible with dense matrices, remain efficient because of the triangular and sparse $\tilde{R}(\mathcal{S})$ (Datta et al., 2016). Furthermore, (8) allows for efficient computation of the gradient for parameters related to the field's marginal variance. For the range, (7) is critical to establish the fact that one range parameter only affects its children in the DAG, allowing for sparse differentiation, in synergy with the computation of the coefficients of $\tilde{R}(\mathcal{S})$. Overall, the cost of gradient computation is linear in the number of observations. The details for computing the gradient of $h_\lambda$ are supplied in Section S6 of the Supplement.

# 5 Applications of the model

Our model and codes are available from the public repository `https://github.com/SebastienCoube/Nonstat-NNGP` and includes an extensive vignette for users.

## 5.1 Synthetic experiments for model specification

We conduct a collection of several synthetic experiments to evaluate whether the benefits of using a richer covariance structure can be confirmed, and to explore the potential pitfalls of the devised architecture with respect to model choice and parameter identification.

We generate data sets that range from stationary specifications to those that exhibit varying degrees of non-stationarity using (10). We randomly sample 12,000 locations on a square with side length 5 using a Uniform distribution. We fix the regression coefficients in (10) $(a)$, $(c)$, and $(e)$ and let $X$ consist only of an intercept. We fix $\beta = 0$ and $\beta_\tau = 0$ in $(10(a))$ and $(10(e))$, respectively. In the stationary configuration, the noise and the latent fields have the same marginal variances equal to $\exp(0) = 1$. For $\beta_\theta$ in $(10(c))$, the intercept coefficient for the range is set to $\log(0.1) \approx -2.3$ while that for smoothness is 0.5. In the anisotropic cases, the intercept is set to 0. These parameters are the same for all runs, including our nonstationary examples. The marginal variance parameters $\gamma_\theta, \gamma_\tau$ from $(10(d))$ and $(10(f))$ change. We set them to 0 to generate data without spatially varying kernels. For nonstationary data, they are set to 1. In the case of data with locally varying anisotropic covariance, $\gamma_A = I_3$. The remaining parameters are generated from respective prior distributions (d), (f), (b) and (a) in (10).

We subsequently fit several candidate models to the synthetic data we generated. We consider four situations: (i) "under-modeling" when the model is less complex than the data generating process; (ii) "over-modeling" which is the opposite of (i); (iii) the so-called "correct" (or "right") modeling, when the specifications of the model and the data are the same; and (iv) models that fall in none of the above three categories. Several indicators were monitored: the Deviance Information Criterion (DIC) (Spiegelhalter et al., 2002), the mean square error (MSE) of the predicted field at unobserved sites, the MSE of the smoothed field at observed sites, and the MSE of the smoothed covariance parameters. Further details are supplied in Section S7 of the Supplement.

Since stationary models emerge as special cases of nonstationary models, we can informally detect over-modeling from MCMC chains without requiring full convergence by looking at the samples of the GP variances $\gamma_\sigma$, $\gamma_\tau$, and $\gamma_\alpha$ or $\gamma_A$. Supplement S7.5 presents a case study with trace plots generated automatically during MCMC runs.

The locally isotropic range and the heteroskedastic marginal variance of $w(\mathcal{S})$ are poorly identified. If we apply a mis-specified model with one nonstationary parameter to a data set generated with the other parameter as nonstationary, then we still obtain inference competitive with the true model. We see this also in data where both parameters are nonstationary and we specify only one as nonstationary. If none are specified as nonstationary when at least one ought to be, the performance deteriorates (see Supplement S7.4).

It is clear that the loss of performance caused by using a stationary model (the bottom rows in the tables in Section S7 of the Supplement) is usually less than the loss caused by a poorer model formulation (in red in the tables) or other under-fitting cases (in gray).

From these results, we proffer that a nonstationary formulation should always accommodate nonstationary noise variance and either a nonstationary variance or range for the latent field. Since the range exhibits slower MCMC convergence than the marginal variance, a "one-size-fits-most" approach would be to model heteroskedasticity in the noise, and either nonstationary range and marginal variance in the latent field. When local anisotropy is suspected, the costlier anisotropic model may prove useful as illustrated in 5.3.

## 5.2 Comparisons with alternate methods on synthetic data

We compared our nonstationary NNGP with three other viable alternatives to analyze spatial data: (i) INLA-SPDE (Lindgren and Rue, 2015); (ii) local GPs (Gramacy and Apley, 2015; Gramacy, 2016; Binois et al., 2018); and (iii) deep Vecchia GPs (Sauer et al., 2023). We implement these methods using the `R` programs available on some simple synthetic nonstationary data sets. Two types of non-stationary data sets were considered: one in which the range of the GP $w(\mathcal{S})$ varies following (5), and one where the Gaussian noise is heteroskedastic. In both cases, the parameter fields were generated from (9) and (10), inducing smoothly varying fields of the covariance parameters. As in Section 5.1, $10,000$ observations were used for training. Fifty experiments were replicated in each of the two cases. Results for the nonstationary models for the noise variance or the latent process range are presented in Tables 1 and 2 respectively, and details are given in Section S8.1 and S8.2 of the Supplement respectively.

We report eight comparative metrics: (i) a smoothing mean squared error (MSE) based upon the values of the process at observed locations in the training data; (ii) a predictive MSE based upon the values of the process at hold out locations; (iii) a parameter-retrieval MSE that evaluates the ability of the methods (except deep GP) to retrieve the values of the field of non-stationary parameters generating the process; (iv) smoothing coverage intervals; (v) predictive coverage intervals; (vi) smoothing empirical log point-wise predictive density (elpd) ; (vii) predictive elpd ; and (viii) the running time for each method. Deep GP precludes retrieval of estimates for the fields of spatially-variable covariance parameters and, hence, is excluded from (iii) while featuring in the remaining comparisons. The elpd is derived from Vehtari et al. (2017), details about the particular implementations to the considered methods being given in Section S9.1 of the Supplement. The elpd being a criterion based on probabilistic density, higher values denote better performance.

In summary, while nonstationary NNGP excels in terms of MSEs and elpds, INLA achieves comparable performances with a smaller running time. While these two methods clock running times in minutes, local GP is able to deliver inference in seconds but with somewhat poorer MSE and elpd scores. The relatively high running time of INLA on data exhibiting heteroskedastic noise is explained by the fact that we had to include all observations in the triangular mesh in order to compute the elpd. Deep GPs, while having commendable accuracy in the retrieval of the latent field, requires considerable computational resources and also exhibits somewhat unreliable MCMC behavior. The deep GP, nonetheless, offers excellent smoothing and prediction for on the model with non-stationary latent range, while being out-performed on the data with heteroskedastic noise, accordingly to its formulation who combines a complex deep GP and a homoskedastic noise. All methods have similar performance regarding coverage intervals, with a slight excess of width on observed data, and good coverage on predicted data.

Table 1: Mean performance of several models on data with heteroskedastic noise.

| | smooth MSE | pred MSE | noise log-var MSE | obs cover | pred cover | obs elpd | pred elpd | time |
|---|---|---|---|---|---|---|---|---|
| INLA | 0.07 | 0.08 | 0.04 | 0.98 | 0.95 | -0.86 | -1.05 | 13 min |
| nonstat NNGP | 0.07 | 0.08 | 0.03 | 0.98 | 0.94 | -0.91 | -1.05 | 23 min |
| deepGP | 0.17 | 0.16 | | 0.97 | 0.94 | -0.93 | -1.19 | 17 h |
| local GP | 0.18 | 0.19 | 0.29 | 0.96 | 0.95 | -1.13 | -1.17 | 10 s |

Table 2: Mean performance of several models on data with nonstationary range.

| | smooth MSE | pred MSE | log range MSE | obs cover | pred cover | obs elpd | pred elpd | time |
|---|---|---|---|---|---|---|---|---|
| INLA | 0.12 | 0.16 | 0.17 | 0.97 | 0.94 | -0.97 | -1.09 | 2 min |
| nonstat NNGP | 0.07 | 0.12 | 0.05 | 0.98 | 0.94 | -0.86 | -1.05 | 25 min |
| local GP | 0.24 | 0.25 | 4.20 | 0.96 | 0.95 | -1.10 | -1.14 | 7 sec |
| deepGP | 0.09 | 0.14 | | 0.98 | 0.95 | -0.90 | -1.07 | 17 h |

## 5.3 Scientific data analysis

We analyze a normalized difference vegetation index (NDVI) data set drawn from Zhang and Banerjee (2022), which contains over a million observations that exhibit anisotropy and also indicate spatially varying (heteroskedastic) noise. We test several models combining homoskedasctic and heteroskedastic noise in $(10(e))$; and stationary and nonstationary, isotropic and anisotropic latent field as in $(10(c))$.

We reduce the effective dimension of the data set by rounding coordinates: we train the model on $920,358$ observations on $124,575$ distinct locations since several locations had multiple observations mapped to them after the rounding of coordinates. The reduced dimension of the latent field allows, as in triangulation of the spatial domain (Lindgren et al., 2011), more elaborate analysis while retaining all observations.

We execute an NNGP with 10 neighbors and the max-min ordering with smoothness $\nu = 1.5$. The nonstationary parameters are modeled using spatial effects as in $(10(e))$ and $(10(f))$ using predictive processes with the Matérn smoothness 1.5, range 0.08 and 300 knots. We also implemented an INLA-SPDE algorithm (Rue and Held, 2005; Lindgren and Rue, 2015; Bakka et al., 2019) and a local GP (Gramacy and Apley, 2015; Gramacy, 2016) as a benchmark, although neither the local GP nor the open version of R-INLA can model anisotropy (see, e.g., Fuglstad et al., 2015a, for some internal version implementations). Table 3 synthesizes the performances of these runs.

Table 3: Model comparisons for the NDVI analysis.

| | a[1] | r[2] | n[3] | train elpd | loo elpd | lump elpd | train cover | loo cover | lump cover | loo MSE | lump MSE | time |
|---|---|---|---|---|---|---|---|---|---|---|---|---|
| nngp | yes | yes | yes | 0.21 | 0.05 | -0.27 | 0.96 | 0.94 | 0.95 | 0.09 | 0.14 | 25h 26min |
| nngp | yes | no | yes | 0.2 | 0.03 | -0.33 | 0.96 | 0.95 | 0.95 | 0.10 | 0.15 | 9h 12min |
| nngp | no | no | yes | 0.2 | 0 | -0.47 | 0.96 | 0.94 | 0.95 | 0.10 | 0.18 | 4h 4min |
| nngp | no | no | no | -0.04 | -0.26 | -0.57 | 0.95 | 0.93 | 0.96 | 0.10 | 0.17 | 3h 29min |
| nngp | no | yes | yes | 0.21 | 0.01 | -0.41 | 0.96 | 0.94 | 0.96 | 0.10 | 0.17 | 7h 55min |
| liGP | no | yes | yes | -0.87 | -0.22 | -0.85 | 0.83 | 0.98 | 0.99 | 0.11 | 0.24 | 0h 2min |
| inla | no | no | no | -0.04 | -0.27 | -0.58 | 0.96 | 0.93 | 0.93 | 0.10 | 0.18 | 0h 8min |

1 : anisotropic range 2 : nonstationary range 3 : nonstationary noise

All runs were executed on the UCLA Hoffman2 computing cluster with 30 threads and an allocated RAM of 120 gigabytes hosted in the Institute of Digital Research and Education at the University of California, Los Angeles. We used Gelman-Rubin-Brooks diagnostics (Gelman et al., 1992; Brooks and Gelman, 1998) and Effective Sample Sizes (ESS) for the high-level parameters (all the parameters of the models except the latent field $w(\mathcal{S})$) to assess convergence of our MCMC implementations for the nonstationary NNGP models. For all models, all ESS for high-level parameters were above 100. These results were based upon 2000 iterations for all models, although some models converged earlier, incurring an effective running time ranging from three hours to a day depending upon the specific covariance structure in use. The stationary INLA delivered, as expected, very expeditious output, but INLA struggled with the nonstationary models and was unable to deliver inferential output even after attempting to use Empirical Bayes integration and restricting the dimension of the PP basis from 300 to 50 in order to reduce the computational burden resulting from the high-dimensional hyperparameter space. This is unexpected in view of the good performance of INLA on synthetic data sets. This problem, we opine, is attributable to the fact that our spatial region is massive, which, in turn, introduces a massive number of hyper-parameters. In addition, we have a very large number of observations encumbering the SPDE for the Gaussian noise as it attempts to model every observation. According to our experience on laptops, this is likely attributable to consumption of (vast) available RAM. To summarize, INLA outperforms our proposed MCMC-based implementation in stationary and isotropic settings, while our method appears more viable for nonstationary modeling with complex covariance patterns (e.g., accommodating anisotropy and heterokedastic noise).

We compared the models by partitioning the data into a training set and test set to account for interpolation using pseudo leave-one-out (LOO) and for mid-range prediction using lump removal. Details are supplied in Section S9 of the Supplement. Several crite-ria are used, including Empirical Logarithm Point-wise Predictive Density (elpd) following Vehtari et al. (2017), where larger values of elpd being preferable, as well as 95 percent posterior coverage intervals and MSE of predictions against real observations. See Section S9.1 of the Supplement for further details. Coverage seems unaffected by the model choice, but we can nonetheless note that the predictive distributions inferred from local GPs tend to depart from the 0.95 coverage objective.

Stationary INLA and NNGP exhibit almost identical performance in terms of elpd , coverage, or MSE. While local GP, which has both nonstationary GP range and noise (it is not easy to specify a simpler model), delivers reasonable predictive performance, its elpd scores are somewhat weaker than the other two methods ; its coverage intervals are too wide, for both observed and predicted data ; and its MSE for the lump test data set is the highest of all methods. Adding some heteroskedasticity in the noise variance of the NNGP gives a boost in all components of the elpd. Introducing anisotropy, in turn, primarily improves the lump elpd and the lump MSE as well, perhaps because the center of a lump is then predicted using the "right" observations laying outside the lump in the direction of the anisotropy, while the training and LOO observations are already predicted reasonably well. Finally, using nonstationarity in the anisotropy improves the performances again, while a model with nonstationary range, without local anisotropy, delivers impaired inferential performance. We refer to Section S9 of the supplement.

The nonstationary anisotropic model is also able to recover the spatially varying range and anisotropy from the observed data. Figure 1 summarizes the results from the full nonstationary model. Its first component, Figure 1(a) presents the raw observations. Figure 1(b) depicts the range ellipses along with the mean latent field. The ellipses are obtained by averaging the matrix logarithms who parametrize the range (see equation (12)) over the

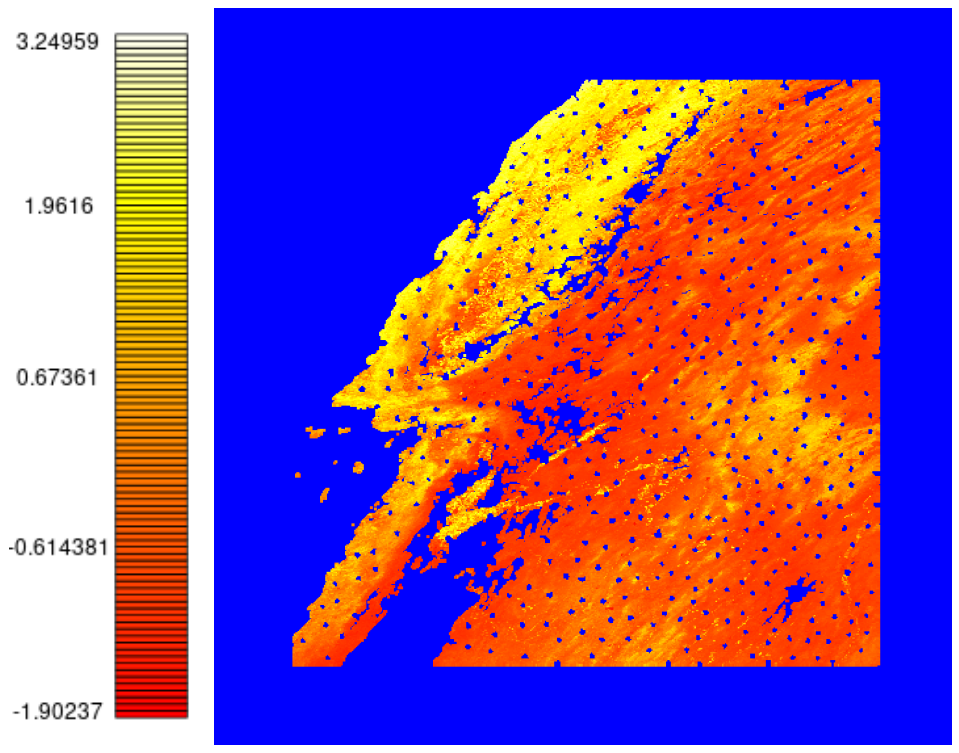

(a) Observed NDVI (gaps in the bulk were colored in blue)

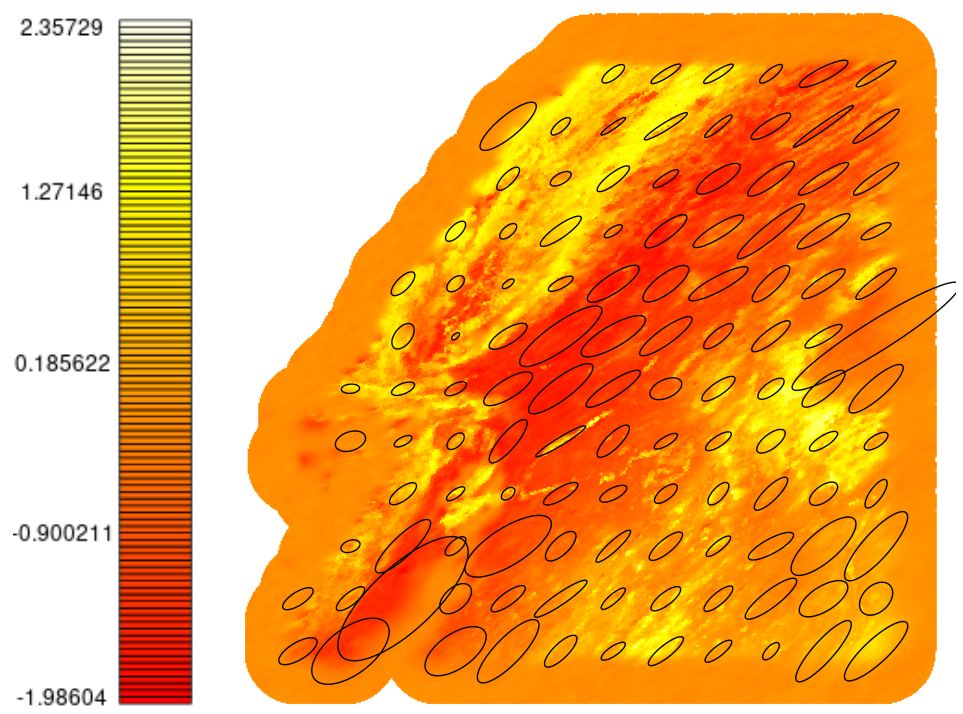

(b) Predicted mean latent field and predicted mean ellipses for correlation = 0.1

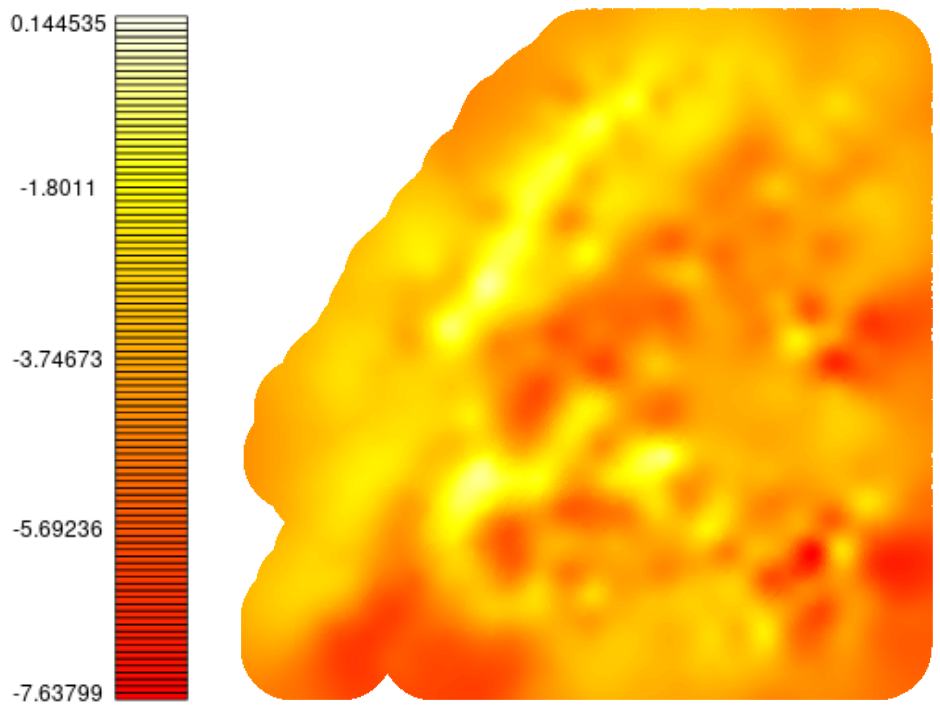

(c) Predicted mean log-noise variance

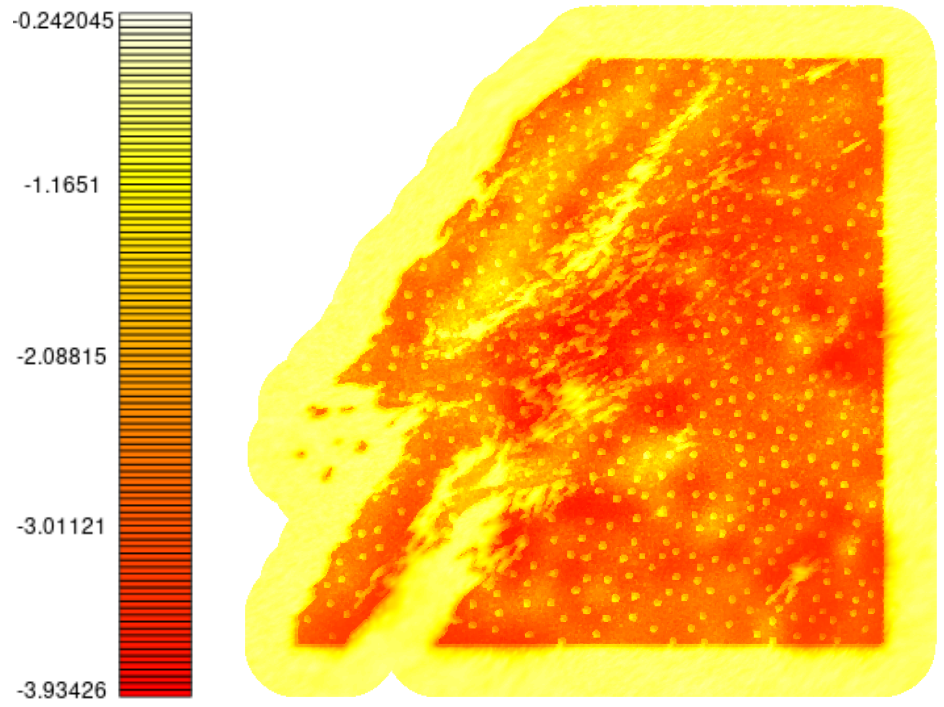

(d) Log standard deviation of the predicted latent field

Figure 1: Maps generated using the nonstationary anisotropic, and heteroskedastic models on the NDVI data set.

posterior density and passing this mean to the matrix exponential. The resulting ellipses are then scaled according the the Matérn smoothness parameter $\nu$ (here 1.5) so that the border of an ellipse corresponds to the zone where the Matérn correlation induced by this ellipse drops to 0.1. It is a 2-dimension version of the $\rho$ parameter (see Lindgren et al. (2011)). The ellipses fit the rays of the top-right corner, while they turn into small circles in the rougher left-hand side. The ellipses are inflated in the smoother regions reflecting increased range. Figure 1(c) maps the mean log variance of the noise revealing that the fuzzier left-hand side of the map is predicted with significantly more noise than the rest. Finally, Figure 1(d) presents the logarithm of the standard deviation of the summed latent field and fixed effects. The clearer dots correspond to the locations left out for validation that incurred a higher variance. Predictive variance and smoothing are clearly impacted by the covariance parameters and, in particular, by the noise variance. An interesting final observation is that the PP variance of the non-stationary range takes smaller values for the anisotropy components than for the determinant component. This is consistent with the response map (Figure 1(a)): anisotropy extends from bottom left to top right with little deviation, while the spatial association range varies between, for example, the calm region of the center and the more troubled region on the upper left corner.

# 6 Summary and future directions

We have developed a class of nonstationary NNGP models that overcome computational bottlenecks while capturing nonstationarity using space-varying parameters in the spatial covariance kernel. The sparsity of NNGP factors along with dimension reduction using low-rank processes allows scaling the analysis to massive data sets. Although low-rank processes tend to inflate estimates of residual variance (Datta et al., 2016; Banerjee, 2017), we find this to be a lesser issue in predictive inference, as they model the range parameters for which information is limited. We also see that nested AS-IS is a viable method for multilayered hierarchical models with large data augmentation and is compatible with Hamiltonian schemes. We exploit derivatives of the NNGP density with respect to spatially indexed covariance parameters and use them in HMC. We compare with local GPs and INLA. Our method is typically slower than INLA and local GPs for stationary fields, but scales better than INLA in “Big Data, Big moMel” settings such as in our NDVI example.

Future work can extend the current offering in different directions. The proposed approach is applicable to any stationary kernel and should be explored beyond the Matérn kernels. Non-Gaussian responses, such as binomial or count data, can be investigated using generalized linear models. Given the "divide and conquer" approach of the Gibbs sampler algorithm we use, it may be possible to address this challenge while maintaining most of the current framework. Non-stationary analysis of multivariate spatial data is also worth undertaking. Another possible extension is modeling in more than two dimensions. In particular, elliptic covariances in 3 dimensions may prove useful to quantify drifts such as rainfall moving across a territory. Ellipses in higher dimensions incur further differentiation since the matrix logarithm of the range parameters will have 6 coordinates instead of 3.

Although the proposed implementation is partially low-level and effective, it can be further improved. In addition to code tuning, alternative options of the differentiation schemes used in the HMC can be explored. Eventually, further elaboration of the MCMC algorithms, such as the use of Riemann manifold methods (Girolami and Calderhead, 2011) instead of HMC, may improve our HMC-within-interweaving approach and deliver even faster model convergence. Other possible explorations include the use of efficient sparse linear system solvers such as conjugate gradients as implemented by Zhang et al. (2019) in stationary NNGP models.

# Supplementary material

An accompanying supplement presents key theoretical results and expanded simulation experiments. Computer programs are accessible from the public repository `https://github.com/SebastienCoube/Nonstat-NNGP` and include an extensive Vignette for users.

# Supplement to "Spatial Process Models with Spatially Varying Covariance Kernels"

Sébastien Coube-Sisqueille[a,1] *, Sudipto Banerjee[b,2], and Benoît Liquet[a,c,3]

[a] Laboratoire de Mathématiques et de leurs Applications, Université de Pau et des Pays de l'Adour, Pau, France

[b] Department of Biostatistics, University of California, Los Angeles, United States of America

[c] School of Mathematical and Physical Sciences, Macquarie University, Sydney, Australia

[1] sebastien.coube@univ.pau.fr [2] sudipto@ucla.edu [3] benoit.liquet-weiland@mq.edu.au

June 3, 2025

# S1 Some results on the nonstationary NNGP

## S1.1 Recursive conditional form of nonstationary NNGP

We begin with the conditional density on the left hand side of (7) and proceed as below:

$$\tilde{f}(w(s_i)\,|\,w(\mathcal{S}_{i-1}),\theta(\mathcal{S})) = f(w(s_i)\,|\,w(\mathrm{pa}(s_i)),\theta(\mathcal{S})) = f(w(s_i\cup\mathrm{pa}(s_i))\,|\,\theta(\mathcal{S}))/f(w(\mathrm{pa}(s_i))\,|\,\theta(\mathcal{S})). \quad \text{(S1)}$$

The joint distributions $f(w(s_i\cup\mathrm{pa}(s_i))\,|\,\theta(\mathcal{S}))$ and $f(w(\mathrm{pa}(s_i))\,|\,\theta(\mathcal{S}))$ are fully specified by $\Sigma(s_i\cup\mathrm{pa}(s_i),\theta(\mathcal{S}))$ and $\Sigma(\mathrm{pa}(s_i),\theta(\mathcal{S}))$. Since the covariance functions given by (4) or (5) specify $\Sigma(s_i,s_j)$ using only $\{\theta(s_i),\theta(s_j)\}$ instead of $\theta(\mathcal{S})$, we obtain $f(w(s_i\cup\mathrm{pa}(s_i))\,|\,\theta(\mathcal{S})) = f(w(s_i\cup\mathrm{pa}(s_i))\,|\,\theta(s_i\cup\mathrm{pa}(s_i)))$ and $f(w(\mathrm{pa}(s_i))\,|\,\theta(\mathcal{S})) = f(w(\mathrm{pa}(s_i))\,|\,\theta(\mathrm{pa}(s_i)))$, which is equal to $f(w(\mathrm{pa}(s_i))\,|\,\theta(s_i\cup\mathrm{pa}(s_i)))$ since $w(\mathrm{pa}(s_i))$ is conditionally independent of $\theta(s_i)$ given $\theta(\mathrm{pa}(s_i))$. Substituting these expressions into the right hand side of (S1) yields (7).

*The authors gratefully acknowledge *E2S: Energy and Environment Solutions*

## S1.2 Marginal variance of nonstationary NNGP

Let $\Sigma(\mathcal{S}) = (K(s_i, s_j))$ and let $\Sigma_0(\mathcal{S}) = (K_0(s_i, s_j))$ be the spatial covariance and correlation matrices, respectively, constructed from the nonstationary covariance function $K(s_i, s_j)$ and the corresponding correlation function $K_0(s_i, s_j)$. Let $\tilde{\Sigma}(\mathcal{S})^{-1} = \tilde{R}(\mathcal{S})^{\mathrm{T}}\tilde{R}(\mathcal{S})$ be the NNGP precision matrix using the nonstationary covariance $K(\cdot)$, where $\tilde{R}(\mathcal{S})$ is the NNGP factor of $\tilde{\Sigma}(\mathcal{S})^{-1}$. Analogously, let $\tilde{R}_0(\mathcal{S})$ be the NNGP precision matrix using the nonstationary correlation $\Sigma_0$ from (3) and either (4) or (5). If $\bar{\sigma}_i = \sqrt{\mathrm{var}(w(s_i) \mid w(\mathrm{pa}(s_i)))}$, then a standard expression is $\bar{\sigma}_i = (\Sigma(s_i, s_i) - \Sigma(s_i, \mathrm{pa}(s_i))\Sigma(\mathrm{pa}(s_i), \mathrm{pa}(s_i))^{-1}\Sigma(\mathrm{pa}(s_i), s_i))^{1/2}$. Therefore, the $i$-th row of $\tilde{R}(\mathcal{S})$ comprises (i) $1/\bar{\sigma}_i$ at index $i$; (ii) $-\Sigma(s_i, \mathrm{pa}(s_i))\Sigma(\mathrm{pa}(s_i), \mathrm{pa}(s_i))^{-1}/\bar{\sigma}_i$ at indices corresponding to $\mathrm{pa}(s_i)$; and (iii) 0 elsewhere. Letting $\bar{\sigma}_{0i}$ be the conditional correlation obtained from $\Sigma_0$ instead of $\Sigma$, it is easily seen that $\bar{\sigma}_i = \sigma(s_i)(\bar{\sigma}_{0i})$ using the elementary observations that $\sigma(s_i)^2 = \Sigma(s_i, s_i)$ (by definition of $\sigma(s_i)$) and that $\Sigma(\mathcal{A}, \mathcal{B}) = \mathrm{diag}(\sigma(\mathcal{A}))\Sigma_0(\mathcal{A}, \mathcal{B})\mathrm{diag}(\sigma(\mathcal{B}))$, where $\mathcal{A}$ and $\mathcal{B}$ are any two non-empty subsets of $\mathcal{S}$. Thus,

$$\begin{aligned}
\bar{\sigma}_i &= \big(\Sigma(s_i, s_i) - \Sigma(s_i, \mathrm{pa}(s_i))\Sigma(\mathrm{pa}(s_i), \mathrm{pa}(s_i))^{-1}\Sigma(\mathrm{pa}(s_i), s_i)\big)^{1/2} \\
&= \sigma(s_i)\left(\Sigma_0(s_i, s_i) - \Sigma_0(s_i, \mathrm{pa}(s_i))\,\Sigma_0(\mathrm{pa}(s_i), \mathrm{pa}(s_i))^{-1}\Sigma_0(\mathrm{pa}(s_i), s_i)\right)^{1/2} \qquad \text{(S2)} \\
&= \sigma(s_i)(\bar{\sigma}_0)_i
\end{aligned}$$

Using this relationship, we can express the coefficients of row $i$ in $\tilde{R}(\mathcal{S})$ as (i) $1/(\bar{\sigma}_{0i})\sigma(s_i))$ at position $i$; (ii) $\big(-\Sigma_0(s_i, \mathrm{pa}(s_i))\Sigma_0(\mathrm{pa}(s_i), \mathrm{pa}(s_i))^{-1}/(\bar{\sigma_0})_i\big) = \tilde{R}_0(\mathcal{S})(i, \mathrm{pa}(i))\mathrm{diag}(\sigma(\mathrm{pa}(s_i)))^{-1}$ at the indices corresponding to $\mathrm{pa}(s_i)$, which means that $\tilde{R}(\mathcal{S})(i, j) = \tilde{R}_0(\mathcal{S})(i, j)/\sigma(s_j)$ for all $s_j \in \mathrm{pa}(s_i)$; and (iii) 0 elsewhere. Comparing elements we obtain $\tilde{R}(\mathcal{S}) = \tilde{R}_0(\mathcal{S})\mathrm{diag}(\sigma(\mathcal{S}))^{-1}$.

# S2 KL divergence between nonstationary NNGP and full nonstationary GP

## S2.1 Spatially indexed variances

The Kullback-Leibler (KL) divergence between two Multi-Variate Normal distributions $\mathcal{N}(\mu_1, \Sigma_1)$ and $\mathcal{N}(\mu_2, \Sigma_2)$ is $KL\left(\mathcal{N}_1 \parallel \mathcal{N}_2\right) = \frac{1}{2}\left(\operatorname{tr}\left(\Sigma_2^{-1}\Sigma_1\right) + \left(\mu_2 - \mu_1\right)^{\mathrm{T}} \Sigma_2^{-1}\left(\mu_2 - \mu_1\right) - k + \ln\left(\frac{|\Sigma_2|}{|\Sigma_1|}\right)\right)$. Recalling that the NNGP precision is $\tilde{\Sigma}(\mathcal{S})^{-1} = \operatorname{diag}(\sigma(\mathcal{S}))^{-1}(\tilde{R}_0(\mathcal{S})^{\mathrm{T}}\tilde{R}_0(\mathcal{S}))\operatorname{diag}(\sigma(\mathcal{S}))^{-1}$, that the full GP's covariance matrix is $\operatorname{diag}(\sigma(\mathcal{S}))\Sigma_0(\mathcal{S})\operatorname{diag}(\sigma(\mathcal{S}))$, and that the NNGP and GP means are equal, we obtain the KL divergence between a nonstationary full GP with zero mean and the NNGP as $\frac{1}{2}\operatorname{tr}\left(\Sigma_0(\mathcal{S})\tilde{R}_0(\mathcal{S})^{\mathrm{T}}\tilde{R}_0(\mathcal{S})\right) + \ln\left(\frac{|\tilde{R}_0(\mathcal{S})^{\mathrm{T}}\tilde{R}_0(\mathcal{S})|}{|\Sigma_0(\mathcal{S})|}\right)$. This is the KL divergence between $N(0, \Sigma_0(\mathcal{S}))$ and $N(0, \tilde{\Sigma}_0(\mathcal{S}))$. It follows that spatially indexed variances do not affect the KL divergence between nonstationary full GPs and NNGPs.

## S2.2 Spatially indexed variances and scalar ranges

Synthetic data sets with $2,500$ observations are simulated on a domain with size $5 \times 5$. The spatially variable log-range has mean $\log(0,05)$. The following 3 factors are tested:

- the extent of nonstationarity corresponding to different values of the marginal variance of the log range as 0.1, 0.3, and 0.5;
- the ordering (coordinate, max-min, random, middle-out); and
- the number of parents (5, 10, 20) in the DAG.

Using a linear model with interactions reveals that the most important factor is the number of parents. The NNGP approximation is further improved using the max-min and random order, corroborating the conclusions in Guinness (2018) for stationary models in 2

Table S1: Summary of linear regression of the KL divergence, in the scalar range case.

The reference case has coordinate ordering, 5 nearest neighbours, and a log-range's marginal variance of 0.05

| | Estimate | Std. Error | t value | Pr(> $\|t\|$) |
|---|---|---|---|---|
| (Intercept) | 41.08 | 0.71 | 58.09 | 0.00 |
| nonstat.intensity 0.3 | 2.56 | 0.46 | 5.53 | 0.00 |
| nonstat.intensity 0.5 | 5.39 | 0.46 | 11.65 | 0.00 |
| ordering max min | -4.18 | 0.93 | -4.51 | 0.00 |
| ordering middle out | -1.90 | 0.93 | -2.06 | 0.04 |
| ordering random | -4.34 | 0.93 | -4.69 | 0.00 |
| 10 nearest neighbours | -31.59 | 0.93 | -34.12 | 0.00 |
| 20 nearest neighbours | -41.27 | 0.93 | -44.58 | 0.00 |
| max min: 10 | 0.83 | 1.31 | 0.63 | 0.53 |
| middle out: 10 | 0.42 | 1.31 | 0.32 | 0.75 |
| random: 10 | 0.99 | 1.31 | 0.76 | 0.45 |
| max min: 20 | 3.19 | 1.31 | 2.43 | 0.02 |
| middle out: 20 | 1.24 | 1.31 | 0.94 | 0.35 |
| random: 20 | 3.35 | 1.31 | 2.56 | 0.01 |

dimensions to nonstationary models. The extent of nonstationarity plays a relatively minor role. Table S1 supplies further details on the effects of the factors.

## S2.3 Elliptic range case

Synthetic data sets with $2,500$ observations are simulated on a domain with size $5 \times 5$. The spatially variable log-matrix range has mean log(0.05). The following three factors are tested:

- the extent of nonstationarity corresponding to different values of the within-site variance of the coordinates of the log-range matrix ($0.1 \times I_3$, $0.3 \times I_3$, and $0.5 \times I_3$);
- the ordering (coordinate, max-min, random, middle-out);
- the number of parents (5, 10, 20) in the DAG.

The conclusion we draw from Table S2 are the same as before.

Table S2: Summary of linear regression of the KL divergence, in the elliptic range case.

The reference case has coordinate ordering, 5 nearest neighbours, and a log-range marginal variance of 0.05

| | Estimate | Std. Error | t value | Pr(> $\|t\|$) |
|---|---|---|---|---|
| (Intercept) | 41.08 | 0.71 | 58.09 | 0.00 |
| nonstat.intensity 0.3 | 2.56 | 0.46 | 5.53 | 0.00 |
| nonstat.intensity 0.5 | 5.39 | 0.46 | 11.65 | 0.00 |
| ordering max min | -4.18 | 0.93 | -4.51 | 0.00 |
| ordering middle out | -1.90 | 0.93 | -2.06 | 0.04 |
| ordering random | -4.34 | 0.93 | -4.69 | 0.00 |
| 10 nearest neighbours | -31.59 | 0.93 | -34.12 | 0.00 |
| 20 nearest neighbours | -41.27 | 0.93 | -44.58 | 0.00 |
| max min: 10 | 0.83 | 1.31 | 0.63 | 0.53 |
| middle out: 10 | 0.42 | 1.31 | 0.32 | 0.75 |
| random: 10 | 0.99 | 1.31 | 0.76 | 0.45 |
| max min: 20 | 3.19 | 1.31 | 2.43 | 0.02 |
| middle out: 20 | 1.24 | 1.31 | 0.94 | 0.35 |
| random: 20 | 3.35 | 1.31 | 2.56 | 0.01 |

# S3 Details and examples for the Predictive Process

## S3.1 Construction of a Predictive Process basis

In Section 3.2, we pursue a truncated Predictive Process (PP) basis (Banerjee et al., 2008), itself obtained from a NNGP. While the PP approximations are prone to over-smoothing (see the discussions in Datta et al., 2016; Banerjee, 2017), this is not a problem here since the hyperprior range and smoothness are supposed to be high. Like in S4, we extend notations and index matrices on spatial locations.

The predictive process spatial basis of size $r$ is $B(\mathcal{S}) = (\tilde{R}_{PP}((\mathcal{S}^*, \mathcal{S}))^{-1}M)_{\mathcal{S}}$, where $\tilde{R}_{PP}((\mathcal{S}^*, \mathcal{S}))$ is an NNGP factor defined as in Section 2 on the joint collection $(\mathcal{S}^*, \mathcal{S})$ of knots $\mathcal{S}^*$ and observed spatial locations $\mathcal{S}$, and $M$ is a matrix of size $(n+r) \times r$ such that $M_{i,j} = 1$ if $i = j$ and $M_{i,j} = 0$ everywhere else. The $r$ spatial locations $\mathcal{S}^*$ are called *knots* and they must be well-spread over the spatial domain to guarantee a good PP approximation. In our implementation, a K-means algorithm on $\mathcal{S}$ sets the knots automatically. Using fast solving relying on the sparsity and triangularity of $\tilde{R}_{PP}((\mathcal{S}^*, \mathcal{S}))$, computing $B(\mathcal{S})u_\theta = (\tilde{R}_{PP}((\mathcal{S}^*, \mathcal{S}))^{-1}(Mu_\theta))_{\mathcal{S}}$ is affordable and avoids large matrix multiplications when $u_\theta$ is a matrix with $r$ rows and few columns (one in the case of Equation (11), and $d(d+1)/2$, in practice three, in the extension of Section 3.3). The procedure is exactly the same for $\tau$ with the same basis $B(\mathcal{S})$ being shared by $\theta(\mathcal{S})$ and $\tau$.

## S3.2 Knot number and placement

We provide a few remarks on the selection of knots. The knots can be found as the centers of a K-means clustering, thereby allowing the user to specify only the desired number of knots. Our package vignette offers a quick look at how the PP samples behave using our spatial plotting functions. We point out that Guhaniyogi et al. (2011) have undertaken a rather thorough investigation of adaptive knot selection in predictive process models.

For now, our recommendation is to make the PP as simple as possible, and to keep in mind that even though we have not solved all our problems, we are able to work with a very large number of high-level parameters, which is an achievement in itself. In view of what we can achieve, the PP should more be seen as a very smooth P-spline rather than a fine approximation of a GP. Lastly, for the sake of identifiability between the GP field $w(\mathcal{S})$ and the fields of covariance parameters $\theta(\mathcal{S})$ and $\tau$, the process used to model a field of covariance parameters must be smooth and long-ranged. In those settings, the over-smoothing induced by the PPs is a minor concern.

In the current layout, only the marginal variance of the PP is estimated, and its Matérn range remains fixed. As a consequence, a satisfying knot placement can be found prior to model estimation and kept without the need for in-run adjustment. The `R` package allows one to compare NNGP and PP samples using spatial field plotting functions in order to detect discrepancies due to PP over-smoothing. The Vignette provided in `https://github.com/SebastienCoube/Nonstat-NNGP` provides hands-on explanations.

### S3.3 Sensitivity with respect to the Predictive Process range

A potential problem for the modeler is to choose the PP range. A larger range should induce a PP leading to a smoother, more regular field of covariance parameters, impacting -positively or negatively- the quality of inference and prediction.

The results of Zhang (2004); Tang et al. (2021) state that the range and marginal variance of a Matérn field cannot be consistently estimated in the context of fill-in asymptotics. In particular, when the Matérn range is large with respect to the spatial domain radius, a ridge-like joint density is observed for the range and the marginal variance: a sample from a "big range, big variance" field is difficult to tell apart from a sample of a "small range, small variance" field. This behavior of communicating vessels is generally considered an issue and is addressed, for example, through Penalize Complexity (PC) Priorities (Fuglstad et al., 2015).

In the context of the Predictive Process prior for covariance parameters, however, this behavior can be used to the advantage of the modeler. The spatial range for the covariance parameters $W_\theta(\mathcal{S})$ and $W_\tau(\mathcal{S})$ must, for the sake of identifiability and interpretability, be larger than the spatial range of the NNGP latent field $w(\mathcal{S})$. As a consequence, the panel of possible ranges for the PP will likely be within the region in which the range and variance cannot be estimated consistently. Therefore, the PP range can be fixed to an arbitrary value within this region and the PP variance will adjust with little effect on the quality of prediction and estimation.

Two cases escape from this convenient procedure: (i) large spatial domains where the possible ranges for the PP include smaller values where the variance is easier to identify; (ii) the range of $W_\tau(\mathcal{S})$, unlike $W_\theta(\mathcal{S})$'s, may not require being much higher than the range of $w(\mathcal{S})$. Those two specific cases are left for future investigation, the ability to work with 100+ covariance parameters in a full Bayesian spatial model being in itself an important contribution.

We carry out an experiment on synthetic data to explore the sensitivity of the model with respect to the PP range. The first part of the experiment investigates the impact of the choice of the PP range when using a model with heteroskedastic noise. Synthetic data with heteroskedastic noise is simulated using the same procedure as S8.1, except for the fact that the value of the "true" Matérn range used to generate the noise variance field is 0.1. The spatial domain is a $1 \times 1$ square. We generate 30 replicas of the synthetic data set using a random seed. We then run the model for each replica of the data, but let the PP range used to fit the model take five values: $0.05, 0.05 \times \sqrt{2}, 0.1$ (the"true" range), $0.1 \times \sqrt{(2)}$, and 0.2. The number of runs amounts to 150 (30 replicas of the data, and 5 PP range values tested each time). Like in S8.1, we monitor the Empirical Log Pointwise Predictive Distribution (elpd) at both observed and unobserved locations, and the retrieval of the latent noise variance through the mean square error (MSE). We also report the estimated mean log-variance of the noise's PP $\gamma_\tau$ in (10). Those indicators of performance, and the estimation of the PP variance, are explored using a simple linear model regressing each indicator upon the chosen PP range and the seed. The seed is used as a control variable, and its effects are not reported here. The PP range is treated as a factor, with the "true" range 0.1 being the reference. The results are reported in Table S3.

The second experiment is similar in the layout of the PP, but features a non-stationary Matérn range, as in S8.2. Its results are presented in Table S4.

In both experiments, the logarithmic variation of PP reported in the first column of Tables S3 and S4 adjusts to the chosen range of PP, confirming the behavior of the communicating vessel. The only case where performance deterioration can be observed is when the PP range is very low (equal to 0.05), inducing a small but significant loss in precision in the retrieval of $\tau$ and $\theta$. In addition to this case, changing the PP range has little effect on the model performance indicators.

Table S3: Impact of the chosen PP range on the performance of the estimation of the nonstationary noise variance.

| | $\tau$ *mean log − variance* | | $\tau$ *MSE* | | *Observed elpd* | | *Prediction elpd* | |
|---|---|---|---|---|---|---|---|---|
| | *Estimate* | *p − value* | *Estimate* | *p − value* | *Estimate* | *p − value* | *Estimate* | *p − value* |
| (*Intercept*) | **-0.36** | **2e-22** | **0.037** | **5.9e-110** | **-0.89** | **3.9e-285** | **-1.1** | **6.2e-286** |
| $PP\ range = 0.05$ | **-0.67** | **6.7e-71** | **0.003** | **5.8e-26** | 0.00076 | $4e-06$ | $-0.00033$ | 0.075 |
| $PP\ range = 0.05\sqrt{2}$ | **-0.47** | **6.2e-55** | 0.00056 | 0.011 | 0.00022 | 0.16 | $9.1e-05$ | 0.62 |
| $PP\ range = 0.1\sqrt{2}$ | **0.68** | **4.6e-72** | 0.00012 | 0.58 | $-0.00064$ | $7.8e-05$ | $-0.00025$ | 0.18 |
| $PP\ range = 0.2$ | **1.5** | **2.7e-111** | 0.00061 | 0.0055 | $-0.00079$ | $1.6e-06$ | $-0.00047$ | 0.013 |

Table S4: Impact of the chosen PP range on the performance of the estimation of the nonstationary range.

| | $\theta$ *mean log − variance* | | $\theta$ *MSE* | | *Observed elpd* | | *Prediction elpd* | |
|---|---|---|---|---|---|---|---|---|
| | *Estimate* | *p − value* | *Estimate* | *p − value* | *Estimate* | *p − value* | *Estimate* | *p − value* |
| (*Intercept*) | **-0.68** | **3.6e-20** | **0.036** | **4e-86** | **-0.62** | **1.3e-263** | **-0.91** | **1.6e-282** |
| $PP\ range = 0.05$ | **-0.4** | **1.9e-22** | 0.0031 | $3.6e-15$ | 0.00078 | $1.7e-05$ | $-0.00034$ | 0.054 |
| $PP\ range = 0.05\sqrt{2}$ | $-0.3$ | $4e-15$ | 0.00073 | 0.037 | 0.00043 | 0.016 | $-1.7e-06$ | 0.99 |
| $PP\ range = 0.1\sqrt{2}$ | **0.51** | **8.9e-30** | 0.00043 | 0.22 | $-0.00052$ | 0.0034 | $-9.5e-05$ | 0.59 |
| $PP\ range = 0.2$ | **1.2** | **3.7e-65** | 0.001 | 0.0033 | $-0.00062$ | 0.00057 | $2.5e-05$ | 0.89 |

### S3.4 Example of range ellipses and GPs generated following the proposed Bayesian hierarchical model

Figure S1(a) presents range ellipses generated with (12), while Figure S1(b) presents range ellipses generated with $(10(d))$. Figures S1(c) and S1(d) represent one of their respective Gaussian process sample paths obtained with $(10(b))$. The ellipse size corresponds to the place where the correlation drops to 0.1, similarly to the $\rho$ parameter in Lindgren et al. (2011).

## S4 Layout of the MCMC algorithm

The proposed algorithm samples the parameters from Equation (9). It relies on Component-Wise Interweaving (Yu and Meng, 2011), a modified Gibbs Sampler method presented in detail in Section S5.1 of this Supplement. It also implements Nested Interweaving (Yu and Meng, 2011), from Section S5.2, and plugs Hybrid Monte-Carlo (HMC) within the Interweaving algorithm. Although HMC is a classical method (see for example Neal et al., 2011), its implementation is far from trivial due to the intrinsic complexity of the gradients

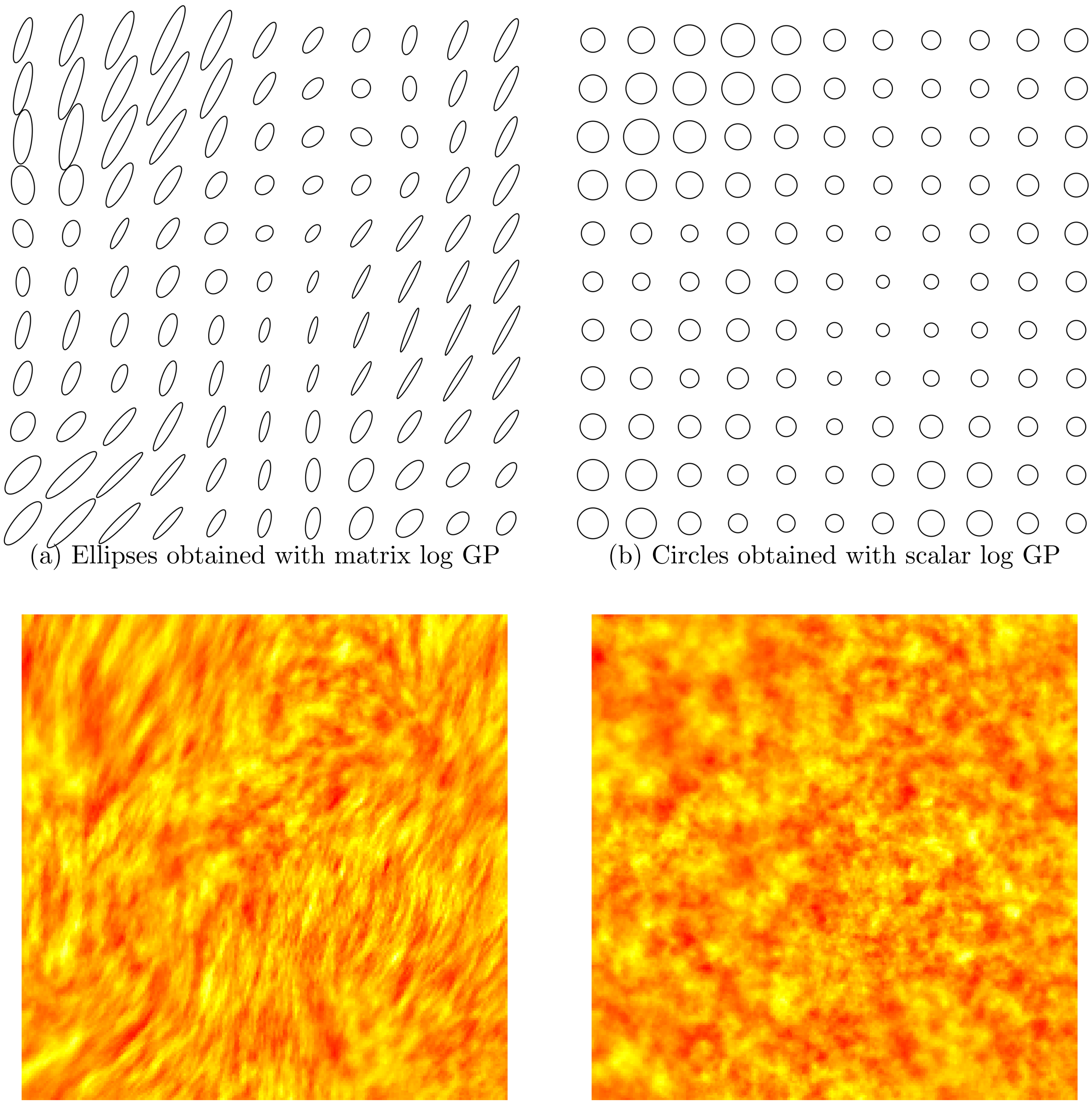

(a) Ellipses obtained with matrix log GP

(b) Circles obtained with scalar log GP

(c) GP samples corresponding to the ellipses

(d) GP samples corresponding to the circles

Figure S1: Range ellipses and GP samples induced by the log-GP and matrix log-GP priors

for the particular application and the coupling of HMC with interweaving. The general rationale of our HMC scheme is expounded in detail in Section 4.3 of the main body of the article, while details are given in Section S6 of this Supplement.

A point worth noting is that the predictive process coefficients $u_\alpha$, $u_{\sigma^2}$, and $u_{\tau^2}$ are used along with their counterparts $W_\alpha(\mathcal{S})$, $W_{\sigma^2}(\mathcal{S})$, and $W_{\tau^2}(\mathcal{S})$, with $W_\cdot(\mathcal{S}) = B(\mathcal{S})u_\cdot$ in the interweaving steps. All these parameters are introduced in Section 3 of the main body of the article. Here, $W_\cdot(\mathcal{S})$ is the *sufficient* parametrization, while $u_\cdot$ is the *ancillary* parametrization. The *whitened* parametrization of the latent field $w^*(\mathcal{S}) = \tilde{R}(\mathcal{S})w(\mathcal{S})$ is also used as an ancillary parameterization of the sufficient $w(\mathcal{S})$.

In the following, we use the bracket notation "$[\cdot|\cdot]$" to denote a Monte Carlo drawing. The parameters on the left are sampled conditionally on the parameters on the right. While all those draws are done following the full conditional distribution, we omit irrelevant parameters for the sake of clarity. Those draws are not always analytic: following the particular circumstances, they can also be HMC, or mere Metropolis-Hastings steps. The sampling is performed by repeating the following sequence:

1. Block update of the latent field using Datta et al. (2016) in Section 4.1. For $\mathcal{S}_1, \ldots, \mathcal{S}_k$ such that $\mathcal{S}_1 \cup \ldots \cup \mathcal{S}_k = \mathcal{S}$, draw

$$[w(\mathcal{S}_1)\,|\,w(\mathcal{S} \setminus \mathcal{S}_1), \beta, \theta(\mathcal{S}), \tau^2(\mathcal{S})] \rightarrow \ldots \rightarrow [w(\mathcal{S}_k)\,|\,w(\mathcal{S} \setminus \mathcal{S}_k), \beta, \theta(\mathcal{S}), \tau^2(\mathcal{S})].$$

   The joint distribution of $w(\mathcal{S})$ arises from the response likelihood presented in Equation $(10(a))$ and the Gaussian process prior presented in Equation $(10(b))$. In a Gaussian response model, exact sampling can be performed using classical results on conditional properties of Gaussian Markov Random Fields (Rue and Held, 2005).

   Like in Section 2 of the Main body of the article, we index matrices by spatial sites, for example, $\tilde{R}(\mathcal{S})_{\mathcal{S}_1,\mathcal{S}_1}$ reads as the square submatrix of $\tilde{R}(\mathcal{S})$ whose rows and columns indices correspond to the indices of $\mathcal{S}_1$ in $\mathcal{S}$.

For $i$ ranging from 1 to $k$, the full conditional distribution of $w(\mathcal{S}_i)$ is Normal, with precision

$$Q(\mathcal{S}_i) \;=\; (\tilde{R}(\mathcal{S})^T\tilde{R}(\mathcal{S}))_{\mathcal{S}_i,\mathcal{S}_i} + (M_{\cdot,\mathcal{S}_i})^{\mathrm{T}}\mathrm{diag}(1/\tau^2)(M_{\cdot,\mathcal{S}_i}),$$

$\tau^2$ being the noise variance of the Gaussian likelihood in $(10(a))$, and $\tilde{R}(\mathcal{S})$ being the NNGP factor from Equation (8), intervening in the prior distribution $(10(b))$. The notation $Q(\mathcal{S}_i)$ is justified by the fact that the conditional precision is merely a submatrix of the joint precision $Q(\mathcal{S}) = \tilde{R}(\mathcal{S})^T\tilde{R}(\mathcal{S}) + M^{\mathrm{T}}\mathrm{diag}(1/\tau^2)M$ (Rue and Held, 2005).

The expectation of the full conditional of $w(\mathcal{S}_i)$ is obtained as

$$\mathbb{E}(w(\mathcal{S}_i)\,|\,\cdot) = -Q(\mathcal{S}_i)(-(M_{\cdot,\mathcal{S}_i})^{\mathrm{T}}\mathrm{diag}(1/\tau^2)(z - X\beta) + (R(\mathcal{S})^T R(\mathcal{S}))_{\mathcal{S}_i,\mathcal{S}\setminus\mathcal{S}_i} w(\mathcal{S}\setminus\mathcal{S}_i)).$$

Computing the conditional expectation and sampling from the Multi-Variate Normal distribution is done using the sparse Cholesky decomposition of $Q(\mathcal{S}_i)$.

2. Interweaving update of $\beta$ using Coube-Sisqueille and Liquet (2021), Section 2. The strategy is outlined in Section 4.2 of the main body of the paper, and its rationale is expounded in more detail in Section S5.1 of this Supplement.

$$\underbrace{[\beta\,|\,w(\mathcal{S}),\tau^2]}_{ancillary} \to \underbrace{[w_{center}(\mathcal{S})\,|\,w(\mathcal{S}),\beta]}_{deterministic} \to \underbrace{[\beta\,|\,w_{center}(\mathcal{S}),\theta(\mathcal{S})]}_{sufficient} \to \underbrace{[w(\mathcal{S})\,|\,w_{center}(\mathcal{S}),\beta]}_{deterministic}$$

The first drawing uses the natural parameterization of the latent field, as presented in (10). In this context, this parameterization is *ancillary*, because $w(\mathcal{S})$ and $\beta$ are *a priori* independent. The procedure corresponds to the usual sampling scheme in a model with a Gaussian response (see, e.g., Datta et al., 2016), adjusted for heteroskedasticity in the Gaussian noise variance. The posterior distribution arises from the data likelihood of (10)$(a)$ and the prior (10)$(g)$, this prior being optional because the posterior distribution is well defined even in the absence of a prior. We expound here the posterior distribution with an improper constant prior:

$$\beta \sim \mathcal{N}((X^{\mathrm{T}}\mathrm{diag}(1/\tau^2)X)^{-1}X^{\mathrm{T}}\mathrm{diag}(1/\tau^2)(z - Mw(\mathcal{S})), (X^{\mathrm{T}}\mathrm{diag}(1/\tau^2)X)^{-1}),$$

the matching matrix $M$ of size $|z| \times |\mathcal{S}|$ being introduced in Section 3.1 of the main.

The formula is found by stating $p(\beta \,|\, z, \cdot) \propto p(z \,|\, \beta, \cdot)$, an improper prior being assumed for $\beta$. Then, the response log-likelihood is re-arranged as

$$\begin{aligned} &log(p(z|\beta, \tau^2, w(\mathcal{S}))) = -(z - Mw(\mathcal{S}) - X\beta)^{\mathrm{T}}\mathrm{diag}(1/\tau^2)(z - Mw(\mathcal{S}) - X\beta)/2 + constant \\ &\quad = X^{\mathrm{T}}\mathrm{diag}(1/\tau^2)X(X^{\mathrm{T}}\mathrm{diag}(1/\tau^2)X)^{-1}X^{\mathrm{T}}\mathrm{diag}(1/\tau^2)(z - Mw(\mathcal{S}) - X\beta) \\ &\qquad -\beta^{\mathrm{T}}X^{\mathrm{T}}\mathrm{diag}(1/\tau^2)X\beta/2 + constant, \end{aligned}$$

allowing to identify the normality of the distribution as well as its parameters.

The centered parametrization of the latent field is defined as

$$w_{center}(\mathcal{S}) = w(\mathcal{S}) + X^s(\mathcal{S})\beta_\kappa,$$

defined by the following procedure. Let $p$ be the number of columns of $X$. Let $\kappa$ be the largest subset of $\{1, \dots, p\}$ who validates $X_{\cdot,\kappa} = MX^s(\mathcal{S})$, $X^s(\mathcal{S})$ being a matrix of size $|\mathcal{S}| \times |\kappa|$. In other words, $X^s(\mathcal{S})$ is obtained by deleting the columns of $X$ whose value changes within the same spatial site, and then deleting all the rows corresponding to redundant spatial locations. Each row of $X^s(\mathcal{S})$ corresponds to one and only one location of $\mathcal{S}$, and one and only one realization of $w(\mathcal{S})$, justifying its indexation on $\mathcal{S}$. This equation allows us to directly derive the two deterministic steps $[w_{center}(\mathcal{S}) \,|\, w(\mathcal{S}), \beta]$ and $[w(\mathcal{S}) \,|\, w_{center}(\mathcal{S}), \beta]$ by, respectively, adding $X^s(\mathcal{S})\beta_\kappa$ to $w(\mathcal{S})$, or removing it from $w_{center}(\mathcal{S})$.

From the Gaussian prior in Equation ($10(b)$), we can derive straightforwardly

$$w_{center}(\mathcal{S}) \overset{a\ priori}{\sim} \mathcal{N}\ (X^s(\mathcal{S})\beta_\kappa, (R(\mathcal{S})^{\mathrm{T}} R(\mathcal{S}))^{-1}).$$

Rearranging this expression yields that $\beta_\kappa$ follows a Normal distribution conditionally on $w_{center}(\mathcal{S})$ and $\theta(\mathcal{S})$, with

$$Var(\beta_\kappa | w_{center}(\mathcal{S}), \theta(\mathcal{S})) \ = \ (X^s(\mathcal{S})^{\mathrm{T}} R(\mathcal{S})^{\mathrm{T}} R(\mathcal{S}) X^s(\mathcal{S}))^{-1},$$

and

$$\mathbb{E}(\beta_\kappa | w_{center}(\mathcal{S}), \theta(\mathcal{S})) \ = \ Var(\beta_\kappa | w_{center}(\mathcal{S}), \theta(\mathcal{S})) X^s(\mathcal{S})^{\mathrm{T}} R(\mathcal{S})^{\mathrm{T}} R(\mathcal{S}) w_{center}(\mathcal{S}).$$

The distribution is found using the same method, but replacing the response likelihood by the Gaussian prior on $w_{center}(\mathcal{S})$:

$$\begin{aligned} &log(p(w_{center}(\mathcal{S})|\beta, \theta(\mathcal{S}))) = \\ &\quad = -(w_{center}(\mathcal{S}) - X^s(\mathcal{S})\beta_\kappa)^{\mathrm{T}} \tilde{R}(\mathcal{S})^T \tilde{R}(\mathcal{S})(w_{center}(\mathcal{S}) - X^s(\mathcal{S})\beta_\kappa)/2 + \ constant \\ &\quad = \beta_\kappa^{\mathrm{T}}\ X^s(\mathcal{S})^{\mathrm{T}} R(\mathcal{S})^{\mathrm{T}} R(\mathcal{S}) X^s(\mathcal{S})\ \ (X^s(\mathcal{S})^{\mathrm{T}} R(\mathcal{S})^{\mathrm{T}} R(\mathcal{S}) X^s(\mathcal{S}))^{-1} \\ &\qquad X^s(\mathcal{S})^{\mathrm{T}} R(\mathcal{S})^{\mathrm{T}} R(\mathcal{S}) w_{center}(\mathcal{S}) - \beta_\kappa^{\mathrm{T}}\ X^s(\mathcal{S})^{\mathrm{T}} R(\mathcal{S})^{\mathrm{T}} R(\mathcal{S}) X^s(\mathcal{S})\beta_\kappa/2 + constant. \end{aligned}$$

Note that this second step does not depend on $z$ being Gaussian.

3. Joint HMC-within-Gibbs update of $u_\tau$ and $\beta_\tau$.

$$[\beta_\tau, u_\tau \,|\, w(\mathcal{S}), \beta, \gamma_\tau]$$

The posterior distribution arises, on the one hand, from the heteroskedastic Gaussian likelihood of the data, in Equation ($10(a)$). The linkage between $\tau^2$, who intervenes in the Gaussian likelihood, and $u_\tau$ and $\beta_\tau$ is enforced by the logarithmic transform of Equation ($10(e)$) and the low-rank decomposition of Equation (11). On the other hand, $\beta_\tau$ has an user-defined prior with expectation $\mu_{\beta_\tau}$, and variance $V_{\beta_\tau}$, while $u_\tau$ follows a standard Normal distribution from Equation (11). The target density is then obtained as

$$N(\beta_\tau, \mu_\tau, V_\tau)N(u_\tau, 0, I_r)N(z, X\beta + Mw(\mathcal{S}), \text{diag}(exp(X_\tau\beta_\tau + \sqrt{\gamma_\tau}MB(\mathcal{S})u_\tau))),$$

$N(\cdot,\cdot,\cdot)$ being the Multi-Variate Normal distribution.

The gradient of the data likelihood from ($10(a)$) with respect to $\tau$ is obtained following the steps presented in Section S6.7. This gradient of size $|z|$ is then plugged into the chain rule described in Section 4.3 in order to obtain the gradient of $N(z, X\beta + Mw(\mathcal{S}), \text{diag}(exp(X_\tau\beta_\tau + \sqrt{\gamma_\tau}MB(\mathcal{S})u_\tau)))$ with respect to $(\beta_\tau, u_\tau)$.

4. Metropolis-Hastings within Interweaving update of $\gamma_\tau$. The strategy is outlined in Section 4.2 of the main body of the article, and explained in more detail in S5.1 of this Supplement. The interweaving step writes as:

$$\underbrace{[\gamma_\tau \,|\, W_\tau(\mathcal{S})]}_{sufficient} \rightarrow \underbrace{[u_\tau \,|\, \gamma_\tau, W_\tau(\mathcal{S})]}_{deterministic} \rightarrow \underbrace{[\gamma_\tau \,|\, u_\tau, \beta_\tau, w(\mathcal{S}), \beta]}_{ancillary} \rightarrow \underbrace{[W_\tau(\mathcal{S}) \,|\, \gamma_\tau, u_\tau]}_{deterministic}$$

Regardless of which parametrization $W_\tau(\mathcal{S})$ or $u_\tau$ is chosen, $\gamma_\tau$ follows a Uniform prior distribution from Equation ($10(g)$).

When the sufficient parametrization is used, $\gamma_\tau$ acts as a marginal variance parameter for $W_\tau(\mathcal{S})$ in Equation ($(10)(f)$), with Equation (11) inducing the low-rank distribution $W_\tau(\mathcal{S}) \overset{\text{a priori}}{\sim} \mathcal{N}(0, \gamma_\tau B(\mathcal{S})B(\mathcal{S})^{\mathrm{T}})$. This distribution being degenerate, we reduce to a full-rank distribution by considering $(B(\mathcal{S})^{\mathrm{T}}B(\mathcal{S}))^{-1}B(\mathcal{S})^{\mathrm{T}}W_\tau(\mathcal{S}) \overset{\text{a priori}}{\sim} \mathcal{N}(0, \gamma_\tau I_r)$. Denoting $U(\cdot, a, b)$ the Uniform density with lower and upper bounds respectively $a$ and $b$, and $N(\cdot,\cdot,\cdot)$ the Multi-variate normal distribution, we get the sufficient target density

$$U(\gamma_\tau, a_{\gamma_\tau}, b_{\gamma_\tau})N((B(\mathcal{S})^{\mathrm{T}}B(\mathcal{S}))^{-1}B(\mathcal{S})^{\mathrm{T}}W_\tau(\mathcal{S}), 0, \gamma_\tau I_r).$$

When the ancillary parametrization is used, $u_\tau$ is multiplied by $\gamma_\tau$ and $B(\mathcal{S})$ from Equation (11), before being plugged into Equation (10($e$)), and eventually evaluated in likelihood in (10($a$)). We combine and obtain the target density

$$U(\gamma_\tau, a_{\gamma_\tau}, b_{\gamma_\tau})N(z, X\beta + Mw(\mathcal{S}), \mathrm{diag}(exp(X_\tau\beta_\tau + \sqrt{\gamma_\tau}B(\mathcal{S})u_\tau))).$$

The two previous steps being univariate, they are addressed through a simple Metropolis update. Eventually, Equation (10($e$)) allows us to perform the deterministic transforms readily.

5. HMC-within-interweaving update of the marginal variance parameters $u_\sigma$ and $\beta_\sigma$. $\sigma(\mathcal{S})$ is one of the two possible cases of $\theta(\mathcal{S})$ in 10, the other being either $\alpha(\mathcal{S})$ or $A(\mathcal{S})$. It corresponds to the non-stationary marginal variance of $w(\mathcal{S})$, as introduced in Equation (3).

$$\underbrace{[u_\sigma, \beta_\sigma \,|\, w(\mathcal{S}), \gamma_\sigma]}_{sufficient} \rightarrow \underbrace{[w^*(\mathcal{S}) \,|\, w(\mathcal{S}), \sigma(\mathcal{S})]}_{deterministic} \rightarrow \underbrace{[u_\sigma, \beta_\sigma \,|\, w^*(\mathcal{S}), \gamma_\sigma, \beta, \tau]}_{ancillary} \rightarrow \underbrace{[w(\mathcal{S}) \,|\, w^*(\mathcal{S}), \sigma(\mathcal{S})]}_{deterministic}$$

This procedure is more complex than its counterpart for the noise variance in the step 3 because it interweaves the ancillary and sufficient parametrization of the latent field, with $w^*(\mathcal{S}) = \tilde{R}_0(\mathcal{S})\mathrm{diag}(1/\sigma(\mathcal{S}))w(\mathcal{S})$, the decomposition of the NNGP factor into $\tilde{R}(\mathcal{S}) = \tilde{R}_0(\mathcal{S})\mathrm{diag}(1/\sigma(\mathcal{S}))$ being defined in Equation (8). This formula allows to perform the two deterministic stages by using sparse triangular solving. User-specified priors for $\beta_\sigma$, and standard Normal prior for $u_\sigma$, are being used as described in 3.1.

In the sufficient step, the Gaussian density of the latent field from (10($b$)) is evaluated,

while $w(\mathcal{S})$ is fixed. The target density is therefore

$$N(\beta_\sigma, \mu_{\beta_\sigma}, V_{\beta_\sigma})N(u_\sigma, 0, I_r)N(w(\mathcal{S}), 0, \tilde{\Sigma}(\mathcal{S}, \beta_\sigma, \gamma_\sigma)),$$

the covariance matrix $\tilde{\Sigma}(\mathcal{S}, \beta_\sigma, \gamma_\sigma)$ being expressed as a "sandwich"

$$\text{diag}(exp(X_\sigma(\mathcal{S})\beta_\sigma + \sqrt{\gamma_\sigma}B(\mathcal{S})u_\sigma)) \left(\tilde{R}_0(\mathcal{S})^{\mathrm{T}}\tilde{R}_0(\mathcal{S})\right)^{-1} \text{diag}(exp(X_\sigma(\mathcal{S})\beta_\sigma + \sqrt{\gamma_\sigma}B(\mathcal{S})u_\sigma)),$$

where $R_0(\mathcal{S}))$ does not depend on the marginal variance $\sigma(\mathcal{S})$, following Equation (8). In the ancillary step, the response likelihood is evaluated instead, $\tilde{R}(\mathcal{S})^{-1}w^*(\mathcal{S})$ being plugged in $(10(a))$. The target distribution is

$$\begin{aligned} &N(\beta_\sigma, \mu_{\beta_\sigma}, V_{\beta_\sigma}) \times N(u_\sigma, 0, I_r) \\ &\qquad \times N(z, X\beta + M\text{diag}(exp(X_\sigma\beta_\sigma + \sqrt{\gamma_\sigma}B(\mathcal{S})u_\sigma))\tilde{R}_0(\mathcal{S})^{-1}w^*(\mathcal{S}), \text{diag}(\tau^2)). \end{aligned}$$

The rationale of the interweaving strategy is expounded in Section 4.2 of the main body of the paper, and explained more in detail in Section S5.1 of this Supplement. Both sampling steps are performed with HMC. The general method for HMC is given in 4.3, while details concerning the computation of the gradients of the two densities are presented in Section S6.2.

6. Metropolis-Hastings within Nested Interweaving update of $\gamma_\sigma$.

$$\left.\begin{aligned} &[\gamma_\sigma \,|\, W_\sigma(\mathcal{S})] \to [u_\sigma \,|\, \gamma_\sigma, W_\sigma(\mathcal{S})] \to [\gamma_\sigma \,|\, u_\sigma, w(\mathcal{S}), \beta_\sigma] \to [W_\sigma(\mathcal{S}) \,|\, \gamma_\sigma, u_\sigma] \to [w^*(\mathcal{S}) \,|\, w(\mathcal{S}), \sigma(\mathcal{S})] \\ &\underbrace{\to [\gamma_\sigma \,|\, W_\sigma(\mathcal{S})] \to [u_\sigma \,|\, \gamma_\sigma, W_\sigma(\mathcal{S})] \to [\gamma_\sigma | u_\sigma, w^*(\mathcal{S}), \beta_\sigma, \beta] \to [W_\sigma(\mathcal{S}) \,|\, \gamma_\sigma, u_\sigma] \to [w(\mathcal{S}) \,|\, w^*(\mathcal{S}), \sigma(\mathcal{S})]}_{\text{Interweaving } W_\sigma(\mathcal{S}) \text{ and } u_\sigma(\mathcal{S})} \end{aligned}\right\} \text{Interweaving } w(\mathcal{S}) \text{ and } w^*(\mathcal{S})$$

This step relies on the *nested interweaving* strategy, and as such, it is especially complex, with 4 distinct sampling operations and 4 deterministic transforms. The rationale of the method is explained in Section S5.2. The main structure is to interweave with respect to $w(\mathcal{S})$ and $w^*(\mathcal{S})$, like in 3. The first line samples $\gamma_\sigma$ conditionally on $w(\mathcal{S})$, while the second samples conditionally on $w^*(\mathcal{S})$. As a consequence, each line is ended by a deterministic transform arising from $w^*(\mathcal{S}) = \tilde{R}(\mathcal{S})w(\mathcal{S})$. Within each line, interweaving is performed with respect to $u_\sigma$ and $W_\sigma(\mathcal{S})$, like in 4. All the steps consisting in sampling $[u_\sigma \,|\, \gamma_\sigma, W_\sigma(\mathcal{S})]$ and $[W_\sigma(\mathcal{S}) \,|\, \gamma_\sigma, u_\sigma]$ are deterministic, in virtue of Equation (11).

While $\gamma_\theta$ is sampled from 4 distinct densities, they all share the same prior: an Uniform prior defined in $(10(g))$.

The problem of sampling $[\gamma_\sigma \,|\, W_\sigma(\mathcal{S})]$ is analogous to that of sampling $[\gamma_\tau \,|\, W_\tau(\mathcal{S})]$, presented in step 4, because $W_\sigma(\mathcal{S})$ is a *sufficient* parametrization: conditionally on $W_\sigma(\mathcal{S})$, $\gamma_\sigma$ is independent from both $w(\mathcal{S})$ or $w^*(\mathcal{S})$.

The two other sampling operations, in turn, are closely related to those presented in the step 5. The density needed to sample from $[\gamma_\sigma \,|\, u_\sigma, w(\mathcal{S}), \beta_\sigma]$ starts from Equation (11), which states that changing $\gamma_\sigma$ affects $W_\sigma(\mathcal{S})$ through multiplication. In turn, $W_\sigma(\mathcal{S})$ is plugged into Equation $(10(c))$, changing $\sigma(\mathcal{S})$. Eventually, the prior of $w(\mathcal{S})$, in $(10(b))$, is affected. Compounding those elements leads to the target distribution

$$U(\gamma_\sigma, a_{\gamma_\sigma}, b_{\gamma_\sigma})N(w(\mathcal{S}), 0, \tilde{\Sigma}(\gamma_\sigma)),$$

$\tilde{\Sigma}(\gamma_\sigma)$ being the NNGP prior matrix from $(10(b))$ evaluated with respect to $\gamma_\sigma$

$$\text{diag}(exp(X_\sigma(\mathcal{S})\beta_\sigma + \sqrt{\gamma_\sigma}B(\mathcal{S})u_\sigma)) \left(\tilde{R}_0(\mathcal{S})^{\mathrm{T}}\tilde{R}_0(\mathcal{S})\right)^{-1} \text{diag}(exp(X_\sigma(\mathcal{S})\beta_\sigma + \sqrt{\gamma_\sigma}B(\mathcal{S})u_\sigma)).$$

Like in the step 5, the "sandwiched" form of the covariance arises from Equation (8), the correlation NNGP factor $R_0(\mathcal{S})$ being impervious to changes in $\sigma(\mathcal{S})$.

The last step $[\gamma_\sigma | u_\sigma, w^*(\mathcal{S}), \beta_\sigma, \beta]$ differs from the step previously presented by the fact that $w^*(\mathcal{S})$ is used. As a consequence, instead of evaluating the GP prior ($10(b)$), a new Gaussian field $\tilde{R}(\mathcal{S})^{-1}w^*(\mathcal{S})$ is plugged into the response likelihood ($10(a)$). The modified target density then writes as:

$$U(\gamma_\sigma, a_{\gamma_\sigma}, b_{\gamma_\sigma})N(z, X\beta + M\text{diag}(exp(X_\sigma\beta_\sigma + \sqrt{\gamma_\sigma}B(\mathcal{S})u_\sigma))\tilde{R}_0(\mathcal{S})^{-1}w^*(\mathcal{S}), \text{diag}(\tau^2)).$$

The low dimension of $\gamma_\sigma$ allows for metropolis-Hastings sampling in all four stages.

7. HMC-within-interweaving update of the range parameters $u_\alpha$ and $\beta_\alpha$.

$$\underbrace{[u_\alpha, \beta_\alpha \,|\, w(\mathcal{S}), \gamma_\alpha]}_{sufficient} \rightarrow \underbrace{[w^*(\mathcal{S}) \,|\, w(\mathcal{S}), \alpha(\mathcal{S})]}_{deterministic} \rightarrow \underbrace{[u_\alpha, \beta_\alpha \,|\, w^*(\mathcal{S}), \gamma_\alpha, \beta]}_{ancillary} \rightarrow \underbrace{[w(\mathcal{S}) \,|\, w^*(\mathcal{S}), \alpha(\mathcal{S})]}_{deterministic}$$

When the Gaussian field is locally isotropic, the local range parameter is a field of scalars $\alpha(\mathcal{S})$ who acts upon the Paciorek covariance, described in Equation (5). This field of scalars is linked to $W_\alpha(\mathcal{S})$ and $\beta_\alpha$ by the mean of the logarithmic transform of ($10(c)$). Eventually, $W_\alpha(\mathcal{S})$ is linked to $u_\alpha$ and $\gamma_\alpha$ by the low-rank process described in Equation (11), leading to

$$\alpha(\mathcal{S}) \;=\; exp(X_\alpha(\mathcal{S})\beta_\alpha + \sqrt{\gamma_\alpha}B(\mathcal{S})u_\alpha).$$

The symbolic layout of the current step is homologous to step 5, the deterministic steps being exactly the same. However, $\alpha(\mathcal{S})$ acts upon $\tilde{R}_0(\mathcal{S})$, while the marginal variance $\sigma(\mathcal{S})$ is fixed. When the sufficient parametrization $w(\mathcal{S})$ is used, the covariance in ($10(b)$) is affected, yielding the target density

$$N(\beta_\alpha, \mu_{\beta_\alpha}, V_{\beta_\alpha})N(u_\alpha, 0, I_r)N(w(\mathcal{S}), 0, \text{diag}(\sigma(\mathcal{S}))\left(\tilde{R}_0(\mathcal{S})^{\mathrm{T}}R_0(\mathcal{S})\right)^{-1}\text{diag}(\sigma(\mathcal{S}))).$$

When the ancillary parametrization is used, ($10(b)$) is replaced by ($10(a)$), giving the

target density

$$N(\beta_\alpha, \mu_{\beta_\alpha}, V_{\beta_\alpha}) N(u_\alpha, 0, I_r) N(z, X\beta + M\text{diag}(\sigma(\mathcal{S}))\tilde{R}_0(\mathcal{S}, \alpha(\mathcal{S}))^{-1} w^*(\mathcal{S}), \text{diag}(\tau^2)).$$

While the same chain rule scheme from Section 4.3 is applied, the practical implementation to find the gradients is more complex, with details given in Sections S6.3 and S6.5.

In the case of anisotropic range parameters from Equation (12), the modified parameters $W_A(\mathcal{S})$ and $\beta_A$ have $d(d+1)/2$ columns, 3 in practice with $d = 2$. The results of Sections S6.3 and S6.5 are applied to the modified covariance parameters. However, due to HMC tuning parameter identification, the HMC step size is tuned separately:

(a) For the first column of both $\beta_A$ and $u_A$. Those two first columns inflate and deflate the ellipses.

(b) For the two remaining columns of both $\beta_A$ and $u_A$, who rotate and squash the ellipses.

Albeit having separate HMC step size, those updates are done in one shared HMC step.

8. Nested Interweaving update of $\gamma_\alpha$ with HMC.

$$\left.\begin{array}{l} [\gamma_\alpha \,|\, W_\alpha(\mathcal{S})] \rightarrow [u_\alpha \,|\, \gamma_\alpha, W_\alpha(\mathcal{S})] \rightarrow [\gamma_\alpha \,|\, u_\alpha, w(\mathcal{S}), \beta_\alpha] \rightarrow [W_\alpha(\mathcal{S}) \,|\, \gamma_\alpha, u_\alpha] \rightarrow [w^*(\mathcal{S}) \,|\, w(\mathcal{S}), \alpha(\mathcal{S})] \\ \underbrace{\rightarrow [\gamma_\alpha \,|\, W_\alpha(\mathcal{S})] \rightarrow [u_\alpha \,|\, \gamma_\alpha, W_\alpha(\mathcal{S})] \rightarrow [\gamma_\alpha \,|\, u_\alpha, w^*(\mathcal{S}), \beta_\alpha, \beta] \rightarrow [W_\alpha(\mathcal{S}) \,|\, \gamma_\alpha, u_\alpha] \rightarrow [w(\mathcal{S}) \,|\, w^*(\mathcal{S}), \alpha(\mathcal{S})]}_{\text{Interweaving } W_\alpha(\mathcal{S}) \text{ and } u_\alpha} \end{array}\right\} \text{Interweaving } w \text{ and } w^*(\mathcal{S})$$

The layout of this step is, conceptually, the same as that of step 6, with the difference that, in the NNGP density defined in Equation (8), the range parameters act upon $\tilde{R}_0(\mathcal{S})$ and not $\text{diag}(\sigma(\mathcal{S}))$, inducing target densities depending on $\alpha(\mathcal{S})$ like in 7.

In the eventuality of anisotropic range parameters, $\gamma_A$ has $(d(d+1)/2)(d(d+1)/2+1)/2$

distinct elements, amounting to 6 for $d = 2$. As a consequence, we use HMC for this step. The chain rule allows to modify the framework of 4.3. Details allowing to compute the relevant gradients are given in Sections S6.3 and S6.5.

# S5 Details about Interweaving

## S5.1 Interweaving

Interweaving is a method introduced by Yu and Meng (2011) to improve the mixing of models relying on data augmentation. Rather than finding a parametrization that "works well", this method exploits the discrepancy between two parametrizations of the latent field. Usually, various parametrizations of the data augmentation are available. For example, in the context of our NNGP model, the latent field $w(\mathcal{S}) \overset{a\ priori}{\sim} \mathcal{N}(0, \tilde{\Sigma}(\mathcal{S}))$ with $\tilde{\Sigma}(\mathcal{S}) = (\tilde{R}(\mathcal{S})^{\mathrm{T}}\tilde{R}(\mathcal{S}))^{-1}$ is reparametrized as $w^*(\mathcal{S}) = \tilde{R}(\mathcal{S})w(\mathcal{S}) \overset{a\ priori}{\sim} \mathcal{N}(0, I_n)$. The component-wise interweaving strategy of Yu and Meng (2011) can be applied when two data augmentations $w_1$ and $w_2$ have a joint distribution $[\theta, w_1, w_2]$ (even if it is degenerate) such that its marginals $[\theta, w_1]$ and $[\theta, w_2]$ correspond to the two models with the different data augmentations. It exploits the discordance between the two parametrizations to construct the following step in order to sample $\theta^{t+1}$:

$$[\theta, w_2' \,|\, w_1^t, \ldots] \to [\theta^{t+1}, w_1^{t+0.5} \,|\, w_2', \ldots],$$

where "..." denotes all the other parameters in the model. Since all draws are from full conditional distributions, the target joint distribution is preserved. Joint sampling of the parameter and data augmentation is much easier to implement when decomposed as

$$\underbrace{[\theta \,|\, w_1^t, \ldots] \to [w_2' \,|\, w_1^t, \theta, \ldots]}_{[\theta, w_2' \,|\, w_1^t, \ldots]} \to \underbrace{[\theta^{t+1} \,|\, w_2', \ldots] \to [w_1^{t+0.5} \,|\, w_2', \theta^{t+1}, \ldots]}_{[\theta^{t+1}, w_1^{t+0.5} \,|\, w_2', \ldots]}.$$

It is possible that the joint distribution is degenerate as long as it is well-defined, so that $[w_2' \mid \theta, w_1^t]$ and $[w_1^{t+0.5} \mid w_2', \theta^{t+1}, \ldots]$ are often deterministic transformations (in our application they are). Therefore, even though the data augmentation is changed at the end of the sampling of $\theta$, $w$ still has to be updated in a separate step in order to have an irreducible chain: that is why we indexed it by $t + 0.5$.

The method builds its efficiency upon the fact that the parameter $\theta$ sampled in the first step, $[\theta, w_2' \mid w_1^t, \ldots]$, is not used later in the algorithm. This first step is therefore equivalent to $[w_2' \mid w_1^t, \ldots]$. If the correlation between the two parametrizations $w_1$ and $w_2$ is small, the subsequent draw $[\theta^{t+1} \mid w_2', \ldots]$ produces a new parameter $\theta^{t+1}$ far from $\theta^t$ even if there is a strong correlation between $\theta$ and both $w_1$ and $w_2$. This is not an alternating scheme that requires $\theta$ to have a weak bond with at least either $w_1$ and $w_2$. Yu and Meng (2011) say, with a witty play on words: *coupling is more promising than compromising.*

This approach is based on the discordance between two parametrizations and it is a good choice to pick an ancillary-sufficient couple, giving an Ancillary-Sufficient Interweaving Strategy (AS-IS), even though any couple is admissible as long as it has a joint distribution. Following the terminology in Yu and Meng (2011), $w(\mathcal{S})$ is sufficient when *a posteriori* $p(\theta \mid w(\mathcal{S}), z) = p(\theta \mid w(\mathcal{S}))$, where $z$ is the observed data and $\theta$ is the target high-level parameter. It is sufficient when it is *a priori* independent from $\theta$, $p(\theta, w(\mathcal{S})) = p(\theta)p(w(\mathcal{S}))$. AS-IS has been shown to be effective for GP models: Filippone et al. (2013) show, empirically, that updating covariance parameters in a Gaussian Process model benefits from interweaving in the natural parametrization $w$ (sufficient) and the whitened parametrization $w^*(\mathcal{S})$ (ancillary), while Coube-Sisqueille and Liquet (2021) use centered and non-centered parametrizations, like in step 2 of Section S4, to efficiently sample the coefficients associated with the fixed effects.

## S5.2 Nested Interweaving for high-level parameters

The problem in our setting is exacerbated by the presence of latent fields on various layers of the hierarchical model. Nested AS-IS is envisioned by Yu and Meng (2011) for such configurations, although the authors do not provide any substantive applications. Let us consider a high-level parameter concerning the log-GP distributions of the covariance parameters, $\gamma_\theta$ from (10), with $\theta(\mathcal{S})$ denoting either the GP range $\alpha(\mathcal{S})$ or marginal standard deviation $\sigma(\mathcal{S})$. As before, $w(\mathcal{S})$ and $w^*(\mathcal{S})$ are the natural and whitened parametrizations, respectively, of the NNGP latent field from (10($a$)).

The vector $u_\theta$ gathers $r$ PP coefficients, so that $W_\theta(\mathcal{S}) = B(\mathcal{S})u_\theta$. Note that $W_\theta(\mathcal{S})$ is the sufficient parametrization and $u_\theta$ is the ancillary parametrization for the corresponding log-GP field of covariance parameters. The role of $u_\theta$ is homologuous to that of $w^*(\mathcal{S})$, with $u_\theta \overset{a\ priori}{\sim} I_r$ and $p(u_\theta, \gamma_\theta) = p(u_\theta)p(\gamma_\theta)$, while $B(\mathcal{S})$ is the low-rank counterpart of $\tilde{R}(\mathcal{S})^{-1}$.

A nested Interweaving step aiming to update $\gamma_\theta$ can be devised as

$$\left.\underbrace{\begin{array}{c} [\gamma_\theta, u_\theta, w^*(\mathcal{S}) \,|\, W_\theta(\mathcal{S}), w(\mathcal{S}), \ldots] \to [\gamma_\theta, W_\theta(\mathcal{S}), w^*(\mathcal{S}) \,|\, u_\theta, w(\mathcal{S}), \ldots] \\ \swarrow \\ [\gamma_\theta, u_\theta, w(\mathcal{S}) \,|\, W_\theta(\mathcal{S}), w^*(\mathcal{S}), \ldots] \to [\gamma_\theta, W_\theta(\mathcal{S}), w(\mathcal{S}) \,|\, u_\theta, w^*(\mathcal{S}), \ldots] \end{array}}_{\text{Interweaving } W_\theta(\mathcal{S})}\right\} \text{Interweaving } w(\mathcal{S}) \tag{S3}$$

Like before, it is much easier to sample sequentially, for example the blocked draw $[\gamma_\theta, u_\theta, w^*(\mathcal{S}) \,|\, W_\theta(\mathcal{S}), w(\mathcal{S}), \ldots]$ writes as

$$[\gamma_\theta \,|\, W_\theta(\mathcal{S}), w(\mathcal{S}), \ldots] \to \underbrace{[u_\theta \,|\, \gamma_\theta, W_\theta(\mathcal{S}), w(\mathcal{S}), \ldots]}_{\text{deterministic}} \to \underbrace{[w^*(\mathcal{S}) \,|\, \gamma_\theta, W_\theta(\mathcal{S}), u_\theta, w(\mathcal{S}), \ldots]}_{\text{deterministic}}.$$

# S6 Gradients for HMC updates of the covariance parameters

This section expounds how to obtain the gradients of the logarithms of the $h$-densities used in Section 4.3 following the role of the parameter in the model and the chosen parametrization. In 4.3, we use the vague notation $h(log(\lambda))$, $\lambda$ being a placeholder for the range $\alpha(\mathcal{S})$ or $A(\mathcal{S})$, the GP marginal variance $\sigma^2(\mathcal{S})$, and the noise variance $\tau^2$. For the sake of clarity, we shall index $h(\cdot)$ by the affected parameter. Moreover, S4 and S5.1 use two parametrizations for the latent field, the *sufficient* $w(\mathcal{S})$ and the *ancillary* $w^*(\mathcal{S})$. As a consequence, we shall also index $h(\cdot)$ by either *sufficient* and *ancillary* when relevant.

In the following, gradients and Jacobian matrices are extensively used. We index a gradient or a Jacobian by the arguments with respect to which differentiation is done. For example, $\nabla_{w(\mathcal{S})}p(z|w(\mathcal{S}), \beta, \tau)$ is a vector of size $n$ whose $i$-th element is given as $\partial p(z|w(\mathcal{S}), \beta, \tau)/\partial w(s_i)$.

## S6.1 Gradient of the log-density of the observations with respect to the latent NNGP field.

This preliminary result is useful to derive $h$-densities when the ancillary augmentation $w^*(\mathcal{S})$ is used. In order to obtain the gradient of the log-density of the observations with respect to the latent field, we reckon with the possibility that there can be several observations at the same spatial site. Consider a site $s \in \mathcal{S}$ and denote the observations there as $obs(s)$. Note that $obs(s) = \{i/M_{i,j} = 1\}$ in $(10(a))$, which are the indices of the observations made at the spatial site $s$. The conditional independence of the observations of $z$ in (10($a$)) imply

$$\frac{\partial p(z \,|\, w(\mathcal{S}), \beta, \tau)}{\partial w(s)} = \frac{\partial \Sigma_{x \in obs(s)} (z_x - X_{x,\cdot}\beta - w(s))^2 / 2\tau_x^2}{\partial w(s)}$$
$$= \Sigma_{x \in obs(s)} (w(s) - (z_x - X_{x,\cdot}\beta))/\tau_x^2,$$

which produces the desired result.

## S6.2 Gradient with respect to the log marginal variance $\log(\sigma^2(\mathcal{S}))$

*Sufficient augmentation.* If sufficient augmentation of the latent field, that is $w(\mathcal{S})$ is used, the marginal variance appears in the NNGP density of the latent field, specifically (10($b$)). Using (8), we can write

$$h_{\sigma^2}^{sufficient}(\log(\sigma^2(\mathcal{S}))) =$$
$$\exp\left(-\sigma^{-1}(\mathcal{S})^{\mathrm{T}} \ \mathrm{diag}(w(\mathcal{S}))\tilde{R}_0(\mathcal{S})^{\mathrm{T}}(\mathcal{S})\tilde{R}_0(\mathcal{S}) \ \mathrm{diag}(w(\mathcal{S}))\sigma^{-1}(\mathcal{S})/2\right) \Pi_{i=1}^n (\tilde{R}_0)_{i,i}/\sigma(s_i).$$

Therefore,

$$-\log\left(h_{\sigma^2}^{sufficient}(\log(\sigma^2(\mathcal{S})))\right) = \mathrm{const} + \textstyle\sum_{i=1}^n \log(\sigma(s_i)) +$$
$$(1/2)\sigma^{-1}(\mathcal{S})^{\mathrm{T}} \ \mathrm{diag}(w(\mathcal{S}))\tilde{R}_0(\mathcal{S})^{\mathrm{T}}(\mathcal{S})\tilde{R}_0(\mathcal{S}) \ \mathrm{diag}(w(\mathcal{S}))\sigma^{-1}(\mathcal{S})$$

is the negated log-density.

We note that

$$\nabla_{\log(\sigma^2(\mathcal{S}))} \sum_{i=1}^n \log(\sigma(s_i)) = \nabla_{\log(\sigma^2)} \sum_{i=1}^n \log((\sigma^2(s_i))^{1/2}) = \nabla_{\log(\sigma^2)} \sum_{i=1}^n \log(\sigma^2(s_i))/2 = (1/2)1_n \ , \tag{S4}$$

where $1_n$ is the $n \times 1$ vector of ones. The Jacobian of $\sigma^{-1}(\mathcal{S})$ with respect to $\log(\sigma^2(\mathcal{S}))$ is $J_{\log(\sigma^2(\mathcal{S}))}\sigma^{-1}(\mathcal{S}) = J_{\log(\sigma^2(\mathcal{S}))} \exp(-\log(\sigma^2(\mathcal{S}))/2) = -\mathrm{diag}(\sigma^{-1}(\mathcal{S})/2)$. We also use the gradient $\nabla_{\sigma^{-1}(\mathcal{S})}\sigma^{-1}(\mathcal{S})^{\mathrm{T}}\ \mathrm{diag}(w(\mathcal{S}))\tilde{R}_0(\mathcal{S})^{\mathrm{T}}\tilde{R}_0(\mathcal{S})\ \mathrm{diag}(w(\mathcal{S}))\sigma^{-1}(\mathcal{S})/2 =$

$\mathrm{diag}(w(\mathcal{S}))\tilde{R}_0(\mathcal{S})^{\mathrm{T}}\tilde{R}_0(\mathcal{S})\ \mathrm{diag}(w(\mathcal{S}))\sigma^{-1}(\mathcal{S}).$

With the Jacobian chain rule, we combine the two previous formulae to find

$$\begin{aligned} &-\nabla_{\log(\sigma^2(\mathcal{S}))}\sigma^{-1}(\mathcal{S})^{\mathrm{T}}\ \mathrm{diag}(w(\mathcal{S}))\tilde{R}_0(\mathcal{S})^{\mathrm{T}}\tilde{R}_0(\mathcal{S})\ \mathrm{diag}(w(\mathcal{S}))\sigma^{-1}(\mathcal{S})/2 \\ &\qquad = (1/2)\sigma^{-1}(\mathcal{S})\ \circ\ \left(\mathrm{diag}(w(\mathcal{S}))\tilde{R}_0(\mathcal{S})^{\mathrm{T}}\tilde{R}_0(\mathcal{S})\mathrm{diag}(w(\mathcal{S}))\sigma^{-1}(\mathcal{S})\right), \end{aligned} \tag{S5}$$

where $\circ$ denotes the Hadamard product. Combining (S4) and (S5) yields

$$\begin{aligned} &-\nabla_{\log(\sigma^2(\mathcal{S}))}log(h_{\sigma^2}^{sufficient}(\log(\sigma^2(\mathcal{S})))) = \\ &\quad (1/2)1_n - (1/2)\sigma^{-1}(\mathcal{S})\ \circ\ \left(\mathrm{diag}(w(\mathcal{S}))\tilde{R}_0(\mathcal{S})^{\mathrm{T}}\tilde{R}_0(\mathcal{S})\ \mathrm{diag}(w(\mathcal{S}))\sigma^{-1}(\mathcal{S})\right). \end{aligned} \tag{S6}$$

*Ancillary augmentation.* When ancillary augmentation is used, the marginal variance affects the observed field likelihood with respect to the latent field, that is the density ($10(a)$), through the relationship $w(\mathcal{S}) = \tilde{R}(\mathcal{S})^{-1}w^*(\mathcal{S})$ as expounded in S5.1. Let us define

$$-h_{\sigma^2}^{ancillary}(log(\sigma^2(\mathcal{S}))) = -p(z\,|\,w(\mathcal{S}) = \tilde{R}(\mathcal{S})^{-1}w^*(\mathcal{S}), \beta, \tau)$$

using ($10(a)$). The marginal variance affects the Gaussian density $p(\cdot)$ through $w(\mathcal{S}) = \tilde{R}(\mathcal{S})^{-1}w^*(\mathcal{S})$. We obtain

$$\nabla_{log(\sigma^2(\mathcal{S}))}p(z\,|\,w(\mathcal{S}), \beta, \tau) = J^{\mathrm{T}}_{\sigma^2(\mathcal{S})}w(\mathcal{S})\nabla_{w(\mathcal{S})}p(z\,|\,w(\mathcal{S}), \beta, \tau)$$

by applying the Jacobian chain rule. The ancillary parametrization and Equation (8) yield

$$J^{\mathrm{T}}_{log(\sigma^2(\mathcal{S}))}w(\mathcal{S}) = J^{\mathrm{T}}_{log(\sigma^2(\mathcal{S}))}(\tilde{R}_0(\mathcal{S})^{-1}w^*(\mathcal{S})) \circ \sigma(\mathcal{S}).$$

Applying the chain rule allows to reformulate as

$$J^{\mathrm{T}}_{log(\sigma^2(\mathcal{S}))}\sigma^2(\mathcal{S})J^{\mathrm{T}}_{\sigma^2(\mathcal{S})}\sigma(\mathcal{S})J^{\mathrm{T}}_{\sigma(\mathcal{S})}(\tilde{R}_0(\mathcal{S})^{-1}w^*(\mathcal{S}))\circ\sigma(\mathcal{S}) = diag(w(\mathcal{S})/2),$$

eventually leading to the result

$$-\nabla_{log(\sigma^2(\mathcal{S}))}h^{ancillary}_{\sigma^2}(log(\sigma^2(\mathcal{S}))) = -\nabla_{w(\mathcal{S})}\ p(z\,|\,w(\mathcal{S}),\beta,\tau)\ \circ(w(\mathcal{S})/2)\,, \quad \text{(S7)}$$

$\nabla_{w(\mathcal{S})}p(z\,|\,w(\mathcal{S}),\beta,\tau)$ being derived in Section S6.1.

## S6.3 Gradient of the negated log-density with respect to $\log(\alpha(\mathcal{S}))$

*Sufficient augmentation.* In sufficient augmentation, the range affects the NNGP prior of the latent field. We consider the negated log density of the latent field with sufficient augmentation arising from the Gaussian prior in $(10(b))$

$$-h^{sufficient}_{\alpha}(\log(\alpha)(\mathcal{S})) = \log\left(|\tilde{R}\,(\mathcal{S})\,|\right) + (1/2)w^{\mathrm{T}}\tilde{R}\,(\mathcal{S})^{\mathrm{T}}\,\tilde{R}\,(\mathcal{S})\,w(\mathcal{S}).$$

$\alpha(\mathcal{S})$ acts upon $\tilde{R}(\mathcal{S})$ through the correlation NNGP $\tilde{R}_0$ introduced in (8). The derivative of the log-determinant $\log(|\tilde{R}(\mathcal{S})|)$ is

$$\begin{aligned}
\frac{\partial}{\partial\log(\alpha(s_j))}\log(|\tilde{R}(\mathcal{S})|) &= \frac{\partial}{\partial\log(\alpha(s_j))}\sum_{i=1}^{n}\log(\tilde{R}(\mathcal{S})_{i,i}) \quad \text{(because } \tilde{R}(\mathcal{S}) \text{ is triangular)}\\
&= \sum_{i=1}^{n}\frac{\partial}{\partial\log(\alpha(s_j))}\log(\tilde{R}(\mathcal{S})_{i,i}) \quad \text{(interchange derivative and sum)}\\
&= \sum_{i:s_i\in\{s_j\cup\mathrm{ch}(s_j)\}}\frac{\partial}{\partial\log(\alpha(s_j))}\log(\tilde{R}(\mathcal{S})_{i,i}) \quad \text{(depends only on rows corr. to } s_j \text{ and ch}(s_j))\\
&= \sum_{i:s_i\in\{s_j\cup\mathrm{ch}(s_j)\}}\frac{1}{\tilde{R}(\mathcal{S})_{i,i}}\frac{\partial}{\partial\log(\alpha(s_j))}\tilde{R}(\mathcal{S})_{i,i} \quad \text{(log-function derivative)}\,.
\end{aligned}$$

The derivative of $(1/2)w(\mathcal{S})^{\mathrm{T}}\tilde{R}(\mathcal{S})^{\mathrm{T}}\tilde{R}(\mathcal{S})w$ is

$$
\begin{aligned}
\frac{\partial}{\partial \log(\alpha(s_j))}\left((1/2)w(\mathcal{S})^{\mathrm{T}}\tilde{R}(\mathcal{S})^{\mathrm{T}}\tilde{R}(\mathcal{S})w\right) &= (1/2)\frac{\partial}{\partial \log(\alpha(s_j))}\left((w(\mathcal{S})^{\mathrm{T}}\tilde{R}(\mathcal{S})^{\mathrm{T}})(\tilde{R}(\mathcal{S})w)\right) \\
&= (1/2)\left(\frac{\partial}{\partial \log(\alpha(s_j))}w(\mathcal{S})^{\mathrm{T}}\tilde{R}(\mathcal{S})^{\mathrm{T}}\right)\left(\tilde{R}(\mathcal{S})w\right)+ \\
&\quad (1/2)\left(w(\mathcal{S})^{\mathrm{T}}\tilde{R}(\mathcal{S})^{\mathrm{T}}\right)\frac{\partial}{\partial \log(\alpha(s_j))}\tilde{R}(\mathcal{S})w(\mathcal{S}) \\
&= \left(w(\mathcal{S})^{\mathrm{T}}\tilde{R}(\mathcal{S})^{\mathrm{T}}\right)\frac{\partial}{\partial \log(\alpha(s_j))}\tilde{R}(\mathcal{S})w(\mathcal{S}) = \left(w(\mathcal{S})^{\mathrm{T}}\tilde{R}(\mathcal{S})^{\mathrm{T}}\right)\frac{\partial}{\partial \log(\alpha(s_j))}\tilde{R}(\mathcal{S})w
\end{aligned}
$$

Combining the above calculations, we obtain the gradient as

$$
\begin{aligned}
-\frac{\partial}{\partial \log(\alpha(s_i))}h_{\alpha}^{sufficient}(\log(\alpha)) &= w(\mathcal{S})^{\mathrm{T}}\tilde{R}(\mathcal{S})^{\mathrm{T}}\left(\frac{\partial}{\partial \log(\alpha(s_i))}\tilde{R}(\mathcal{S})\right)w(\mathcal{S}) \\
&\quad + \sum_{j:s_j\in\{s_i\cup \mathrm{ch}(s_i)\}}\frac{1}{\tilde{R}(\mathcal{S})_{j,j}}\left(\frac{\partial}{\partial \log(\alpha(s_i))}\tilde{R}(\mathcal{S})_{j,j}\right). \qquad \text{(S8)}
\end{aligned}
$$

*Ancillary Augmentation.* When ancillary augmentation is used, the covariance parameters appear in the data likelihood given the latent field expressed in $(10(a)$. Let

$$
-h_{\alpha}^{ancillary}(\log(\alpha(\mathcal{S}))) = -p\left(z\,|\,w(\mathcal{S}) = \tilde{R}(\mathcal{S})^{-1}w^{*}(\mathcal{S}),\beta,\tau\right),
$$

Applying differentiation and using conditional independence, we obtain, for $1\leq j\leq n$

$$
\begin{aligned}
&\frac{\partial}{\partial \log(\alpha(s_j))}\left(-h_{\alpha}^{ancillary}(\log(\alpha(\mathcal{S})))\right) \\
&\overset{\text{cond. ind.}}{=} \sum_{i=1}^{n}\sum_{x\in obs(s_i)} -\frac{\partial}{\partial \log(\alpha(s_j))}p\left(z_x\,|\,w(s_i) = \left(\tilde{R}(\mathcal{S})^{-1}w^{*}(\mathcal{S})\right)_i,\beta,\tau\right) \\
&\overset{\text{chain rule}}{=} \sum_{i=1}^{n} -\frac{\partial}{\partial \log(\alpha(s_j))}\left(\tilde{R}(\mathcal{S})^{-1}w^{*}(\mathcal{S})\right)_i\left(\sum_{x\in obs(s_i)}\frac{\partial}{\partial w(s_i)}p\left(z_x\,|\,w(s_i) = \left(\tilde{R}(\mathcal{S})^{-1}w^{*}(\mathcal{S})\right)_i,\beta,\tau\right)\right) \\
&\overset{w^{*}(\mathcal{S})\text{ not changed by }\theta}{=} \sum_{i=1}^{n} -\left(\frac{\partial\tilde{R}(\mathcal{S})^{-1}}{\partial \log(\alpha(s_j))}w^{*}(\mathcal{S})\right)_i\left(\sum_{x\in obs(s_i)}\frac{\partial}{\partial w(s_i)}p\left(z_x\,|\,w(s_i) = \left(\tilde{R}(\mathcal{S})^{-1}w^{*}(\mathcal{S})\right)_i,\beta,\tau\right)\right) \\
&\overset{\text{differentiation of inverse}}{=} \sum_{i=1}^{n}\left(\tilde{R}(\mathcal{S})^{-1}\frac{\partial\tilde{R}(\mathcal{S})}{\partial \log(\alpha(s_j))}\tilde{R}(\mathcal{S})^{-1}w^{*}(\mathcal{S})\right)_i \\
&\qquad\qquad \left(\sum_{x\in obs(s_i)}\frac{\partial}{\partial w(s_i)}p\left(z_x\,|\,w(s_i) = \left(\tilde{R}(\mathcal{S})^{-1}w^{*}(\mathcal{S})\right)_i,\beta,\tau\right)\right) \\
&\overset{\text{using gradient of }p(\cdot)\text{ in }w}{=} \left(\nabla_{w(\mathcal{S})}p\left(z\,|\,\tilde{R}(\mathcal{S})^{-1}w^{*}(\mathcal{S}),\beta,\tau\right)\right)^{\mathrm{T}}\tilde{R}(\mathcal{S})^{-1}\left(\frac{\partial}{\partial \log(\alpha(s_j))}\tilde{R}(\mathcal{S})\right)\tilde{R}(\mathcal{S})^{-1}w^{*}(\mathcal{S})\,.
\end{aligned}
$$

These calculations produce the final result,

$$\frac{\partial}{\partial \log(\alpha(s_i))} h_\alpha^{ancillary}(\log(\alpha)) = \\ \left(\nabla_{w(\mathcal{S})} p(z(s_i) \,|\, w(\mathcal{S}), \beta, \tau)\right)^{\mathrm{T}} \tilde{R}(\mathcal{S})^{-1} \left(\frac{\partial}{\partial \log(\alpha(s_i))} \tilde{R}(\mathcal{S})\right) \tilde{R}(\mathcal{S})^{-1} w(\mathcal{S}) \;, \tag{S9}$$

the gradient $\nabla_{w(\mathcal{S})} p(z(s_i) \,|\, w(\mathcal{S}), \beta, \tau)$ being obtained in S6.1.

## S6.4 Computational cost of the gradient of the negated log-density with respect to $\log(\alpha)$

Both sufficient and ancillary formulations have a partial derivative with a term under the shape $u^{\mathrm{T}} \partial \tilde{R} / \partial(\log(\alpha(s_j)))v$, where $u$ and $v$ are two vectors with affordable computational cost whose value does not depend on $\log(\alpha(s_j))$. Our proposed construction ensures that $\partial \tilde{R} / \partial(\log(\alpha(s_j)))$ has non-null rows only at the rows that correspond to $s_j$ and $\mathrm{ch}(s_j)$, and each of those rows has itself at most $m+1$ non-null coefficients. Sparse matrix-vector multiplication is used to evaluate $(\partial \tilde{R}(\mathcal{S}) / \partial(\log(\alpha(s_j))))v$, which, therefore, costs $O((m+1) \times (1 + |\mathrm{ch}(s_j)|))$ floating point operations. Given the fact that $\sum_{j=1}^n |\mathrm{ch}(s_j)| = \sum_{j=1}^n |\mathrm{pa}(s_j)| = n \times m$, we can expect that the computational cost needed to compute $(\partial \tilde{R}(\mathcal{S}) / \partial(\log(\alpha(s_j))))v$ for $j \in 1, \ldots, n$ will be $O(n \times (m+1)^2)$ operations, which is linear in $n$ and, hence, significantly more affordable.

Moreover, since $\partial \tilde{R} / \partial(\log(\alpha(s_j)))$ has non-null rows only at the rows that correspond to $s_j$ and $\mathrm{ch}(s_j)$, we can deduce that $(\partial \tilde{R}(\mathcal{S}) / \partial(\log(\alpha(s_j))))v$ has non-null terms only at positions corresponding to location $s_i$ and its children. This implies that $u^{\mathrm{T}}(\partial \tilde{R}(\mathcal{S}) / \partial(\log(\alpha(s_j))))v$ costs $O(\mathrm{ch}(s_j) + 1)$ operations. Again, using $\sum_{j=1}^n |\mathrm{ch}(s_j)| = \sum_{j=1}^n |\mathrm{pa}(s_j)| = n \times m$, we find that (if we know already $(\partial \tilde{R}(\mathcal{S}) / \partial(\log(\alpha(s_j))))v$) computing $u^{\mathrm{T}}(\partial \tilde{R}(\mathcal{S}) / \partial(\log(\alpha(s_j))))v$ for $j \in 1, \ldots, n$ costs $O(n(m+1))$ floating point operations.

## S6.5 General derivative of $\tilde{R}(\mathcal{S})$ with respect to nonstationary range parameters

Our aim is to find $\partial \tilde{R}/\partial \log(\alpha(s_j))$ with $j \in 1, \ldots, n$. Let us focus on the $i^{th}$ row of $\tilde{R}(\mathcal{S})$, denoted as $\tilde{R}(\mathcal{S})_{i,\cdot}$. The index of row $i$ can be different from $j$. To find the derivative of $\tilde{R}(\mathcal{S})_{i,\cdot}$ with respect to $\log(\alpha(s_j))$, we use the covariance matrix between $s_i$ and its parents $\text{pa}(s_i)$. Let $\Sigma^i$ be the covariance matrix corresponding to $(\text{pa}(s_i), s_i)$, which we express as $\Sigma^i = \left[\begin{array}{c|c} \Sigma^i_{11} & \Sigma^i_{12} \\ \hline \Sigma^i_{21} & \Sigma^i_{22} \end{array}\right]$, where $\Sigma^i_{11}$ is the $m \times m$ covariance matrix corresponding to $\text{pa}(s_i)$, where $m = |\text{pa}(s_i)|$, and $\Sigma^i_{22}$ is $1 \times 1$ corresponding to $s_i$. From its construction, $\tilde{R}(\mathcal{S})_{i,\cdot}$ has non-null coefficients only for the column entries that correspond to $s_i$ and its parents $\text{pa}(s_i)$. Hence, there is no need to compute the gradient but for those coefficients. The diagonal element $\tilde{R}(\mathcal{S})_{i,i}$ has value $1/\bar{\sigma}_i$, where $\bar{\sigma}_i$ is the standard deviation of $w(s_i)$ conditional on $w(\text{pa}(s_i))$. The elements that correspond to $\text{pa}(s_i)$ has the value $-\Sigma^i_{21}(\Sigma^i_{11})^{-1}/\bar{\sigma}_i$.

We now consider the derivative of the diagonal coefficient $\tilde{R}(\mathcal{S})_{i,i}$.

$$
\begin{aligned}
\partial(\tilde{R}(\mathcal{S})_{ii})/\partial \log(\alpha(s_j)) &= \partial((\bar{\sigma}_i^2)^{-1/2})/\partial \log(\alpha(s_j)) \quad \text{(chain rule)} \\
&= -(\bar{\sigma}_i^{-3}/2) \times \partial(\bar{\sigma}_i^2)/\partial \log(\alpha(s_j)) \quad \text{(using conditional variance formula)} \\
&= -(\bar{\sigma}_i^{-3}/2) \times \partial(\Sigma^i_{22} - \Sigma^i_{21}(\Sigma^i_{11})^{-1}\Sigma^i_{12})/\partial \log(\alpha(s_j)) \quad \text{(product rule)} \\
&= -(\bar{\sigma}_i^{-3}/2) \times \partial(\Sigma^i_{22})/\partial \log(\alpha(s_j)) \\
&\qquad + (\bar{\sigma}_i^{-3}) \times \partial(\Sigma^i_{21})/\partial \log(\alpha(s_j))(\Sigma^i_{11})^{-1}\Sigma^i_{12} \\
&\qquad + (\bar{\sigma}_i^{-3}/2) \times \Sigma^i_{21}\partial\left((\Sigma^i_{11})^{-1}\right)/\partial \log(\alpha(s_j))\Sigma^i_{12} \quad \text{(derivative of inverse)} \\
&= -(\bar{\sigma}_i^{-3}/2) \times \partial(\Sigma^i_{22})/\partial \log(\alpha(s_j)) \\
&\quad + (\bar{\sigma}_i^{-3}) \times \partial(\Sigma^i_{21})/\partial \log(\alpha(s_j))(\Sigma^i_{11})^{-1}\Sigma^i_{12} \\
&\quad - (\bar{\sigma}_i^{-3}/2) \times \Sigma^i_{21}(\Sigma^i_{11})^{-1}\left(\partial(\Sigma^i_{11})/\partial \log(\alpha(s_j))\right)(\Sigma^i_{11})^{-1}\Sigma^i_{12}\,.
\end{aligned}
$$

Differentiating coefficients corresponding to $\text{pa}(s_i)$ to the left of the diagonal in $\tilde{R}(\mathcal{S})_{i,\cdot}$, yields

$$
\begin{aligned}
\partial(-\Sigma^i_{21}(\Sigma^i_{11})^{-1}/\bar{\sigma}_i)/\partial\log(\alpha(s_j)) &= -\partial(\Sigma^i_{21}(\Sigma^i_{11})^{-1}\times\tilde{R}(\mathcal{S})_{ii})/\partial\log(\alpha(s_j)) \quad \text{(product rule)} \\
&= -\left(\partial\Sigma^i_{21}/\partial\log(\alpha(s_j))\right)(\Sigma^i_{11})^{-1}\times\tilde{R}(\mathcal{S})_{ii} \\
&\quad - \Sigma^i_{21}\left(\partial\left((\Sigma^i_{11})^{-1}\right)/\partial\log(\alpha(s_j))\right)\times\tilde{R}(\mathcal{S})_{ii} \\
&\quad - \Sigma^i_{21}(\Sigma^i_{11})^{-1}\partial\tilde{R}(\mathcal{S})_{ii}/\partial\log(\alpha(s_j)) \quad \text{(derivative of inverse)} \\
&= -\left(\partial\Sigma^i_{21}/\partial\log(\alpha(s_j))\right)(\Sigma^i_{11})^{-1}\times\tilde{R}(\mathcal{S})_{ii} \\
&\quad + \Sigma^i_{21}(\Sigma^i_{11})^{-1}\left(\partial\Sigma^i_{11}/\partial\log(\alpha(s_j))\right)(\Sigma^i_{11})^{-1}\times\tilde{R}(\mathcal{S})_{ii} \\
&\quad - \Sigma^i_{21}(\Sigma^i_{11})^{-1}\times\underbrace{\partial\tilde{R}(\mathcal{S})_{ii}/\partial\log(\alpha(s_j))}_{\text{already known}}\,.
\end{aligned}
$$

These derivatives reveal that we only need $(\Sigma^i)^{-1}$ and $\partial\Sigma^i/\partial\log(\alpha(s_j))$ (with $s_j \in s_i \cup \text{pa}(s_i)$) to compute the derivative of $\tilde{R}(\mathcal{S})_{i,\cdot}$. The former can be reused in order to obtain $\tilde{R}(\mathcal{S})$. The latter can be approximated using finite differences,

$$
\frac{\partial\Sigma^i}{\partial\log(\alpha(s_j))} \approx \left(\Sigma^i(\log(\alpha(s_j)) + d\log(\alpha(s_j))) - \Sigma^i(\log(\alpha(s_j)))\right)/d\log(\alpha(s_j))\,.
$$

## S6.6 Computational cost of the derivative of $\tilde{R}(\mathcal{S})$ with respect to nonstationary range parameters

The derivatives of $\tilde{R}(\mathcal{S})_{i,\cdot}$ are non-null only for $\alpha(\text{pa}(s_i)\cup s_i)$ because the entries of $\Sigma^i$ are $K(s,t,\alpha(s),\alpha(t))$, where $s,t \in s_i \cup \text{pa}(s_i)$. Conversely, if $\log(\alpha(s_j))$ moves, only the rows of $\tilde{R}(\mathcal{S})$ corresponding to $s_i$ and its children on the DAG move. Therefore, to compute the derivative of $\tilde{R}(\mathcal{S})$ with respect to $\alpha_j$, we execute $|\text{ch}(s_j)|+1$ (and not $n$) number of row differentiation operations. Since $\sum_{j=1}^n |\text{ch}(s_j)| = \sum_{j=1}^n |\text{pa}(s_j)| = m\times n$ ($m$ is the number of nearest neighbors used in the NNGP), row differentiation must be executed $(m+1)\times n$ times in order to get all the derivatives of $\tilde{R}(\mathcal{S})$ with respect to $\alpha(s_1,\dots,s_n)$. Furthermore, one row has $m+1$ non-null terms and $(m+1)\times n$ rows are differentiated. Therefore, the cost in RAM to store the differentiation of $\tilde{R}(\mathcal{S})$ is $O(m+1)^2 n$.

While the flop cost of differentiation itself may seem daunting, the problem is considerably mitigated because spatially-varying covariance parameters only affect pairwise covariance. In the derivatives, there are only 3 terms that depend on $\alpha(s_j)$, they are $\partial(\Sigma^i_{22})/\partial\log(\alpha(s_j))$, $\partial(\Sigma^i_{12})/\partial\log(\alpha(s_j))$, and $\partial(\Sigma^i_{11})/\partial\log(\alpha(s_j))$. It is instructive to separate the cases.

1. When $i \neq j$

   (a) $\partial(\Sigma^i_{12})/\partial\log(\alpha(s_j))$ has only one non-null coefficient.

   (b) $\partial(\Sigma^i_{11})/\partial\log(\alpha(s_j))$ is $m \times m$ with cross structure (non-null coefficients only for the row and the column corresponding to $s_j$).

   (c) $\partial(\Sigma^i_{22})/\partial\log(\alpha(s_j))$ is a null $1 \times 1$ matrix.

2. When $i = j$

   (a) $\partial(\Sigma^i_{12})/\partial\log(\alpha(s_j))$ is a dense vector of length $m$.

   (b) $\partial(\Sigma^i_{11})/\partial\log(\alpha(s_j))$ is null.

   (c) $\partial(\Sigma^i_{22})/\partial\log(\alpha(s_j))$ is null because a change in $\log(\alpha)(s_i)$ does not affect the marginal variance of $w(s_i)$ (a change in $w_{\sigma^2(s_i)}$ does).

The costliest task is to compute $(\Sigma^i_{11})^{-1}$. However, this needs to be computed only once because it is not affected by differentiation. Moreover, $(\Sigma^i_{11})^{-1}$ and $\Sigma^i_{21}(\Sigma^i_{11})^{-1}$ are used to compute $\tilde{R}(\mathcal{S})$ and then recycled on the fly to compute the derivatives. The computational effort required here can then be subtracted from the cost of the derivative. Applying the above remarks produces Table S5.

Table S5: costs to compute $\partial\tilde{R}(\mathcal{S})_{i,\cdot}/\partial(\log(\alpha(s_j)))$

| | (a) | (b) | (c) | (d) | (e) | (f) |
|---|---|---|---|---|---|---|
| $i = j$ | $O(1)$ | $O(m)$ | 0 | $O(m^2)$ | 0 | 0 |
| $s_i \in \text{ch}(s_j)$ | 0 | $O(1)$ | $O(m)$ | $O(m)$ | $O(m)$ | 0 |

Using Table S5 and, again, since $\sum_{j=1}^{n}|\text{ch}(s_j)| = \sum_{j=1}^{n}|\text{pa}(s_j)| = m \times n$, the matrix operations should have a total cost of $O(m^2 \times n)$. The cost of the finite difference approximation to $\partial\Sigma^i/\partial\log(\alpha(s_j))$ must be added to this. The cost of computing the finite differences in one coefficient of $\Sigma^i$ depends on whether isotropic or anisotropic range parameters are used. In the case of isotropic range parameters, only only needs to compute the covariance function (5) with range $\exp(\log(\alpha(s) + \phi)), \phi << 1$ instead of $\exp(\log(\alpha(s)))$. In the other case, the SVD of $\log(A)$ is computed again. What is more, the covariance function (4) involves the Mahalanobis distance instead of the Euclidean distance. The cost here will, then, depend on $d$, and be higher than in the case with isotropic covariance parameters. However, (3) reveals that if $\log(\alpha)(s_j)$ moves, only the row and column of $\Sigma^i$ that correspond to $s_j$ is affected. Moreover, due to the symmetry of $\Sigma^i$, the row and the column is changed exactly the same way. Therefore, computing $\partial\Sigma^i/\partial\log(\alpha(s_j))$ involves only $m+1$ finite differences since $\Sigma^i$ is of size $(m+1)\times(m+1)$.

The finite difference $\partial\Sigma^i/\partial\log(\alpha(s_j))$ must be computed $m+1$ times for each row of $\tilde{R}(\mathcal{S})$, and there are $n$ rows. Therefore, the total cost of the finite differences is $O(m+1)^2n$ flops. This suggests that carefully executing the differentiation $\partial\tilde{R}(\mathcal{S})/\partial(\alpha(s_1,\ldots,s_n))$ costs $O(n(m+1)^2)$ operations, in the same order as computing $\tilde{R}(\mathcal{S})$ itself (Guinness, 2018).

## S6.7 Gradient of the negated log-density with respect to $\log(\tau^2)$

$\tau^2$ acts upon the Gaussian likelihood of the observations laid out in $(10(a))$, inducing

$$-h_\tau(log(\tau^2)) = \Sigma_{i=1}^{|z|} log(\tau_i^2)/2 + (z_i - (Mw(\mathcal{S}))_i - X_{i,\cdot}\beta)^2/2\tau_i^2.$$

Differentiation is done with respect to the $i$-th element of $\tau^2$, with $1 \leq i \leq |z|$, leading to :

$$-\frac{\partial}{\partial\log(\tau_i^2)} h_\tau(\log(\tau^2)) = 1/2 + (z_i - (Mw(\mathcal{S}))_i - X_{i,\cdot}\beta)^2/2\tau_i^2. \tag{S10}$$

# S7 Experiments with synthetic data

## S7.1 Objectives

We would like to investigate the improvements caused by using nonstationary modeling when it is relevant, the problems caused by using nonstationary modeling when it is irrelevant, and the potential identification and overfitting problems of the model we devised. Our general approach to find answers to those questions is to run our implementation on synthetic data sets and analyze their results. Following the nonstationary process and data model we defined using (10) and (12), there is 12 possible configurations counting the full stationary case: 2 marginal variance models, 2 noise variance models, 3 range models. In order to keep the Section readable, we use the following notation for the different models:

- ($\emptyset$) is the stationary model.
- ($\sigma^2$) is a model with nonstationary marginal variance of the GP $w$.
- ($\tau^2$) is a model with heteroskedastic noise variance.
- ($\alpha$) is a model with nonstationary range and isotropic range parameters.
- ($A$) is a model with nonstationary range and elliptic range parameters.
- Complex models are noted using "+". For example, a model with nonstationary marginal variance and heteroskedastic noise variance is noted ($\sigma^2 + \tau^2$).

## S7.2 Method

Our approach here is to use a possibly misspecified model and see what happens. Four cases are possible:

- The “right” model, in the sense it matches perfectly the process used to generate the data (however, potential identification and overfitting problems may cause it to be a bad model in practice).
- “Wrong” models, where some parameters that are stationary in the data are non-stationary in the model, and some parameters that are stationary in the model are nonstationary in the data.
- Under-modeling, where some parameters that are stationary in the model are nonstationary in the data, but all parameters that are stationary in the data are stationary in the model.
- Over-modeling, where some parameters that are stationary in the data are nonstationary in the model, but all parameters that are stationary in the model are stationary in the data.

If a nonstationary model is needed for data analysis, we should see if the “right” model does better than “under-modeling”. The problem of overfitting will be assessed by comparing over-modeling, under-modeling, and the “right” model. If there is some overfitting, over-modeling or even “right”modeling would have worse performances than simpler models. Identification problems will be monitored by looking at the “wrong” models and under-modeling. If some model formulations are interchangeable, then some of the “wrong” models should perform as good as the “right” model. Also, if two parametrizations are equivalent, then using either parametrization should do as good as using both, therefore under-modeling should do as good as the “true” model.

The models are compared using several indicators. The first is the Deviance Information Criterion (DIC) (Spiegelhalter et al., 1998). The latent field retrieval is assessed using the smoothing MSE and the prediction MSE. The covariance parameter retrieval is assessed by treating these parameters as latent fields and using smoothing MSE, no prediction was done because of the very high spatial coherence of those latent fields. Eventually, credible interval coverage is studied for the high-level parameters, that is the response, range, noise, and latent field marginal variance intercept, and the covariance parameters marginal variance. The MSEs are derived as follows. The observations of the model at the observed sites $\mathcal{S}$ are disrupted by the white noise $\epsilon$. The true latent field, used to generate the data, is named $w_{true}$. The estimated latent field is named $\hat{w}$. The smoothing MSE is

$$MSE_{smooth} = \frac{1}{\#\mathcal{S}} \sum_{s \in \mathcal{S}} (\hat{w}(s) - w_{true}(s))^2.$$

The smoothing MSE is also used to assess the estimation of the spatially-indexed covariance parameters $\theta$

$$MSE_{\theta} = \frac{1}{\#\mathcal{S}} \sum_{s \in \mathcal{S}} (\log(\hat{\theta}(s)) - \log(\theta_{true}(s)))^2,$$

and log noise variance $\tau^2$

$$MSE_{\theta} = \frac{1}{\#\mathcal{S}} \sum_{s \in \mathcal{S}} (\log(\hat{\tau}^2(s)) - \log(\tau^2_{true}(s)))^2.$$

While several observations are allowed for one spatial site, $\tau^2$ does not vary within one spatial site for this particular model specification where $X_\tau$ is only an intercept, allowing to compute the MSE for $\tau$ on spatial sites.

The field is also predicted at unobserved sites $\mathcal{P}$, giving the prediction MSE

$$MSE_{pred} = \frac{1}{\#\mathcal{P}} \sum_{s \in \mathcal{P}} (\hat{w}(s) - w_{true}(s))^2.$$

The results are analyzed using simple linear models. In an experiment, for each type of data, each measure of performance is regressed on the seed and the model formulation. The seed is an useful control variable, some data sets being "trickier" than others. We do not report the associated coefficients. In each case, we choose the "right" model to be the reference factor. The p-values are reported. For example, in the first experiment, there are eight types of data(($\emptyset$), ($\sigma^2$), ($\tau^2$), ($\alpha$), ($\sigma^2+\tau^2$), ($\tau^2+\alpha$), ($\sigma^2+\alpha$) and ($\sigma^2+\tau^2+\alpha$)) and six measures of the performance (the DIC, the smoothing MSE, the prediction MSE, the noise, range, and marginal variance MSE). Therefore, forty-eight linear models are run. That is, in Table S6, each column of the sub-tables correspond to the coefficients of a linear model.

The following method was used to create one synthetic data set.

1. 12000 locations are drawn uniformly on a square whose sides have length 5.

2. The 10000 first locations are kept for training. 20000 observations are done at these locations. First, each location is granted an observation. Then, each of the 10000 remaining observations is assigned to a location chosen following an uniform multinomial distribution.

3. Predictive Process bases $B(\mathcal{S})$ (see S3.1) for the parameter latent fields are defined. The marginal variance is unitary, the range is equal to 0.5, and the Matérn smoothness is equal to 1.5. There are 49 knots.

4. The parameter fields are sampled following (10):

   - The spatial range $\log(\alpha(\mathcal{S})) = \log(.1) + \sqrt{\gamma}_\alpha B(\mathcal{S})u_\alpha$, with $u_\alpha \overset{ind}{\sim} \mathcal{N}(0,1)$, following Equations (10($c$)), (10($d$)), and (11).

   - The process marginal variance $\log(\sigma(\mathcal{S})) = \log(10)+\sqrt{\gamma}_\sigma B(\mathcal{S})u_\sigma$, with $u_\sigma \overset{ind}{\sim} \mathcal{N}(0,1)$, following Equations (10($c$)), (10($d$)), and (11).

   - The noise variance $\log(\tau) = \log(10) + \sqrt{\gamma}_\tau MB(\mathcal{S})u_\tau$, with $u_\tau \overset{ind}{\sim} \mathcal{N}(0,1)$, following Equations (10($e$)), (10($f$)), and (11).

   If $\gamma_\cdot = 0$ then the parameter field is stationary.

5. The nonstationary NNGP latent field $w_{true}(\mathcal{S})$ is sampled using $\alpha(\mathcal{S})$ and $\sigma(\mathcal{S})$, using a Matern function with smoothness 1.5 as the isotropic kernel of (4).

6. The response variable z is sampled by adding a Gaussian noise with variance $\text{diag}(\tau^2)$ to $Mw_{true}(\mathcal{S})$, following Equation $(10(a))$.

We started with the eight models obtained by combining $(\sigma), (\alpha)$, and $(\tau)$, giving us $(\emptyset)$, $(\sigma^2)$, $(\tau^2)$, $(\alpha)$, $(\sigma^2 + \tau^2)$, $(\tau^2 + \alpha)$, $(\sigma^2 + \alpha)$, and $(\sigma^2 + \tau^2 + \alpha)$. We tested each data-model configuration, yielding 64 situations in total. 49 spatial basis functions were used to sample the latent parameter fields. Each case was replicated 60 times. Second, we focused on the case of elliptic range parameters with the three models obtained by combining $(\alpha)$ and $(A)$, giving us $(\emptyset)$, $(\alpha)$, and $(A)$. Like before, we tested the 9 data-model configurations 60 times each.

## S7.3 Results

The results are presented in Tables S6, S7, S8, and S9.

Table S6 presents results for settings combining nonstationary, locally isotropic correlation, nonstationary marginal variance of the latent field, and heteroskedasticity of the noise. We can see that in general under-modeling (in gray) or "wrong" modeling (in brown) impairs the model performance in terms of smoothing MSE (Table S6a), prediction MSE (Table S6b), and DIC (Table S6c), usually with good significance. We also can see that over-modeling, in blue has little effect on performance. The same conclusions also seem to hold for the estimation of the fields of covariance parameters (Tables S6d, S6e, S6f), which is less surprising.

The 95% *a posteriori* coverage of the high-level parameters are presented in the isotropic runs S7. An important feature is that picking the "right model" (in green) or over-modeling (in blue) seem to provide satisfying coverage. Under-modeling or picking a "wrong" model, on the other hand, adversely affects coverage. An important point to take into account is that in Tables S7d, S7e and S7f, some of the zeroes and ones are due to model formulation, while others are due to genuine model performance. In Table S7f, the bottom-right corner is filled with ones, because the data sets are generated without nonstationarity in the noise and the model being stationary can only correctly estimate the marginal variance of the noise PP. On the other hand, the top-right corner is filled with zeroes, because the models are nonstationary and even if their estimate of the marginal variance of the PP is extremely low, it will never include the "true" null variance in their *posterior* intervals. The bottom-left corner is filled with zeroes because a stationary model does not retrieve a non-null marginal variance for the noise variance PP. Eventually, the ones on the top-left quadrant are informative and tell, along with other high coefficients, that there is some over-coverage.

The analysis of the performance of the anisotropic runs, in Table S8, shows modest yet significant improvements of the model to predict the latent field in relevant cases. The DIC is also improved. On the other hand, over-modelling does not increase the error criteria, or very slightly and with little significance in general. This is due to the ability of the model to regress towards a stationary state. Eventually, the parameter MSE show interesting results in term of parameter estimation. Table S8d shows that the locally anisotropic model is able to capture, to some extent, the parameters driving the anisotropy. What is more surprising is that S8c shows that it also improves the retrieval of the ellipse diameter, with respect to the locally isotropic model especially. Like before, over-modeling causes no problem. As a conclusion, using the locally anisotropic model will -slightly- improve prediction, but greatly boost the estimation of the covariance structure.

The coverage results of the anisotropic runs, presented in Table S9, need some clarification. The 6 parameters for the variance of the range correspond to the half-vectorization of $Var(W(s), W(s))$, referred to in Section 3 of the main body as the *within-site multi-variate variance parameter*. In the stationary case, this reuces to is a $3 \times 3$ null matrix. In the locally isotropic case, only its upper-left corner is non-null. In the anisotropic case, its diagonal is non-null. The ones in the six tables ranging from S9d to S9i are "false" due to model formulation. It is worth noticing that the upper-left coefficients of Tables S9g and S9i are below the target 0.95. Those parameters control the intensity of the anisotropy in the range latent field. This is quite coherent with the trace plots such as Figure S4(c): there, the true log marginal variance is 0, and we can see that the two PP marginal variance parameters associated with anisotropy are mixing between $-2$ and 0. Other runs confirmed this behavior. The three range intercepts, from Table S9a to S9c, correspond to the three components of the range parameter. The first component retrieves the ellipse determinant, while the other two parametrize the anisotropy. The ones are, once again, "false". The notable point is that, in Table S9a, we can see that using the right model formulation considerably improves the coverage of the determinant of the range ellipse. The same remark goes for the latent field's variance intercept (Figure S9j) and noise variance intercept (Figure S9k). Overall, even if the elliptic model tends to undershoot its anisotropy within-site variance parameters, it improves the coverage of the other parameters.

## S7.4 Identification between range and variance

Even though it seems difficult to identify nonstationary range and variance, a troubling observation shows that there is some kind of identification: when given the possibility, the model is able to make the right choice between the two parameters. In Figure S2, we used boxplots to summarize results of the models that estimate both nonstationary marginal variance and range. On the left (Figure S2(a)), we can see estimates for the log within-site variance of $W_\alpha(\mathcal{S})$'s log-GP prior. On the right (Figure S2(b)), we see its counterpart for $W_\sigma(\mathcal{S})$. In both subfigures, the boxplots are separated following the type of the data, $(\emptyset)$ being stationary data, $(\alpha)$ being data with nonstationary range, $(\sigma^2)$ being data with nonstationary variance, and $(\alpha + \sigma^2)$ being data with both nonstationarities. Recall that when the log marginal variance is low, the corresponding field is practically stationary. Then we can see that the right kind of nonstationarity is detected for all four configurations: when data is stationary, both log marginal variances are very low, when the data is $(\sigma^2)$, then only a Latent field smoothing.

Table S6: Performance of the models with nonstationary range, noise, and field marginal variance, in isotropic settings

| Type of model \ Type of data | $\alpha+\sigma+\tau$ | $\sigma+\tau$ | $\alpha+\tau$ | $\tau$ | $\alpha+\sigma$ | $\sigma$ | $\alpha$ | $\emptyset$ |
|---|---|---|---|---|---|---|---|---|
| $\alpha+\sigma+\tau$ | 0 | 0 | 0 | 0.01 | 0 | 0 | 0 | 0 |
| $\sigma+\tau$ | 0.01 | 0 | 0.02 | 0.01 | 0.02* | 0 | 0.02** | 0 |
| $\alpha+\tau$ | 0.02 | 0.09** | 0 | 0.01 | 0.01 | 0.08*** | 0 | 0 |
| $\tau$ | 0.13** | 0.05* | 0.02 | 0 | 0.13*** | 0.06** | 0.03** | 0 |
| $\alpha+\sigma$ | 0.13** | 0.14*** | 0.15*** | 0.14*** | 0 | 0 | 0 | 0 |
| $\sigma$ | 0.14** | 0.1** | 0.11** | 0.14*** | 0.02 | 0 | 0.03** | 0 |
| $\alpha$ | 0.12** | 0.2*** | 0.14*** | 0.18*** | 0.01 | 0.07*** | 0 | 0 |
| $\emptyset$ | 0.13** | 0.07** | 0.06** | 0.04 | 0.08*** | 0.03** | 0.03** | 0 |

| Type of model \ Type of data | $\alpha+\sigma+\tau$ | $\sigma+\tau$ | $\alpha+\tau$ | $\tau$ | $\alpha+\sigma$ | $\sigma$ | $\alpha$ | $\emptyset$ |
|---|---|---|---|---|---|---|---|---|
| $\alpha+\sigma+\tau$ | 0 | −0.01 | 0.01 | 0.01 | −0.06 | 0.01 | 0.01 | 0 |
| $\sigma+\tau$ | 0.05 | 0 | 0.04 | 0.01 | 0.08 | −0.01 | 0.02 | 0 |
| $\alpha+\tau$ | 0.03 | 0.08* | 0 | 0.01 | −0.1 | 0.06 | 0 | 0 |
| $\tau$ | 0.02 | 0.01 | 0.01 | 0 | −0.03 | 0.06 | 0.04 | 0 |
| $\alpha+\sigma$ | 0.04 | 0.04 | 0.06** | 0.06*** | 0 | −0.02 | 0 | 0 |
| $\sigma$ | 0.09 | 0.04 | 0.08** | 0.07*** | 0 | 0 | 0.04* | 0 |
| $\alpha$ | 0.05 | 0.13** | 0.06** | 0.08*** | −0.05 | 0.09* | 0 | 0 |
| $\emptyset$ | 0.18** | 0.04 | 0.04 | 0.02** | 0.05 | 0.02 | 0.02 | 0 |

b Latent field prediction.

| Type of model \ Type of data | $\alpha+\sigma+\tau$ | $\sigma+\tau$ | $\alpha+\tau$ | $\tau$ | $\alpha+\sigma$ | $\sigma$ | $\alpha$ | $\emptyset$ |
|---|---|---|---|---|---|---|---|---|
| $\alpha+\sigma+\tau$ | 0 | −606 | 186 | 2634 | −165 | 18 | 18 | −12 |
| $\sigma+\tau$ | −1916 | 0 | −1090 | 2038 | 103 | 5 | 114 | 3 |
| $\alpha+\tau$ | −666 | 1245 | 0 | 2398 | −59 | 43 | −5 | −11 |
| $\tau$ | 70 | 159 | −252 | 0 | 1370*** | 961*** | 443** | −31 |
| $\alpha+\sigma$ | 3887** | 5861** | 5362** | 6081** | 0 | 23 | 18 | 10 |
| $\sigma$ | 4994** | 5142** | 4238** | 8155** | 15 | 0 | 242* | 22 |
| $\alpha$ | 3361* | 6344** | 4417** | 7264** | −3 | 211* | 0 | −9 |
| $\emptyset$ | 6684** | 7183** | 6532** | 8250** | 1185*** | 658** | 472** | 0 |

c DIC.

Type of data

| Type of model | $\alpha+\sigma+\tau$ | $\sigma+\tau$ | $\alpha+\tau$ | $\tau$ | $\alpha+\sigma$ | $\sigma$ | $\alpha$ | $\emptyset$ |
|---|---|---|---|---|---|---|---|---|
| $\alpha+\sigma+\tau$ | 0 | 0 | 0.01 | 0.01 | 0.02 | 0 | 0 | 0 |
| $\sigma+\tau$ | 0.72 ** | 0 | 0.8 ** | 0 | 0.74 ** | 0 | 0.83 *** | 0 |
| $\alpha+\tau$ | 3.15 *** | 2.62 *** | 0 | 0 | 2.38 *** | 2.92 *** | 0 | 0 |
| $\tau$ | 0.78 ** | 0.01 | 0.82 ** | 0 | 0.78 ** | 0 | 0.84 *** | 0 |
| $\alpha+\sigma$ | 0.68 ** | 0.65 ** | 0.78 ** | 0.59 ** | 0 | 0 | 0 | 0 |
| $\sigma$ | 0.74 ** | 0.01 | 0.81 ** | 0.02 | 0.77 ** | 0 | 0.83 *** | 0 |
| $\alpha$ | 4.49 *** | 3.8 *** | 2.04 *** | 2.04 *** | 2.86 *** | 3.36 *** | 0 | 0 |
| $\emptyset$ | 0.77 ** | −0.01 | 0.79 ** | 0 | 0.77 ** | 0 | 0.78 *** | 0 |

d Log-range smoothing MSE.

Type of data

| Type of model | $\alpha+\sigma+\tau$ | $\sigma+\tau$ | $\alpha+\tau$ | $\tau$ | $\alpha+\sigma$ | $\sigma$ | $\alpha$ | $\emptyset$ |
|---|---|---|---|---|---|---|---|---|
| $\alpha+\sigma+\tau$ | 0 | 0 | 0.01 | 0 | 0 | 0 | 0 | 0 ** |
| $\sigma+\tau$ | 0.17 | 0 | 0.19 ** | 0 | 0.18 | 0 | 0.21 *** | 0 ** |
| $\alpha+\tau$ | 1.95 *** | 1.79 *** | 0 | 0 | 1.48 *** | 2.03 *** | 0 | 0 |
| $\tau$ | 0.66 ** | 0.88 *** | 0.01 | 0 | 0.68 ** | 0.91 ** | 0.01 | 0 |
| $\alpha+\sigma$ | 0.19 | 0.15 | 0.18 ** | 0.14* | 0 | 0 | 0 | 0 ** |
| $\sigma$ | 0.7 ** | 0.43 ** | 0.55 *** | 0.51 *** | 0.22 | 0 | 0.2 *** | 0 ** |
| $\alpha$ | 2.16 *** | 1.95 *** | 0.35 ** | 0.5 *** | 1.88 *** | 2.38 *** | 0 | 0 |
| $\emptyset$ | 0.82 ** | 0.92 *** | 0.01 | 0 | 0.85 ** | 0.88 ** | 0.01 | 0 |

e Log marginal variance of the latent field smoothing MSE.

Type of data

| Type of model | $\alpha+\sigma+\tau$ | $\sigma+\tau$ | $\alpha+\tau$ | $\tau$ | $\alpha+\sigma$ | $\sigma$ | $\alpha$ | $\emptyset$ |
|---|---|---|---|---|---|---|---|---|
| $\alpha+\sigma+\tau$ | 0 | 0 | 0 | 0 | 0 | 0 | 0 ** | 0 *** |
| $\sigma+\tau$ | 0 | 0 | 0 | 0 | 0 | 0 | 0 *** | 0 *** |
| $\alpha+\tau$ | 0 | 0 | 0 | 0 | 0 | 0 | 0 ** | 0 *** |
| $\tau$ | 0.09 | 0.02 | 0.01 | 0 | 0.08 *** | 0.03 *** | 0.01 *** | 0 *** |
| $\alpha+\sigma$ | 0.85 *** | 0.79 *** | 0.8 *** | 0.83 *** | 0 | 0 | 0 | 0 |
| $\sigma$ | 0.82 *** | 0.81 *** | 0.73 *** | 0.84 *** | 0 | 0 | 0 | 0 |
| $\alpha$ | 0.85 *** | 0.74 *** | 0.84 *** | 0.83 *** | 0 | 0 | 0 | 0 |
| $\emptyset$ | 0.84 *** | 0.88 *** | 0.85 *** | 0.88 *** | 0 | 0 | 0 ** | 0 |

f Log-variance of the noise smoothing MSE.

Legend: "right model" ■; "wrong model" ■; "over-modeling" ■; "under-modeling" ■

$***: p < 2e-16; **: p < 0.001; *: p < 0.01$

$\alpha$: (locally isotropic) nonstationary range. $\sigma$: nonstationary latent field variance. $\tau$: nonstationary noise variance. $\emptyset$: full stationarity.

*Read: table S6a shows the results for latent field smoothing. The second column summarizes the experiments made with data sets presenting nonstationary variance in the latent field and noise, as indicates its column name "$\sigma+\tau$". The reference, in green, is the "right model" with the same specification. The associated coefficient is therefore* 0*. Just below, we can see the coefficient associated with the model $\alpha+\tau$. Its brown color tells that it is a "wrong" model. It adds, in average,* 0.09 *to the smoothing MSE with respect to the "right" model. The three p-value stars indicate $p < 2e-16$.*

Table S7: Coverage of the parameters by 95% credibility intervals in the locally isotropic experiment.

| Type of model \ Type of data | $\alpha+\sigma+\tau$ | $\sigma+\tau$ | $\alpha+\tau$ | $\tau$ | $\alpha+\sigma$ | $\sigma$ | $\alpha$ | $\emptyset$ |
|---|---|---|---|---|---|---|---|---|
| $\alpha+\sigma+\tau$ | 1 | 1 | 1 | 0.92 | 1 | 0.95 | 0.95 | 0.89 |
| $\sigma+\tau$ | 0.46 | 1 | 0.36 | 0.97 | 0.45 | 0.97 | 0.29 | 0.81 |
| $\alpha+\tau$ | 0.16 | 0.07 | 1 | 0.93 | 0.02 | 0 | 1 | 0.86 |
| $\tau$ | 0.52 | 0.46 | 0 | 0.64 | 0.61 | 0.79 | 0.19 | 0.91 |
| $\alpha+\sigma$ | 0.98 | 0.98 | 1 | 1 | 0.98 | 1 | 1 | 0.88 |
| $\sigma$ | 0.52 | 0.88 | 0.43 | 0.85 | 0.53 | 0.93 | 0.33 | 0.88 |
| $\alpha$ | 0.25 | 0.02 | 0.83 | 0.71 | 0.03 | 0 | 1 | 0.91 |
| $\emptyset$ | 0.36 | 1 | 0.22 | 0.85 | 0.12 | 0.98 | 0.2 | 0.83 |

a Range intercept

| Type of model \ Type of data | $\alpha+\sigma+\tau$ | $\sigma+\tau$ | $\alpha+\tau$ | $\tau$ | $\alpha+\sigma$ | $\sigma$ | $\alpha$ | $\emptyset$ |
|---|---|---|---|---|---|---|---|---|
| $\alpha+\sigma+\tau$ | 0.93 | 0.91 | 1 | 1 | 0.88 | 1 | 1 | 1 |
| $\sigma+\tau$ | 0.8 | 0.9 | 0.98 | 1 | 0.78 | 0.97 | 1 | 1 |
| $\alpha+\tau$ | 0 | 0 | 1 | 1 | 0.02 | 0 | 1 | 0.98 |
| $\tau$ | 0.15 | 0.03 | 0.3 | 1 | 0.17 | 0 | 0.41 | 0.98 |
| $\alpha+\sigma$ | 0.75 | 0.85 | 1 | 1 | 0.88 | 0.92 | 1 | 1 |
| $\sigma$ | 0.8 | 0.89 | 1 | 1 | 0.9 | 0.97 | 1 | 1 |
| $\alpha$ | 0 | 0 | 0.02 | 0 | 0 | 0 | 1 | 1 |
| $\emptyset$ | 0.02 | 0 | 0.5 | 1 | 0.02 | 0 | 0.69 | 1 |

b Field variance intercept

| Type of model \ Type of data | $\alpha+\sigma+\tau$ | $\sigma+\tau$ | $\alpha+\tau$ | $\tau$ | $\alpha+\sigma$ | $\sigma$ | $\alpha$ | $\emptyset$ |
|---|---|---|---|---|---|---|---|---|
| $\alpha+\sigma+\tau$ | 0.97 | 0.95 | 0.9 | 1 | 1 | 1 | 1 | 1 |
| $\sigma+\tau$ | 0.86 | 0.93 | 0.8 | 0.98 | 1 | 1 | 1 | 1 |
| $\alpha+\tau$ | 0.97 | 1 | 0.82 | 0.98 | 1 | 1 | 1 | 1 |
| $\tau$ | 0.87 | 0.88 | 0.85 | 0.78 | 1 | 1 | 1 | 1 |
| $\alpha+\sigma$ | 0 | 0 | 0 | 0.02 | 0.98 | 1 | 1 | 1 |
| $\sigma$ | 0 | 0.02 | 0 | 0 | 1 | 1 | 0.93 | 1 |
| $\alpha$ | 0 | 0 | 0 | 0 | 0.83 | 0.86 | 1 | 1 |
| $\emptyset$ | 0 | 0 | 0 | 0 | 0.85 | 1 | 0.85 | 1 |

c Noise variance intercept

Type of data

| Type of model | $\alpha+\sigma+\tau$ | $\sigma+\tau$ | $\alpha+\tau$ | $\tau$ | $\alpha+\sigma$ | $\sigma$ | $\alpha$ | $\emptyset$ |
|---|---|---|---|---|---|---|---|---|
| $\alpha+\sigma+\tau$ | 1 | 0 | 1 | 0 | 1 | 0 | 1 | 0 |
| $\sigma+\tau$ | 0 | 1 | 0 | 1 | 0 | 1 | 0 | 1 |
| $\alpha+\tau$ | 0 | 0 | 1 | 0 | 0.02 | 0 | 1 | 0 |
| $\tau$ | 0 | 1 | 0 | 1 | 0 | 1 | 0 | 1 |
| $\alpha+\sigma$ | 0.46 | 0 | 0.26 | 0 | 1 | 0 | 1 | 0 |
| $\sigma$ | 0 | 1 | 0 | 1 | 0 | 1 | 0 | 1 |
| $\alpha$ | 0 | 0 | 0.25 | 0 | 0 | 0 | 1 | 0 |
| $\emptyset$ | 0 | 1 | 0 | 1 | 0 | 1 | 0 | 1 |

d Marginal variance of the range's PP

Type of data

| Type of model | $\alpha+\sigma+\tau$ | $\sigma+\tau$ | $\alpha+\tau$ | $\tau$ | $\alpha+\sigma$ | $\sigma$ | $\alpha$ | $\emptyset$ |
|---|---|---|---|---|---|---|---|---|
| $\alpha+\sigma+\tau$ | 0.98 | 0.88 | 0 | 0 | 0.88 | 0.93 | 0 | 0 |
| $\sigma+\tau$ | 0.34 | 0.98 | 0 | 0 | 0.32 | 0.95 | 0 | 0 |
| $\alpha+\tau$ | 0 | 0 | 1 | 1 | 0 | 0 | 1 | 1 |
| $\tau$ | 0 | 0 | 1 | 1 | 0 | 0 | 1 | 1 |
| $\alpha+\sigma$ | 0.68 | 0.92 | 0 | 0 | 0.91 | 0.95 | 0 | 0 |
| $\sigma$ | 0.2 | 0.4 | 0 | 0 | 0.52 | 0.88 | 0 | 0 |
| $\alpha$ | 0 | 0 | 1 | 1 | 0 | 0 | 1 | 1 |
| $\emptyset$ | 0 | 0 | 1 | 1 | 0 | 0 | 1 | 1 |

e Marginal variance of the field variance's PP

Type of data

| Type of model | $\alpha+\sigma+\tau$ | $\sigma+\tau$ | $\alpha+\tau$ | $\tau$ | $\alpha+\sigma$ | $\sigma$ | $\alpha$ | $\emptyset$ |
|---|---|---|---|---|---|---|---|---|
| $\alpha+\sigma+\tau$ | 1 | 1 | 0.98 | 0.98 | 0 | 0 | 0 | 0 |
| $\sigma+\tau$ | 1 | 1 | 1 | 1 | 0 | 0 | 0 | 0 |
| $\alpha+\tau$ | 1 | 0.98 | 1 | 1 | 0 | 0 | 0 | 0 |
| $\tau$ | 1 | 1 | 1 | 1 | 0 | 0 | 0 | 0 |
| $\alpha+\sigma$ | 0 | 0 | 0 | 0 | 1 | 1 | 1 | 1 |
| $\sigma$ | 0 | 0 | 0 | 0 | 1 | 1 | 1 | 1 |
| $\alpha$ | 0 | 0 | 0 | 0 | 1 | 1 | 1 | 1 |
| $\emptyset$ | 0 | 0 | 0 | 0 | 1 | 1 | 1 | 1 |

f Variance of the noise variance's PP

Legend: "right model" ■; "wrong model" ■; "over-modeling" ■; "under-modeling" ■

$*** : p < 2e-16; ** : p < 0.001; * : p < 0.01$

$\alpha$: (locally isotropic) nonstationary range. $\sigma$: nonstationary latent field variance. $\tau$: nonstationary noise variance. $\emptyset$: full stationarity.

*Read: table S7e shows the coverage of the log-variance of the PP controlling the latent field's variance. The sixth column is named "$\sigma$", which means that the data was generated with a latent field whose variance is nonstationary. We can see that the "right" model, in green, has a coverage of 88 %. Models who have a nonstationary field variance and other(s) type(s) of nonstationarity as well, in blue on the same column, have a satisfying coverage, but the "wrong" models (in brown) or the under-modeling stationary model (in gray) model have a 0% coverage.*

Table S8: Performance in anisotropic settings

a smooth field MSE (Type of model; Type of data)

| | $A$ | $\alpha$ | $\emptyset$ |
|---|---|---|---|
| $A$ | 0 | 0 | $0**$ |
| $\alpha$ | $0.01***$ | 0 | 0 |
| $\emptyset$ | $0.03***$ | $0.02***$ | 0 |

b pred field MSE (Type of model; Type of data)

| | $A$ | $\alpha$ | $\emptyset$ |
|---|---|---|---|
| $A$ | 0 | 0 | $0*$ |
| $\alpha$ | $0.04***$ | 0 | 0 |
| $\emptyset$ | $0.06***$ | $0.02***$ | 0 |

c log range det MSE (Type of model; Type of data)

| | $A$ | $\alpha$ | $\emptyset$ |
|---|---|---|---|
| $A$ | 0 | 0.02 | $0**$ |
| $\alpha$ | $0.5***$ | 0 | $0*$ |
| $\emptyset$ | $0.85***$ | $0.66***$ | 0 |

d log range aniso MSE (Type of model; Type of data)

| | $A$ | $\alpha$ | $\emptyset$ |
|---|---|---|---|
| $A$ | 0 | $0***$ | $0***$ |
| $\alpha$ | $0.69***$ | 0 | 0 |
| $\emptyset$ | $0.69***$ | 0 | 0 |

e DIC (Type of model; Type of data)

| | $A$ | $\alpha$ | $\emptyset$ |
|---|---|---|---|
| $A$ | 0 | 2 | $-16**$ |
| $\alpha$ | $218***$ | 0 | $-14**$ |
| $\emptyset$ | $612***$ | $354***$ | 0 |

Legend: "right model"; "over-modeling"; "under-modeling"

$***: p < 2e-16; **: p < 0.001; *: p < 0.05$

A: nonstationary anisotropic. $\alpha$: nonstationary isotropic. $\emptyset$: stationary

*Read: in table a, giving results for the smoothing MSE, the left-hand column corresponds to nonstationary anisotropic data. The "right" model is then the nonstationary anisotropic model, called "A", and it is the reference of the column, explaining that the upper-left coefficient is* 0 *and its color is green. The bottom-left coefficient is* 0.03*, meaning that using a stationary model instead of the "right" model will add* 0.03 *to the smoothing MSE. Its color is gray, indicating under-modeling. The three stars indicate that* $p < 0.001$.

Table S9: Coverage of the parameters by 95% credibility intervals in the anisotropic experiment.

a range intercept 1

| Model \ Type of data | $A$ | $\alpha$ | $\emptyset$ |
|---|---|---|---|
| $A$ | 1 | 0.91 | 0.91 |
| $\alpha$ | 0.68 | 0.93 | 0.93 |
| $\emptyset$ | 0 | 0.2 | 0.88 |

b range intercept 2

| Model \ Type of data | $A$ | $\alpha$ | $\emptyset$ |
|---|---|---|---|
| $A$ | 0.92 | 0.96 | 0.98 |
| $\alpha$ | 1 | 1 | 1 |
| $\emptyset$ | 1 | 1 | 1 |

c range intercept 3

| Model \ Type of data | $A$ | $\alpha$ | $\emptyset$ |
|---|---|---|---|
| $A$ | 0.92 | 0.96 | 0.98 |
| $\alpha$ | 1 | 1 | 1 |
| $\emptyset$ | 1 | 1 | 1 |

d range log within-site var 1

| Model \ Type of data | $A$ | $\alpha$ | $\emptyset$ |
|---|---|---|---|
| $A$ | 0.86 | 0.89 | 0 |
| $\alpha$ | 0.85 | 0.92 | 0 |
| $\emptyset$ | 0 | 0 | 1 |

e range log within-site var 2

| Model \ Type of data | $A$ | $\alpha$ | $\emptyset$ |
|---|---|---|---|
| $A$ | 0.92 | 1 | 1 |
| $\alpha$ | 1 | 1 | 1 |
| $\emptyset$ | 1 | 1 | 1 |

f range log within-site var 3

| Model \ Type of data | $A$ | $\alpha$ | $\emptyset$ |
|---|---|---|---|
| $A$ | 0.9 | 0.94 | 0.98 |
| $\alpha$ | 1 | 1 | 1 |
| $\emptyset$ | 1 | 1 | 1 |

g range log within-site var 4

| Model \ Type of data | $A$ | $\alpha$ | $\emptyset$ |
|---|---|---|---|
| $A$ | 0.71 | 0 | 0 |
| $\alpha$ | 0 | 1 | 1 |
| $\emptyset$ | 0 | 1 | 1 |

h range log within-site var 5

| Model \ Type of data | $A$ | $\alpha$ | $\emptyset$ |
|---|---|---|---|
| $A$ | 0.9 | 1 | 1 |
| $\alpha$ | 1 | 1 | 1 |
| $\emptyset$ | 1 | 1 | 1 |

i range log within-site var 6

| Model \ Type of data | $A$ | $\alpha$ | $\emptyset$ |
|---|---|---|---|
| $A$ | 0.67 | 0 | 0 |
| $\alpha$ | 0 | 1 | 1 |
| $\emptyset$ | 0 | 1 | 1 |

j marginal variance intercept

| Model \ Type of data | $A$ | $\alpha$ | $\emptyset$ |
|---|---|---|---|
| $A$ | 0.94 | 0.96 | 0.89 |
| $\alpha$ | 0.92 | 0.93 | 0.9 |
| $\emptyset$ | 0.65 | 0.67 | 0.95 |

k noise intercept

| Model \ Type of data | $A$ | $\alpha$ | $\emptyset$ |
|---|---|---|---|
| $A$ | 0.98 | 0.96 | 0.89 |
| $\alpha$ | 0.95 | 0.93 | 0.93 |
| $\emptyset$ | 0.6 | 0.7 | 0.9 |

Legend: "right model" ■; "over-modeling" ■; "under-modeling" ■

$* * * : p < 2e - 16; ** : p < 0.001; * : p < 0.05$

Read: *In table b, we can see that the second component of the range Intercept was covered by the 95% credibility intervals in 92% of the cases, when the data and the model are both nonstationary anisotropic. The green color tells us that this combination of model and data is "the right" formulation.*

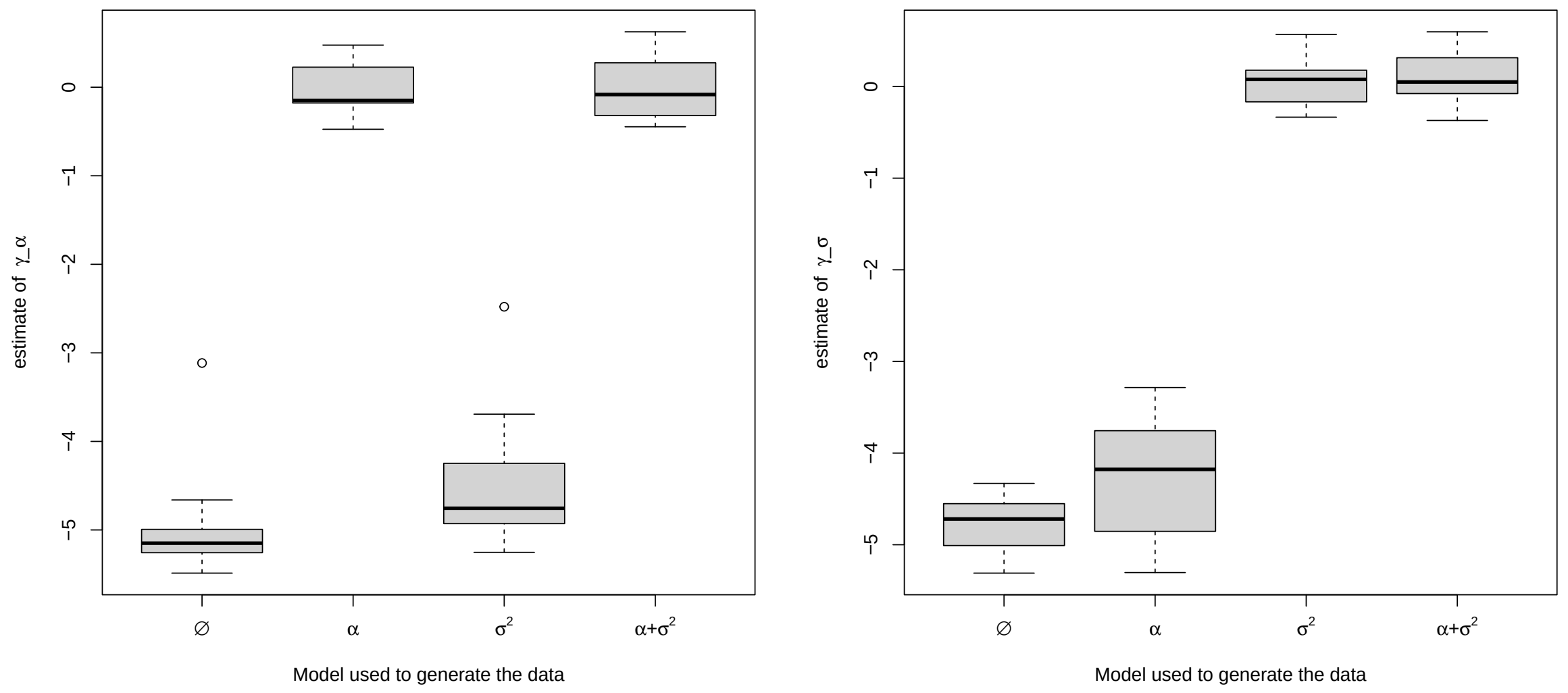

(a) Estimates of $\gamma_\alpha$, the log marginal variance for $W_\alpha(\mathcal{S})$

(b) Estimates of $\gamma_\sigma$, the log marginal variance for $W_\sigma(\mathcal{S})$

Figure S2: Estimates of the log marginal variance of $W_\alpha(\mathcal{S})$ and $W_\sigma(\mathcal{S})$ in the model $(\alpha + \sigma^2)$ following the type of the data

the log-variance of $W_\sigma(\mathcal{S})$ is high, etc.

## S7.5 In-run assessment of over-modeling by tracking MCMC samples

Assessing an overly refined model choice can be done informally done using trace plots who are generated automatically during the MCMC run, presented in Figures S3 and S4, where each curve corresponds to one Markov chain. If the data is generated from a stationary process, but the fitted model is nonstationary, $log(\gamma)$ drops to very small values as seen in Figures S3(a) and S4(a), effectively inducing a stationary latent field. If the range is locally isotropic, but nonstationary, we observe the behavior in Figure S3(b), where the log-variance of the range is high. In Figure S4, the first component, labeled "determinant", controls the variance of the determinant of the range ellipses, allowing them to inflate or deflate over space, such as the parameters in Figure S3. Two other parameters, labeled "anisotropy", allow the range ellipses to squash (with area preservation) and change orientation. In Fig-ure S4(b) only “determinant” is high. Hence, the ellipses may change their size, but not their shape, practically retrieving the locally isotropic covariance. However, here a simpler model is preferable. In Figure S4(c), all three components are high so the size and orientation of the ellipse varies over space.

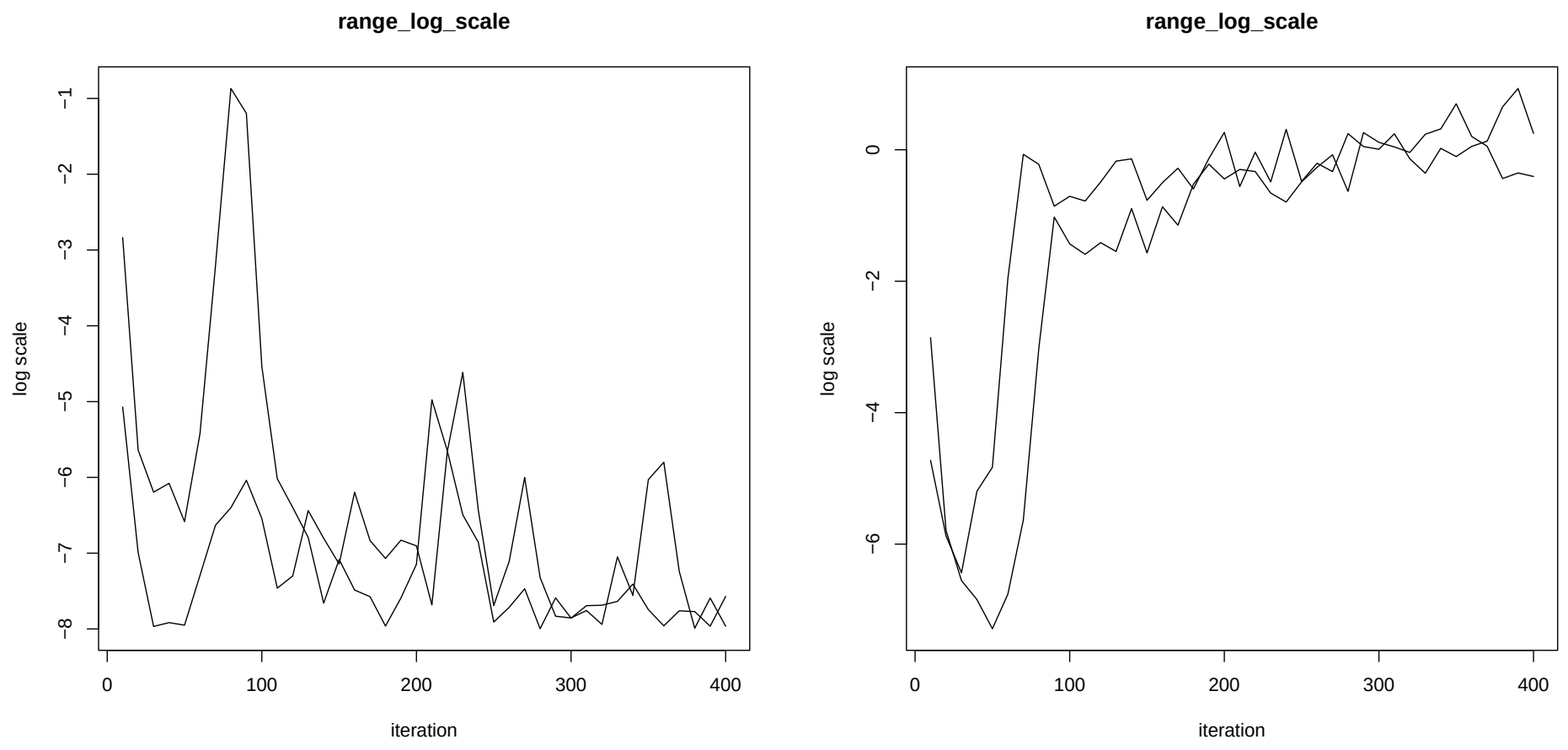

(a) Data with stationary range ($\gamma_\alpha = 0$)

(b) Data with nonstationary range ($\gamma_\alpha = 1$)

Figure S3: MCMC samples of the range PP log-variance parameter $\log(\gamma_\alpha)$ following the type of range model used in the data.

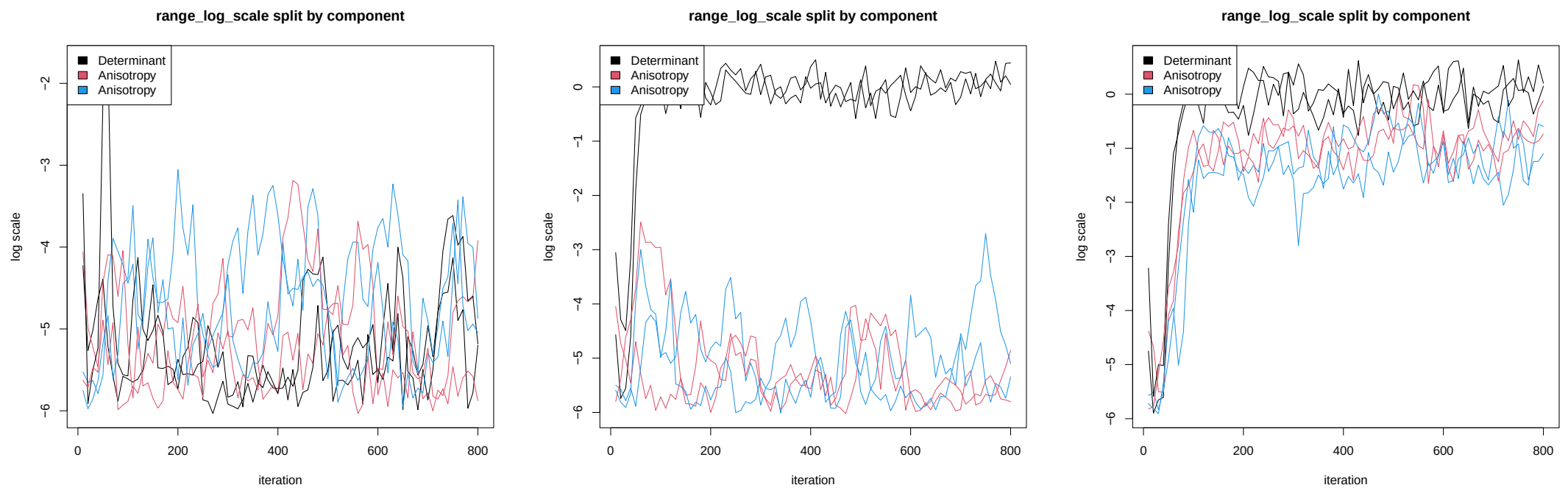

(a) Data with stationary isotropic range ($\log(\gamma_A) = \text{diag}(0, 0, 0)$).

(b) Data with nonstationary locally isotropic range ($\log(\gamma_A) = \text{diag}(1, 0, 0)$).

(c) Data with nonstationary locally isotropic range ($\log(\gamma_A) = \text{diag}(1, 1, 1)$).

Figure S4: MCMC samples of the diagonal coefficients of the range PP log-variance parameter $\log(\gamma_A)$, following the type of range model used in the data.

# S8 Comparison with state-of-the-art methods

This section expounds further details concerning the experiments presented in Section 5.2 of the main body of the paper. The experiments are conducted on a computational cluster with 12 cores for each run, mimicking the settings of a laptop. In particular we compare with the following methods:

- INLA-SPDE with the R package INLA (Rue et al., 2009; Lindgren et al., 2015; Krainski et al., 2018).
- Local Gaussian Processes using R packages laGP, liGP, and hetGP (Gramacy and Apley, 2015; Gramacy, 2016; Binois et al., 2018).
- Vecchia-Approximated Deep Gaussian Processes with the R package deepGP (Sauer et al., 2023b,a).

The aim here is not to trick any method into failing, so the competitors receive any tip available. However, those data sets are simulated following the model of our article, so our implementation benefits from a "home advantage", INLA too in some extent. The results must therefore be taken with a pinch of salt. Two types of nonstationary data sets are tested.

## S8.1 Heteroskedastic noise

The results of this experiment are presented in Table 1 of the main body of the article. We simulate spatial data sets with heteroskedastic noise using a simplified version of (10),

$$z(\mathcal{S}) = w(\mathcal{S}) + \epsilon(\mathcal{S}), \quad \epsilon(\mathcal{S}) \sim N(0, \text{diag}(\tau^2(\mathcal{S}))) \,, \tag{S11}$$

where $\log \tau^2(\mathcal{S}) = -1_{|\mathcal{S}|} + W_\tau(\mathcal{S})$, $1_{|\mathcal{S}|}$ is the $\mathcal{S} \times 1$ vector of ones with $|\mathcal{S}|$ denoting the number of locations in $\mathcal{S}$, $w(\mathcal{S})$ and $W_\tau(\mathcal{S})$ represent realizations from zero-centered Gaussian

processes over locations $\mathcal{S}$. Note that while we accommodate multiple observations per spatial site in (10) using the matching matrix $M$, we reduce here to the case $M = I_n$, allowing to index $z$ on $\mathcal{S}$. We also remove the fixed effects to focus on nonstationary random effects estimation. We generate data by executing the following steps.

- We draw and fix $11,000$ spatial locations uniformly inside a unit square.
- We use the R package GpGp (Guinness, 2018) to simulate the latent spatial process at the spatial locations, $w(\mathcal{S})$, using a Gaussian process specified by a Matérn covariance kernel with range 0.025, marginal variance 1.0 and smoothness 1.5.
- We simulate another latent process, $W_\tau(\mathcal{S})$, from a Gaussian process specified by another Matérn covariance kernel with range 0.2 (roughly ten times the range of $w$), marginal variance 1, and smoothness 1.5.
- We set a noise variance with pseudo-intercept $-1$ as $\tau^2(\mathcal{S}) = \exp(W_\tau(\mathcal{S}) - 1)$.
- We draw $\epsilon \sim \mathcal{N}(0, \text{diag}(\tau^2(\mathcal{S})))$ and then set $z(\mathcal{S}) = w(\mathcal{S}) + \epsilon(\mathcal{S})$.
- We retain the first $10,000$ elements of $z(\mathcal{S})$ as observations to train the model (training data) and hold out the remaining $1,000$ points as test data set.

The models that we fit are specified as follows.

- The nonstationary NNGP in (10) is specified using the Matérn smoothness parameters set to their true values and a heteroskedastic noise model. We use 30 Predictive Process spatial basis functions to approximate $W_\tau(\mathcal{S})$ and estimate the marginal variance of the process $W_\tau(\mathcal{S})$.
- We followed the SPDE formulation prescribed for INLA, where the smoothness parameter for $w(s)$ is set to 1 instead of 1.5. Here, too, we work with the heteroskedastic noise model using the Predictive Process with 30 spatial basis functions. Unlike in the nonstationary NNGP, the "true" prior precision for $W_\tau(\mathcal{S})$ is passed to INLA. This produces the following hierarchical model for spatial locations $s$ in any finite set.

$$z(s) = w(s) + \epsilon(s) \ , \quad \epsilon(s) \sim \mathcal{N}(0, \exp(\beta_\tau^0 + W_\tau(s)))$$

$$w(s) \,|\, \theta \sim \mathcal{N}(0, \Sigma SPDE(\theta)) \ ; \quad \theta \sim INLAprior(\theta) \ ;$$

$$\beta_\tau^0 \sim \mathcal{N}(0,1) \ , \quad W_\tau(\mathcal{S}) \sim \mathcal{N}(0, B(\mathcal{S})^T B(\mathcal{S})) \ ,$$

where $\Sigma SPDE$ is the covariance implicitly induced by the SPDE, $INLAprior(\cdot)$ is the default prior used by INLA, $\beta_\tau^0$ is the intercept of the log-noise, and $B(\mathcal{S})$ is the PP basis giving a degenerate covariance $B(\mathcal{S})^T B(\mathcal{S})$.

- The local GPs are formulated using a heteroskedastic noise model with the underlying GP range used to simulate the data. The prediction at site $s$ is obtained through

$$\tilde{w}(s) \sim \mathcal{N}(E(w(s)|w(N(s))),\ Var(w(s)|w(N(s)))),$$

  $N(s)$ being neighboring locations. This prediction is computed using the "true" range parameter $\theta$ and a local nugget estimated using maximum likelihood.

- The deep GPs had depth 2. Note that the current implementation of deepGP does not accommodate heteroskedasticity in the noise variable. The complexity of the deep GP makes it difficult to lay out using a few equations.

We reported the following eight criteria.

- The smoothing mean squared error (MSE) defined as

$$\frac{1}{10000} \sum_{i=1}^{10000} (w(s_i) - E(w(s_i) \,|\, z_{1:10000}, \text{model}))^2 \ ,$$

  where $w(s_i)$ are the values of the latent process at the locations included in the training data.

- The predictive MSE

$$\frac{1}{1000}\sum_{i=10001}^{11000}(w(s_i) - E(w(s_i)\,|\,z_{1:10000}, \text{model}))^2 ,$$

which is the MSE computed over the hold out locations.

- The MSE for the noise variance retrieval,

$$\frac{1}{10000}\sum_{i=1}^{10000}(\log(\tau(s_i)) - E(\log(\tau(s_i))\,|\,z_{1:10000}, \text{model}))^2 ,$$

is calculated for nonstationary NNGP, local GP and INLA-SPDE. Deep GP is excluded from this criterion as it does not offer heteroskedastic noise retrieval.

- Empirical Log Pointwise Predictive Distribution scores (elpd) from Vehtari et al. (2017) at observed locations. The approach is the same as in the real data analysis in Section S9.1. This criterion, like the smoothing MSE, may be prone to over-fitting.

- Elpd at predicted locations.

- Coverage intervals at observed locations, whose rationale is explained in Section S9.1 along with the elpd.

- Coverage intervals at predicted locations.

- The running time for each method.

We repeat the above experiment over 50 replicated datasets. The results from each analysis are averaged and presented in Table 1 of the main body. First, we note that all methods deliver coverage intervals above the 0.95 target for observed data. The intervals are too wide, or, in other words, the models are being a little bit too pessimistic when smoothing. We could not explain this issue. The intervals for unobserved locations are satisfying for all the methods.

- Local GPs is the fastest of the methods that we considered and performs commendably in retrieving the "true" parameters, but high MSEs compared to the other methods.
- INLA demonstrates excellent retrieval and fast running time, although its running time of 13 minutes in average is an order of magnitude slower than local GP's.
- Nonstationary NNGPs' performance are very similar to INLA's, for, however, a slightly longer running time of 23 minutes in average.
- While deep GPs enjoy good predictive performances for the latent field, their predictive performance in terms of MSE and elpd are not as good as INLA's and nonstationary NNGPs'. In addition, they demand significantly more computational resources; even long runs with over $25,000$ iterations led to dubious mixing of their MCMC chains.

## S8.2 Nonstationary range

Our next simulation experiment concerns spatial data analysis with nonstationary range parameters. The results of this experiment are presented in Table 2 of the main body of the article. We now generate data from the model

$$z(\mathcal{S}) = w(\mathcal{S}) + \epsilon(\mathcal{S}), \quad \epsilon(\mathcal{S}) \sim N(0, \tau^2 I_{|\mathcal{S}|}) , \tag{S12}$$

where $|\mathcal{S}|$ is the number of locations in $\mathcal{S}$ and $w(\mathcal{S})$ is the realization over the locations in $\mathcal{S}$ from a nonstationary Gaussian process with the covariance kernel (5). Like in the previous experiment, we present a simplified version of the model presented in Equation (10), removing $X$ and allowing indexation of $z$ on $\mathcal{S}$. More specifically, we model $\log(\alpha(\mathcal{S})) = -3.7 * 1_{|\mathcal{S}|} + W_\alpha(\mathcal{S})$, where $W_\alpha(\mathcal{S})$ denotes the realizations from a zero-centered Gaussian process with a Matérn spatial covariance with range 0.2, marginal variance 1 and smoothness 1.5. The data is generated by executing the following steps.

- We draw and fix $11,000$ spatial locations uniformly in a unit square.

- We simulate a latent process $W_\alpha(\mathcal{S})$ with range 0.2, marginal variance 1, and smoothness 1.5.

- We define the spatial field of local range parameters as $\alpha(\mathcal{S}) = \exp(W_\alpha(\mathcal{S}) - 3.7)$, where $-3.7$ is a pseudo-intercept with $\exp(-3.7) = 0.025$, which corresponds to the spatial range in the previous experiment. This implies that the latent process $w(s)$ has a range much smaller than $W_\alpha(s)$.

- We simulate the latent Matérn-Paciorek process realizations $w(\mathcal{S})$ with the spatial field of the range $\alpha(\mathcal{S})$, marginal variance 1, and smoothness 1.5 using our R package Bidart.

- We draw $\epsilon(\mathcal{S}) \overset{iid}{\sim} \mathcal{N}(0, \tau^2 I_{|\mathcal{S}|})$, where $\tau^2 = \exp(-1)$ so that $-1$ acts as the intercept for the noise log-variance as in the previous experiment.

- We set the observations as $z(\mathcal{S}) = w(\mathcal{S}) + \epsilon(\mathcal{S})$.

- We retain the the first $10,000$ elements of $z(\mathcal{S})$ as observations to train the model and hold out the remaining $1,000$ to comprise the test data.

The model formulations and the comparison criteria metrics are analogous, *mutatis mutandis*, to those in Section S8.1. The noise variance is replaced by the process range whenever applicable. In the formulation of the INLA model, the spatially variable parameter now goes into $\Sigma SPDE(\cdot)$. Local GPs are provided the "true" data generating noise variance and process smoothness but must now guess the local range parameter. We also adjusted the range parameters in INLA to account for the smoothness difference with $\nu_{INLA} = 1$ and $\nu_{NNGP} = 1.5$. Following Lindgren et al. (2011) we use effective range, which is the distance $\rho$ where the correlation drops to 0.1, as the premise for converting $\log(\kappa_{INLA})$ into $\log(\alpha_{NNGP})$. In particular, we equate the effective range for nonstationary NNGP and INLA, i.e. $\rho_{NNGP} = \rho_{INLA}$ and $\alpha_{NNGP}\sqrt{8\nu_{NNGP}} = \sqrt{8\nu_{INLA}}/\kappa_{INLA}$, which yields

$$\begin{aligned}\log(\alpha_{NNGP}) + 0.5 \times \log(8\nu_{NNGP}) &= -\log(\kappa_{INLA}) + 0.5 \times \log(8\nu_{INLA}) \\ &= -\log(\kappa_{INLA}) + 0.5 \times (\log(\nu_{INLA}) - \log(\nu_{NNGP})) \\ &= -\log(\kappa_{INLA}) + 0.5 \times (\log(1) - \log(1.5)) \\ &= -\log(\kappa_{INLA}) - 0.5 \times \log(1.5).\end{aligned}$$

We remark that had we specified $\nu_{INLA} = \nu_{NNGP}$, we would obtain $\alpha = 1/\kappa$, which is coherent with the parametrization of Lindgren and Rue (2015).

Local GPs use a squared exponential kernel with a length scale $l$, $K_{sqexp}(d, l) = \exp(-d^2/l^2)$, where $d$ is the Euclidean distance. We approximate this with a Matérn function with large smoothness $\nu_{large}$, $K_{sqexp}(d, l) \approx K_{Matern}(d, \alpha_{large} = l/2\sqrt{\nu}, smoothness = \nu_{large})$. This relationship implies $\log(\alpha_{large}) = \log(l) - \log(2) - 0.5 \times \log(\nu_{large})$. Applying the same reasoning as for INLA we obtain $\log(\alpha_{NNGP}) = \log(\alpha_{large}) + 0.5 \times (\log(\nu_{large}) - \log(\nu_{NNGP}))$. Plugging in, we obtain

$$\begin{aligned}\log(\alpha_{NNGP}) &= \log(l) - \log(2) - 0.5 \times \log(\nu_{\text{large}}) + 0.5 \times (\log(\nu_{\text{large}}) - \log(\nu_{NNGP})) \\ &= \log(l) - \log(2) - 0.5 \times \log(\nu_{NNGP}).\end{aligned}$$

The above approximations imply that the results concerning the retrieval of the range parameters should be considered with some caution. As in Section S8.1, the results in Table 2 are averaged over 50 replicates of the data.

A brief summary of observations from our experiments is provided in the following. Like in the previous experiment, all methods are too pessimistic with coverage intervals for observed data, while being very adequate with the intervals at unobserved locations.

- The local GP range estimation appears somwehat erratic. The range maps appear unclear and do not seem to capture the "true" range. Nevertheless, the method delivers inference expeditiously and performs adequately well in smoothing and predicting the latent field $w$.

- Deep GPs provide competitive smoothing and predictions in terms of elpd or MSE. However, the needed computational power remains daunting, and the convergence of the chains is dubious, which is problematic for a MCMC-based method.
- INLA is fast and accurate.
- Nonstationary NNGPs take somewhat longer than INLA to run, but excels in terms of the MSE and elpd based metrics. This may, in part, be attributed to the advantage over INLA resulting from the latter's specification of the smoothness parameter.

# S9 The log point-wise predictive density

## S9.1 Empirical logarithmic pointwise predictive density and intervals

Following (Vehtari et al., 2017) we use the empirical logarithmic point-wise predictive density, $elpd()$, as a score reflecting a model's inferential performance. This is defined as

$$elpd(z_{test}) = \sum_{i=1}^{k} \log(p(z_{test}^i \mid z_{obs})) = \sum_{i=1}^{k} \log\left(\int p(z_{test}^i \mid x)p(x \mid z_{obs})dx\right) , \tag{S13}$$

where $x$ represents all the parameters in the hierarchical model (9), including $w(\mathcal{S})$. If the test locations are the same as the observed locations, then $p(z_{obs} \mid x) = p(z_{obs} \mid w(\mathcal{S}), \tau^2, \beta)$ corresponds to the likelihood in $(10(a))$. The integral is typically computed using MCMC samples from the chain. This "training error" is prone to over-fitting.

If the test locations $\mathcal{S}_{test}$ are new with $\mathcal{S}_{test} \cap \mathcal{S} = \emptyset$, then one can write for the observations

$z_{test}$ done at the new sites

$$p(z_{test}\,|\,x) = \int p(z_{test}\,|\,w(\mathcal{S}_{test}), \tau^2_{test}, x)p(w(\mathcal{S}_{test}), \tau^2_{test}\,|\,x)dw(\mathcal{S}_{test})d\tau^2_{test}, \tag{S14}$$

$\tau^2_{test}$ being the noise variance at the new observations, and $w(\mathcal{S}_{test})$ the latent field at the new locations, respectively. Plugging (S14) into (S13) produces

$$elpd(z_{test}) = \sum_{i=1}^{k} \log\left(\int p(z_{test}\,|\,w(\mathcal{S}_{test}), \tau^2_{test}, x)p(w(\mathcal{S}_{test}), \tau^2_{test}\,|\,x)p(x\,|\,z_{obs})dx dw_{test} d\tau^2_{test}\right) \tag{S15}$$

and we simplify the likelihood $p(z_{test}\,|\,w(\mathcal{S}_{test}), \tau^2_{test}, x) = p(z_{test}\,|\,w(\mathcal{S}_{test}), \tau^2_{test}, \beta)$ by introducing a matching matrix between $\mathcal{S}_{test}$ and its observations like in 3.1, and the predictive density $p(w(\mathcal{S}_{test}), \tau^2_{test}\,|\,x) = p(w(\mathcal{S}_{test})\,|\,x)p(\tau^2_{test}\,|\,x)$ using (9) and (10). For each drawn sample of $x \sim p(x\,|\,z_{obs})$ using MCMC, we draw one instance of $w(\mathcal{S}_{test})$ and $\tau^2_{test}$ from $p(w(\mathcal{S}_{test}), \tau^2_{test}\,|\,x)$ given each sampled value of $x$ to compute the Monte Carlo estimate of (S15).

Local GPs and deep GPs output a latent conditional expectation $\bar{\mu}$, a latent conditional variance $\bar{\sigma}^2$, and a noise variance $\tau^2$. The predictive distribution is $\mathcal{N}(\bar{\mu}, \bar{\sigma}^2 + \bar{\tau}^2)$.

We use the same scheme for INLA. When an empirical Bayes method is used, the predictive distribution is Gaussian and exact computation is possible. This is a simplification for the stationary model that integrates over several values for the SPDE parameters.

Coverage intervals can be derived from the predictive density.

## S9.2 Train and test partition for NDVI data analysis.

In the illustration on NDVI data of Section 5.3, we have $1,020,000$ observations over $138,146$ distinct spatial coordinates. Figure S5 shows the partition of the data into training and test data, which is obtained by the following method.

- A k-means algorithm with 6,000 clusters is applied on the 138,146 distinct spatial

coordinates. Subsequently, 500 clusters are removed from the data set. These removed clusters are selected using a max-min heuristic (Guinness, 2018) so that the clusters spread out over the space. These are the red lumps seen in Figure S5.

- The 500 centers of the clusters, indicated by blue in Figure S5, are kept as a validation set. This set aims to allow for mid-range prediction.

- We remove 2,000 locations from the remaining coordinates using max-min heuristics. These points are viewed as a pseudo leave-one-out (LOO) validation set. Here, prediction at close range is evaluated. They are the yellow dots.

- The remaining 124,575 distinct locations, indicated in black in Figure S5, are used as a training set and comprise a total of 920,358 observations.

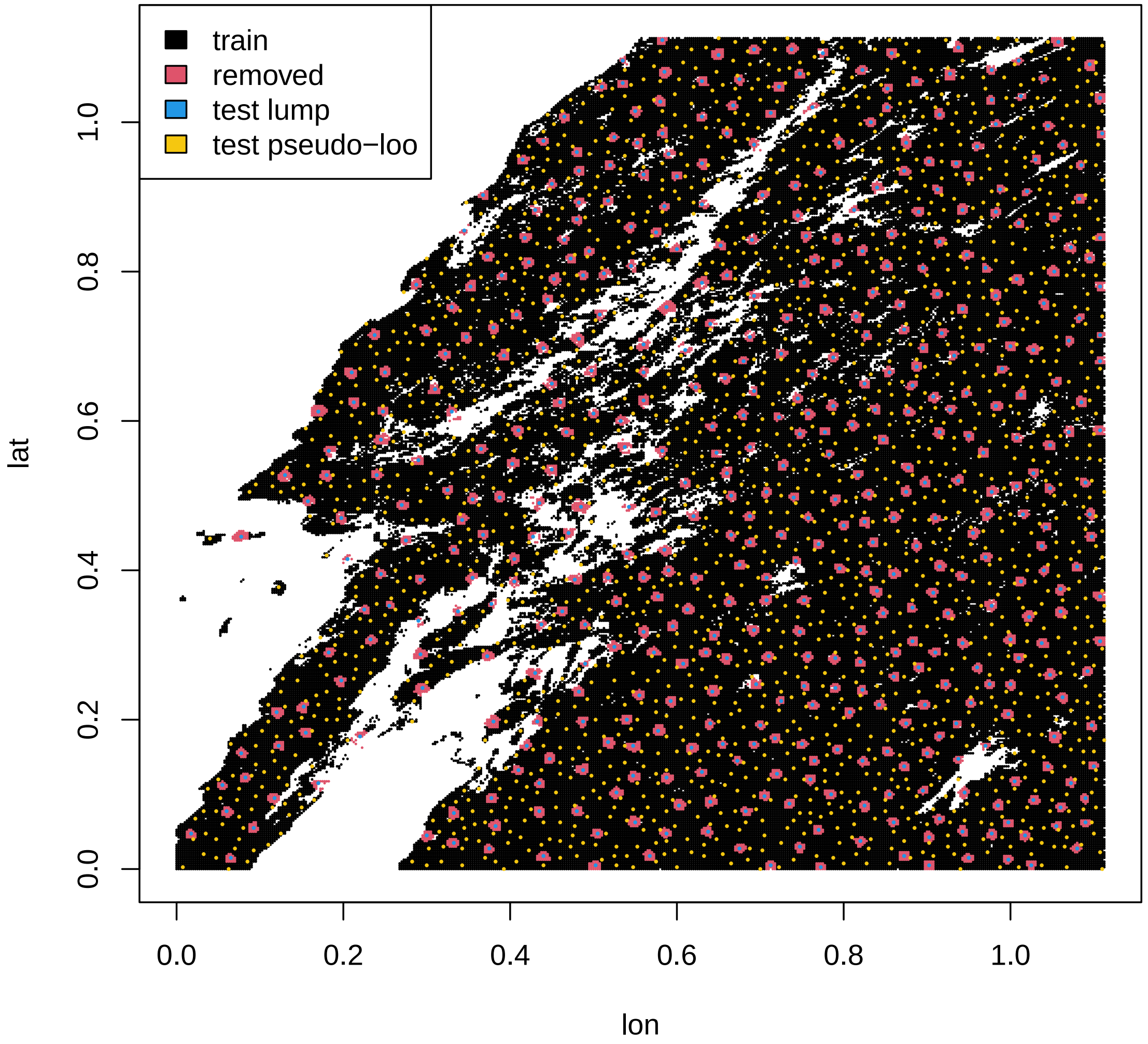

Figure S5: Train-test split for NDVI data